\documentclass[preprint,floatfix] {revtex4} 
\usepackage{graphicx}
\usepackage{subfigure}
\usepackage{xcolor}
\usepackage{amsmath}
\usepackage{enumerate}
\usepackage{tabularx}

\newenvironment{rcases}
  {\left.\begin{aligned}}
  {\end{aligned}\right\rbrace}

\usepackage{color, soul}
\definecolor{darkblue}{rgb}{0,0,0.5}
\setulcolor{darkblue}

\begin{document}

\title{Information entropic measures of a quantum harmonic oscillator in symmetric and asymmetric confinement 
within an impenetrable box}

\author{Abhisek Ghosal, Neetik Mukherjee and Amlan K.~Roy}
\altaffiliation{Corresponding author. Email: akroy@iiserkol.ac.in, akroy6k@gmail.com}
\affiliation{Department of Chemical Sciences\\
Indian Institute of Science Education and Research (IISER) Kolkata, 
Mohanpur-741246, WB, India}

\begin{abstract}
 Information-based uncertainty measures like Shannon entropy, Onicescu energy and Fisher information (in position and 
 momentum space) are employed to understand the effect of \emph{symmetric and asymmetric} confinement in a quantum 
 harmonic oscillator. Also, the transformation of Hamiltonian into a dimensionless form gives an idea of the composite 
 effect of force constant and confinement length ($x_c$). In symmetric case, a wide range of $x_{c}$ has 
 been taken up, whereas asymmetric confinement is dealt by shifting the minimum of potential from origin keeping box length 
 and boundary fixed. Eigenvalues and eigenvectors for these systems are obtained quite accurately via an imaginary time 
 propagation scheme. For asymmetric confinement, a variation-induced exact diagonalization procedure is also introduced, 
 which produces very high-quality results. One finds that, in symmetric confinement, after a certain characteristic 
 $x_{c}$, all these 
 properties converge to respective values of free harmonic oscillator. In asymmetric situation, excited-state 
 energies always pass through a maximum. For this potential, the classical turning-point decreases, whereas well depth 
 increases with the strength of asymmetry. Study of these uncertainty measures reveals that, localization increases 
 with an increase of asymmetric parameter. 


\vspace{3mm}

\end{abstract}
\maketitle

\section{Introduction}
In recent years, instancy in studying spatially confined quantum systems has increased momentously. In such small
spatial dimensions, they exhibit many fascinating physical and chemical phenomena, in contrast to their corresponding
\emph{free unconfined} counterparts. This occurs mostly due to their complex energy spectra. These 
have potential applications in a wide range of problems, e.g., cell-model of liquid state, high-pressure physics, 
study of impurities in semiconductor materials, matrix isolated molecules, endohedral complexes of fullerenes, 
zeolites cages, helium droplets, nanobubbles, etc.  Recent progress in nanotechnology has also inspired extensive 
research activity to explore and understand confined quantum systems (on a scale comparable to their de Broglie wave 
length). Their unique properties have been realized in a large array of quantum systems such as quantum wells, 
quantum wires, quantum dots as well as nanosized circuits (as in a quantum computer), and so forth, employing a 
wide variety of confining potentials. The literature is quite vast; interested reader is referred to a special 
issue and a book \cite{sabin09, sen14}, and references therein. 

A significant amount of confinement work exists for model systems such as particle in a box (PIB), harmonic oscillator, 
as well as real systems like H, He and other atoms; H$_2^+$, H$_2$ and other molecules. At this
point, it may be worthwhile to mention a few theoretical methods \cite{vawter68,vawter73,navarro80,fernandez81,
arteca83,taseli93,vargas96,sinha99,aquino01,campoy02,montgomery07,montgomery10,roy15} employed for a 1D quantum 
harmonic oscillator (QHO) enclosed inside an impenetrable box, which is the subject matter of our current work. Some 
prominent ones are: a semi-classical WKB approximation \cite{vawter68, sinha99}, a series analytical solution 
\cite{vawter73}, perturbative, asymptotic and Pad{\'e} approximant solutions \cite{navarro80}, diagonal hypervirial 
\cite{fernandez81}, hypervirial perturbative \cite{arteca83}, Rayleigh-Ritz variation method with trigonometric 
basis functions \cite{taseli93}, numerical method \cite{vargas96}, perturbation method \cite{aquino01}, 
power-series expansion \cite{campoy02}, exact \cite{montgomery07, montgomery10} as well as highly accurate 
power-series solution \cite{montgomery10}, imaginary-time propagation (ITP) technique \cite{roy15}, etc. While 
most of the works deal with effect of boundary on energy levels, several other important properties such as 
dipole moment \cite{aquino01}, Einstein's A, B coefficients \cite{aquino01, campoy02}, magnetic effects 
\cite{li1991, buchholz2006}, effect of 
confinement size on non-classical statistical properties of coherent states \cite{harouni2008} under high pressure, 
etc., were also considered. Most of these works focus largely on \emph{symmetrically} confined 
harmonic oscillator (SCHO); while the \emph{asymmetrically} confined harmonic oscillator (ACHO) situation is treated 
only in a few occasions (such as \cite{campoy02, roy15}).  

It is well-known that, information entropy (IE)-based uncertainty measures, such as, $S_x+S_p \geq 1+\ln \pi$, with $S_x$, 
$S_p$ denoting Shannon entropies in position ($x$) and momentum ($p$) space, provide more rigorous, stronger 
bound than conventional uncertainty product, $\Delta x  \Delta p \geq \frac{\hbar}{2}$ (symbols have usual 
significance). In quantum mechanics, $x$ and $p$ spaces are connected through uncertainty relation.  
Localization in $x$ space leads to delocalization in $p$ space and \emph{vice versa}. In case of IE, 
measurements are carried out in both $x$ and $p$ space. The most striking point in this respect, 
however, is that extent of localization is not exactly same as delocalization in $p$ space and 
\emph{vice versa}. Thus an inspection of IE in composite space gives a more complete information of net 
localization-delocalization in a quantum system. 

In the last decade, much light has been shed on the topic of IE in a multitude of systems. This continues 
to go unabated, as evidenced by a large volume of literature available on this topic. For example, information
measures, especially, Shannon entropy ($S$) and Fisher information ($I$) in a decent number of physically, 
chemically important potentials have
been reported lately. Some notable ones are: P\"oschl-Teller \cite{sun2013quantum}, Rosen-Morse 
\cite{sun2013quantum1}, squared tangent well \cite{dong2014}, hyperbolic \cite{valencia2015}, infinite circular 
well \cite{song2015}, hyperbolic double-well (DW) \cite{sun2015} potential, position-dependent mass Schr\"odinger systems 
\citep{falaye16,serrano16,hua15} etc. In a recent publication 
\citep{mukherjee15}, two of the present authors employed entropy measures like $S$, $I$, Onicescu 
energy ($E$) and Onicescu-Shannon information ($OS$) to analyze competing effect of localization-delocalization in 
a \emph{symmetric} DW potential, represented by, $V(x)= \alpha x^4 + \beta x^2 + \frac{\beta^2}{4 \alpha}$. 
It was found that quasi-degeneracy exists for certain values of parameters $\alpha, \beta$. Further, it was realized
that, while traditional uncertainty relation and $I$ were unable to explain such dual effects, measures like $S$, 
$E$ were quite successful. Such measures in a 1D ``Landau" system undergoing phase transition, have been studied 
recently \cite{song16}. In an analogous work \cite{mukherjee16}, oscillation of a particle between larger and 
smaller wells were followed through information analysis in an \emph{asymmetric} DW potential, given by, 
$V(x)=\alpha x^4- \beta x^2 + \gamma x$. In this case, it was possible to frame some simple rules to predict 
quasi-degeneracy occurring only for some characteristic values of parameters present in potential. 

A vast majority of IE-related works, referred above or otherwise, deal with a respective \emph{free or unconfined} 
system. Such reports on quantum \emph{confined} systems have been rather scarce and hence, it would be highly 
desirable to inspect these quantities in an enclosed system as they may open up some new windows to explore. In a 
recent publication \cite{laguna2014}, $S$ and traditional uncertainty relations were calculated and contrasted 
(significant differences were found in their behavior) in a SCHO model in $x$, $p$ and phase space. In this
follow-up, we wish to extend our previous IE analysis to two celebrated model \emph{confined quantum} systems, 
\emph{viz.,} SCHO and ACHO inside an impenetrable 1D box. Moreover, we have transformed our original Hamiltonian 
into a dimensionless form \citep{patil07} to make the results more universal and appealing from the perspectives of 
experimentalists \citep{zawadzki87, buttiker88}. As it turns out, this leads to a dimensionless parameter 
($\eta = \frac{mkx_c^4}{\hbar^2}$), which depends on the product of force constant and forth power of 
confinement length; this is derived later in Sec.~III. Clearly, the variation of $x_c$ predominates over $k$. 
Thus, at first, we examine variation of energy as well as $S, I$, $E$ in a SCHO for small, intermediate, large 
regions of $\eta$ and $x_c$, in both $x$ and $p$ space. Next we monitor analogous changes in an ACHO,
by varying $\eta$ and shifting the potential minimum ($d_m$), while keeping the box length stationary. 
As mentioned earlier, the only IE analysis for a CHO, to the best of our knowledge, is $S$
in case of SCHO. No such attempts are known for $I$, $E$, however. And so far, no work has been reported for analogous 
entropic analysis in an ACHO. So the present communication aims to provide a more satisfactory picture of 
localization of bound stationary states of an enclosed QHO in $x$ space and delocalization of same in $p$ space,
and \emph{vice versa}. Additionally, classical turning points are calculated for ACHO as functions of $d_m$ invoking 
the well-known semi-classical concept $|V(x)-\epsilon_n(d_m)|=0$ \cite{griffiths2004}, which can throw some 
light on the localization of particle in $x$ space. 

In order to facilitate further understanding, companion calculations are performed for phase-space area ($A_n$),
defined as, 
\begin{equation} 
A_n = \int \sqrt{ (V(x)-\epsilon_n)} dx, 
\end{equation}
as a function of $x_c$ in SCHO and $d_m$ in case of an ACHO. While IE gives a purely quantum 
mechanical viewpoint, this presents a semi-classical picture correlating IE and phase-space results. This may enable us to 
explain how $x_c$ and $d_m$ impact tunneling and shape (nature) of phase space in such potentials.  

In both cases, accurate eigenvalues and eigenfunctions of ground and excited states are calculated by employing an 
imaginary-time evolution method \cite{roy02,roy02a,roy05,roy14,roy15}. It has been found to be quite successful for a
variety of problems, including the confinement situation. The paper is organized as follows. In section II, a brief 
outline of ITP method and its implementation is presented; Section III offers a detailed discussion on IE for a
boxed-in SCHO and ACHO in 1D. Finally Section V makes a few concluding remarks and future prospect. 

\section{Method of calculation}
This section provides a brief account of the ITP method, which is applied here to obtain eigenvalues and eigenfunctions
of a caged-in quantum system inside an 1D hard, impenetrable box. It involves a transformation of respective time-dependent
Schr\"odinger equation in imaginary time, to a non-linear diffusion equation. The latter is then solved numerically
in conjunction with minimization of an expectation value to reach the lowest-energy state. By maintaining the orthogonality
requirement with all lower states with same space and spin symmetry, higher states could be generated sequentially. Since 
the method has been discussed at length earlier, here we summarize only essential details. For various other features, the 
reader may consult references \cite{dey99,roy99,roy02,roy02a,gupta02,wadehra03,roy05,roy14,roy15} and therein.

The method is \emph{in principle, exact} and was originally proposed several decade ago. It was first utilized in the 
context of a few physical, chemical problems, e.g., random-walk simulation for \emph{ab initio} Schr\"odinger equation for electronic 
systems, like H $^2P$, H$_3^+$ (D$_{3h}$) $^1A_1$, H$_2$ $^3\Sigma_u^+$, H$_4$ $^1\Sigma_g^+$, Be $^1S$ \cite{anderson75, 
garmer87}, etc. Later, it was invoked for Morse potential, H\'enon-Heiles system and weakly bound states of He on a Pt 
surface, by representing the Hamiltonian in grid through a relaxation method \cite{kosloff86}. Thereafter, it
was applied in direct calculation of \emph{ground-state} densities and 
other properties of rare gas atoms, molecules (H$_2$, HeH$^+$, He$_2^{++}$) through a time-dependent quantum fluid dynamical density 
functional theory \cite{dey99,roy99,roy02a}, \emph{ground, excited} states of various 1D anharmonic DW \cite{roy02}, 
multiple-well \cite{gupta02}, self-interacting \cite{wadehra03}, 2D DW potentials \cite{roy05}, etc. Also a finite difference
time domain approach was proposed for numerical solution of diffusion equation, which was applied to infinite
square potential, harmonic oscillators in 1D, 2D, 3D; H atom, a charged particle in magnetic field, etc., 
\cite{sudiarta07,sudiarta08,sudiarta09}. It was also successfully used in discretization of linear and 
non-linear Schr\"odinger equation by means of split-operator method \cite{lehtovaara2007}, large-scale 2D eigenvalue problems 
in presence of magnetic field \cite{luukko2013}, and in several other situations \cite{aichinger2005,chin2009,strickland2010}. 

Without any loss of generality, the non-relativistic time-dependent Schr\"odinger equation for a particle trapped inside an 
1D impenetrable box (atomic unit employed unless mentioned otherwise) can be written as, 
\begin{equation}
i\frac{\partial}{\partial t}\psi(x,t) = H\psi(x,t) = \left[-\frac{1}{2}\frac{d^2}{dx^2}+v(x)+ v_c (x)\right]\psi(x,t).
\end{equation}
All the terms have usual significance. In present work, we mainly focus on case of a harmonic oscillator, 
$v(x)=\frac{1}{2}k(x-d_m)^2$ (force constant $k$ being kept fixed at unity throughout). For a SCHO, $d_m=0$, whereas for
an ACHO, $d_m \neq 0$. Our desired confinement is achieved by enclosing the system inside two infinitely high hard walls:
\begin{equation} v_c(x) = \begin{cases}
0,  \ \ \ \ \ \ \ -x_c < x < +x_c   \\
+\infty, \ \ \ \ |x| \geq x_c.  \\
 \end{cases} 
\end{equation}
Figure~(1), panel (a) illustrates a SCHO scenario for two box lengths 1, 2 respectively. It is well-known that, SCHO can be 
understood as an intermediate between a PIB and a QHO \cite{gueorguiev06}. Next, we note that asymmetric confinement in a 
QHO can be accomplished in two ways: (i) by changing the \emph{box boundary}, keeping \emph{box length} and $d_m$ 
fixed at zero (ii) the other way is to adjust $d_m$ by keeping \emph{length} and \emph{boundary} fixed. We have adopted 
the second option;
where a rise in $d_m$ shifts the minimum towards right of origin, keeping box length, $b_2-b_1=2$ constant, while 
left, right boundaries are located at $b_1=-1$ and $b_2=1$. Furthermore, since $\frac{1}{2}k(x\pm d)^2$ 
represents a mirror-image pair potential confined within $-b_1$ and $b_2$, their eigenspectra, expectation values 
are same for all states. As respective wave functions are mirror images of each other, in this occasion, it suffices to 
consider the behavior of any one of them; other one automatically follows from it. Right panel (b) of Fig.~(1) shows a schematic 
representation of an ACHO potential, at five specific $d_m$ values. 

\begin{figure}                         
\begin{minipage}[c]{0.45\textwidth}\centering
\includegraphics[scale=0.80]{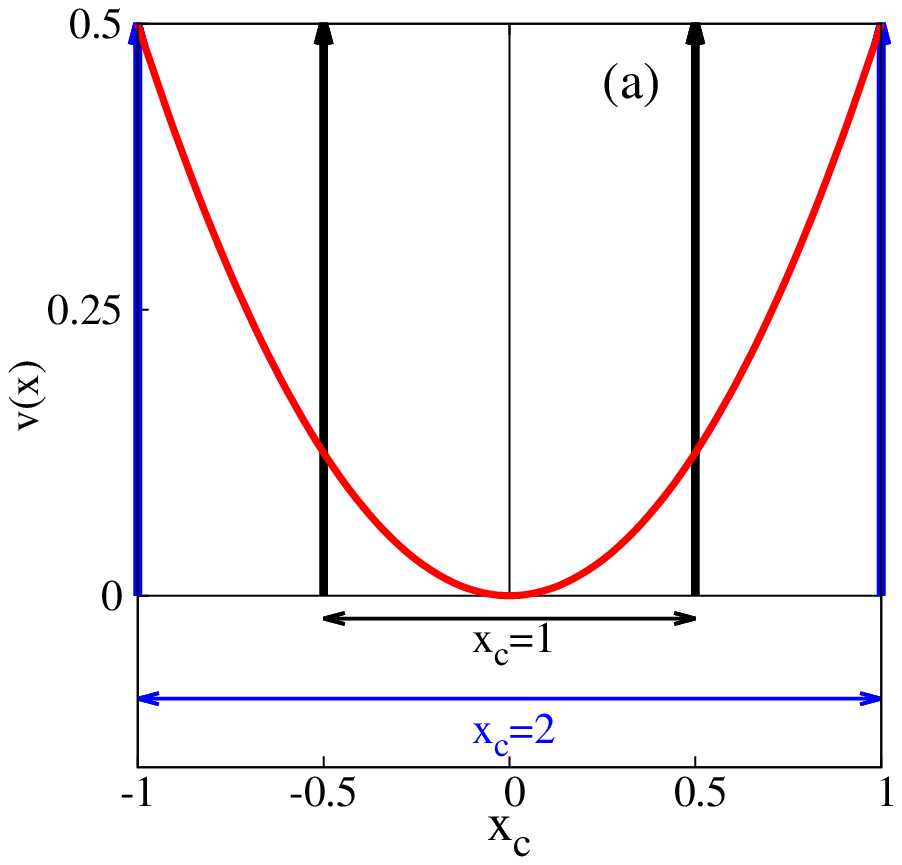}
\end{minipage}%
\hspace{0.1in}
\begin{minipage}[c]{0.45\textwidth}\centering
\includegraphics[scale=0.80]{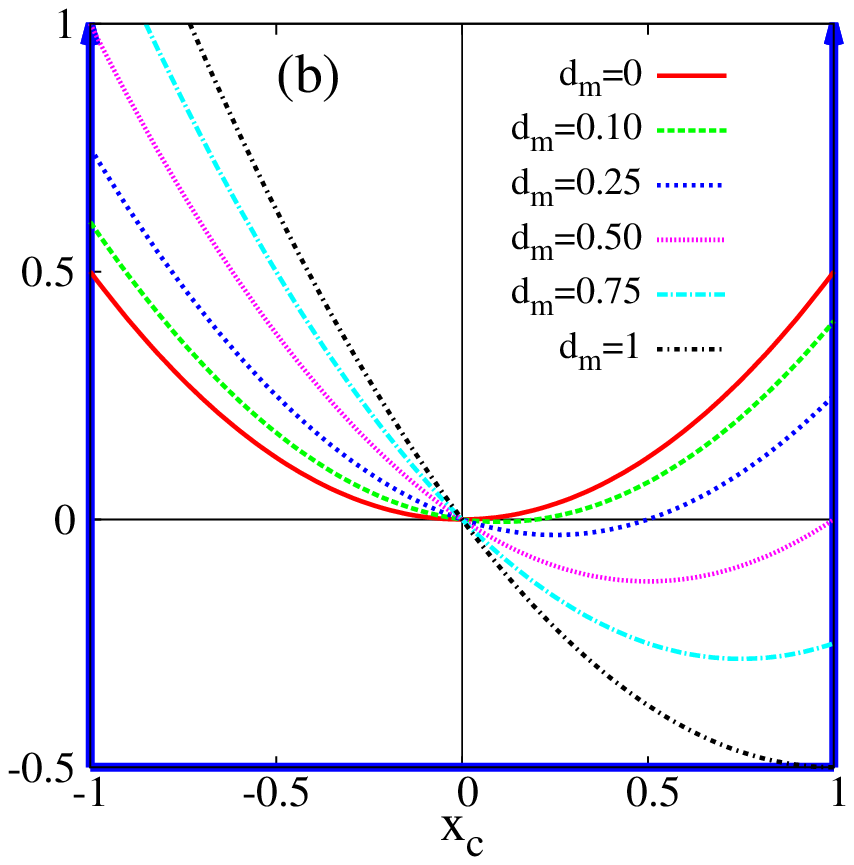}
\end{minipage}%
\caption{Schematic representation of a confined QHO potential: (a) SCHO at two separate box lengths 1 and 2 
(b) ACHO at six chosen values of $d_m$.}
\end{figure}

One can introduce a Wick rotational transformation from real time to imaginary time ($\tau = i t$, where $t$ denotes 
real time), to write Eq.~(2) in a form given below, 
\begin{equation}
-\frac{\partial \psi(x, \tau) }{\partial \tau} =  H \psi(x, \tau),
\end{equation}
whose formal solution can be written as below, 
\begin{equation}
\psi(x, \tau) = \sum_{\kappa=0}^{\infty} c_{\kappa} \phi_{\kappa} (x) \exp{(-\epsilon_{\kappa} \tau)}.
\end{equation}
Thus, taking $\epsilon_0 < \epsilon_1 < \epsilon_2 < \cdots$, for large imaginary time, wave function $\psi(x, \tau)$ 
will contain the lowest-energy (ground) state as dominating, i.e.,   
\begin{equation}
\lim_{\tau \rightarrow \infty} \psi(x, \tau) \approx c_0 \psi_0 (x) e^{-\epsilon_0 \tau}.
\end{equation}
Hence if a initial trial wave function $\psi(x, \tau)$ at $\tau=0$ is evolved for sufficiently long $\tau$, one can 
reach the desired lowest-state energy. In other words, provided $c_0 \neq 0$, apart from a normalization constant, the 
global minimum corresponding to an expectation value $\langle \psi(x, \tau) |H| \psi(x,\tau) \rangle$ could be attained. 

For numerical solution of Eq.~(4), one needs to follow the time propagation of $\psi(x, \tau)$; this is achieved by 
invoking a Taylor series expansion of $\psi(x, \tau+\Delta \tau)$ around time $\tau$, 
\begin{equation}
\psi(x, \tau+\Delta \tau)= e^{-\Delta \tau H} \psi(x, \tau).
\end{equation}
This gives a prescription to advance the diffusion function $\psi(x, \tau)$ at an initial time $\tau$ to a future function 
$\psi(x, \tau+\Delta \tau)$ at time $\tau+\Delta \tau$. This is accomplished through the right-hand side evolution operator 
$e^{-\Delta \tau H}$, which is non-unitary and thus normalization of $\psi(x, \tau)$ at a given instant does not 
necessarily preserve the same at a forward time $\tau + \Delta \tau$. At this stage, it is convenient to write above
equation in to an equivalent, symmetrical form, 
\begin{equation}
e^{(\Delta \tau/2) H_j} \ \psi_j^{'(l+1)}= e^{-(\Delta \tau/2)H_j} \ \psi_j^l, 
\end{equation}
where $j, l$ signify space and time indices, whereas a prime indicates an \emph{unnormalized} diffusion 
function. Now one can (i) expand the exponentials in both sides (ii) truncate them after second term (ii) make use of 
five-point finite-difference formula for spatial derivative, to obtain a set of $N$ simultaneous equations, 
\begin{equation}
\alpha_j \psi_{j-2}^{'(l+1)} + \beta_j \psi_{j-1}^{'(l+1)} + \gamma_j \psi_{j}^{'(l+1)} + 
\delta_j \psi_{j+1}^{'(l+1)} + \zeta_j \psi_{j+2}^{'(l+1)} = \xi_j^l. 
\end{equation}
Quantities $\alpha_j, \beta_j, \gamma_j, \delta_j, \zeta_j, \xi_j^l$ can be derived from straightforward algebra
\cite{roy15}; so not repeated here. Primes in left-hand side signify unnormalized diffusion
function at $(l+1)$th time. There may be some cancellation of errors as discretization and truncation occurs 
in both sides. Equation~(9) can be recast in to a pentadiagonal matrix eigenvalue problem, 
\begin{equation}
\begin{bmatrix}
\gamma_{1} ~ & \delta_{1} ~ & \zeta_{1} ~ & \hspace{1in} & ~ & ~& (0) \\
\beta_{2} ~ & \gamma_{2} ~ & \delta_{2} ~ & \zeta_{2} ~ &       \\
\alpha_{3} ~ & \beta_{3} ~ & \gamma_{3} ~ & \delta_{3} ~ & \zeta_{3} ~ &    \\
           ~ & \ddots ~ & \ddots ~ & \ddots ~ & \ddots ~ & \ddots ~ &     \\
          ~ & ~ & \ddots ~ & \ddots ~ & \ddots ~ & \ddots ~ & \zeta_{N-2}   \\
          ~ & ~ & ~ & \alpha_{N-1} ~ & \beta_{N-1} ~ & \gamma_{N-1} ~ & \delta_{N-1} \\
(0)       ~ & ~ & ~ & ~ & \alpha_{N} ~ & \beta_{N} ~ & \gamma_{N}          
\end{bmatrix}
\begin{bmatrix}
\psi^{\prime (l+1)}_1 \\
\psi^{\prime (l+1)}_2 \\
\psi^{\prime (l+1)}_3 \\
\vdots \\
\psi^{\prime (l+1)}_{N-2} \\
\psi^{\prime (l+1)}_{N-1} \\
\psi^{\prime (l+1)}_{N} \\
\end{bmatrix}
=
\begin{bmatrix}
\xi^{l}_1 \\
\xi^{l}_2 \\
\xi^{l}_3 \\
\vdots \\
\xi^{l}_{N-2} \\
\xi^{l}_{N-1} \\
\xi^{l}_{N} \\
\end{bmatrix}
\end{equation}
which can be readily solved by standard routines to obtain $\{ \psi'^{(l+1)} \}$; this work uses an algorithm provided in 
\cite{pentaroutine}. Thus, an initial trial function $\psi^{0}_{j}$ at zeroth time step is guessed to launch the 
calculation. Then the desired propagation is completed via a series of instructions through (as detailed in 
\cite{roy15}) Eq.~(7) and satisfying boundary condition that $\psi_1^l= \psi_N^l=0$ for all $l$. Further at any given
time step also, another sequence of operations need to be carried out, \emph{viz.}, (a) normalization of $\psi'^{(l+1)}$ to 
$\psi^{(l+1)}$ (b) maintaining orthogonality requirement with all lower states (c) computation of relevant 
expectation value, such as, $\epsilon_{(l+1)}=\langle \psi^{(l+1)}|\hat{H}|\psi^{(l+1)} \rangle $, etc. Initial
trial functions are chosen as $H_m (x) e^{-x^2}$ for $m$-th state. In principle, one can choose arbitrary functions; 
however a good guess reduces computational time and reaches convergence in lesser propagation steps. For more
details, see \cite{roy15}. 

The $p$-space wave functions are obtained from Fourier transform of their $x$-space counterparts; this is 
accomplished through standard, available routines \cite{fftw3}, 
\begin{equation}
\phi(p)=\frac{1}{\sqrt{2\pi}}\int \psi(x) e^{-ipx/\hbar} dx. 
\end{equation} 
In the current work, we deal with three information measures. First of them is Shannon entropy \citep{shannon1951}, 
that signifies the expectation value of logarithmic probability density function. 
In $x$ and $p$ space, this is given by, 
\begin{equation}
S_{x} =  -\int \rho(x) \mbox{ln}\left[\rho(x)\right] dx,   \ \ \ \ \ \ \ \ \
S_{p} =  -\int \rho(p) \mbox{ln}\left[\rho(p)\right] dp. 
\end{equation}
Total Shannon entropy ($S$), defined below, obeys the bound given below, 
\begin{equation}
S=S_{x}+S_{p} \geq (1+ \mbox{ln}\pi). 
\end{equation}
Second one is Fisher information \citep{cover2006}, which in $x$ and $p$ space, reads as, 
\begin{equation}
I_{x} =  \int \left[\frac{|\nabla\rho(x)|^2}{\rho(x)}\right] dx,  \ \ \ \ \ \ \ \ \
I_{p} =  \int \left[\frac{|\nabla\rho(p)|^2}{\rho(p)}\right] dp. 
\end{equation}
Net Fisher information, $I$, given as product of $I_x$ and $I_p$, satisfies the following bound,  
\begin{equation}
I=I_{x}I_{p} \geq 4.
\end{equation}
The last one, Onicescu energy \citep{chatzisavvas2005, alipour2012, agop2015} is a rather recent development. This is 
expressed in $x$ and $p$ space as, 
\begin{equation}
E_{x} =  \int \left[|\rho(x)|^2\right] dx,  \ \ \ \ \ \ \ \ \ \ \
E_{p} =  \int \left[|\rho(p)|^2\right] dp. 
\end{equation}
The corresponding total quantity, $E$ is defined as, 
\begin{equation}
E=E_{x}E_{p} \leq \frac{1}{2\pi}. 
\end{equation}

\begingroup     
\squeezetable
\begin{table}
\caption{Energy spectrum for first five energy states of SCHO potential at low $\eta$ values, namely 0.001,~0.01,~0.1. PT
stands for Perturbation Theory. See text for details.} 
\begin{ruledtabular}
\begin{tabular}{c|cc|cc|cc} 
  Energy & \multicolumn{2}{c}{$\eta=0.001$}  & \multicolumn{2}{c}{$\eta=0.01$}   & \multicolumn{2}{c}{$\eta=0.1$}   \\
    \cline{2-7}  
     & PT$^{\S}$  & Exact$^\dagger$  & PT$^{\S}$ &  Exact$^\dagger$ & PT$^{\S}$ &  Exact$^\dagger$  \\
   \hline
$\epsilon_{0}$ & 1.2337658956 & 1.2337658950 & 1.23435400 & 1.23435394  & 1.240235 & 1.240229   \\
$\epsilon_{1}$ & 4.9349435369  & 4.9349435363 & 4.93621556 & 4.93621550  & 4.948935  & 4.948930   \\
$\epsilon_{2}$ & 11.103460359  & 11.103460360  & 11.10485904  & 11.10485905 & 11.118846  & 11.118847   \\
$\epsilon_{3}$ & 19.7393691362  & 19.7393691364  & 19.74081214 & 19.74081215  & 19.755242  & 19.755243  \\
$\epsilon_{4}$ & 30.8426763672 & 30.8426763673 & 30.84413989  & 30.84413990  & 30.868775  & 30.858776   \\
\end{tabular}
\end{ruledtabular}
\begin{tabbing}
$^{\S}$1st-order perturbation theory, Eq.~(21).    \hspace{50pt} \=
$^\dagger$Calculated from exact analytical wave function, given in {\cite{montgomery07}}. 
\end{tabbing}
\end{table} 
\endgroup  

\section{Results and Discussion}
\subsection{Symmetric confinement}
The original SCHO Hamiltonian can be represented as,
\begin{align}
\begin{rcases}
-\frac{\hbar^2}{2m} \frac{d^2 \psi}{dx^2} + \frac{1}{2} kx^2\psi + V\theta(x^2-x_{c}^{2}) \psi = E \psi, \\
\theta(x^2-x_{c}^{2}) = 0, \ \ \     \mathrm{at}  \ \ \  x \le \abs{x_c} \\
\theta(x^2-x_{c}^{2}) = 1, \ \ \  \mathrm{at} \ \ \  x > \abs{x_c}. 
\end{rcases}
\end{align}

Here, $\theta(x^2-x_{c}^{2}) $ is a Heaviside theta function and $V$ is a constant, having very large value.
The effect of localization and delocalization depends on $x_c$ and $k$. It has been observed 
\citep{zawadzki87, buttiker88} that the above Hamiltonian can be generalized into a dimensionless form, 
so that one can correlate experimental observation with theoretical results. Further, it is established \citep{zawadzki87}
that $k$ is proportional to square root of magnetic field parallel to the gradient of confining potential. Now, 
it seems more appropriate to study the composite effect of $x_c$ and $k$ with the aid of a single dimensionless 
parameter. This will make our current observation more interesting and appealing from an experimental viewpoint. 
It follows that, 
\begin{align*}
 \epsilon \equiv \epsilon \bigg(\frac{\hbar^2}{m}, k, x_c\bigg) \quad \mathrm{and} \quad \psi \equiv \psi 
 \bigg(\frac{\hbar^2}{m}, k, x_c, x\bigg).
\end{align*}

After substitution of $x=\lambda x'$ and $x_c = \lambda$, into Eq.~(18), the modified dimensionless Hamiltonian can be written as,
\begin{equation}
-\frac{1}{2} \frac{d^2 \psi(x')}{d{x'}^2} + \frac{1}{2} \frac{m}{\hbar^2} k x_c^{4} {x'}^2 \psi(x') + 
V\theta({x'}^2-1) \psi(x') = \frac{m x_c^{2}}{\hbar^2} \epsilon \psi(x'), 
\end{equation}
where $x'$ is a dimensionless variable. The above conversion leads to, 
\begin{align}
\begin{rcases}
\epsilon \bigg(\frac{\hbar^2}{m}, k, x_c\bigg)  =  \frac{\hbar^2}{mx_c^{2}} \ \epsilon 
\bigg(1, \frac{mkx_c^{4}}{\hbar^2}, 1 \bigg) \\ 
 \psi \bigg(\frac{\hbar^2}{m}, k, x_c, x\bigg)  =  \frac{1}{\sqrt{\lambda}}  \ \psi \bigg(1, \eta, 1, x' \bigg)  \\
 \eta   =  \frac{mkx_c^{4}}{\hbar^2}.      
\end{rcases}
\end{align}
Equation~(20) implies that $\eta$ depends on product of $k$, $m$ and quartic power of $x_c$. However, if we choose $m=\hbar=1$, 
then \emph{effective} dependence remains on product of $x_c^{4}$ and $k$.

\begingroup        
\squeezetable
\begin{table}
\centering
\caption{$S_{x^{\prime}}$ and $S_{p^{\prime}}$ for first three states of SCHO potential, at six different $\eta$, 
namely, 0.001,~0.01,~0.1,~25,~50,~100. See text for details.} 
\begin{ruledtabular}
\begin{tabular}{c|cc|cc|cc}
$\eta$  & $S_{x^{\prime}}^{0}$ & $S_{p^{\prime}}^{0}$ & $S_{x^{\prime}}^{1}$ & $S_{p^{\prime}}^{1}$ & 
$S_{x^{\prime}}^{2}$ & $S_{p^{\prime}}^{2}$ \\
   \hline
0.001 & 0.386286 &1.825737 & 0.386293 & 2.220674 & 0.386294 & 2.366773 \\
0.01 & 0.386215 & 1.825769 & 0.386289 & 2.220676 & 0.386294 & 2.366793 \\
0.1 & 0.385506 &1.826088 & 0.386244  & 2.220701 & 0.386291 & 2.366998 \\
25 & 0.209026 & 1.946894 & 0.353513  & 2.258814 & 0.382511 & 2.425027 \\
50 & 0.080943 & 2.065966 & 0.296874  & 2.337153 & 0.374755 & 2.488682 \\
100 & -0.080239 & 2.225139 & 0.179801 & 2.488431 & 0.350346 & 2.630012 \\
\end{tabular}
\end{ruledtabular}
\end{table} 
\endgroup

It is well-known that SCHO represents an intermediate situation between a PIB and QHO model. The two limiting
values of smaller ($\ll 1$) and larger ($ \gg 1$) $\eta$, would correspond to these two model systems respectively. So, at 
small-$\eta$ region, actually one can use Rayleigh-Schr\"{o}dinger perturbation theory to get an approximate analytical 
expression for energy. The Hamiltonian is constructed by choosing the PIB as unperturbed system, and 
quadratic potential ($\frac{1}{2} \eta {x'}^2$) as perturbed term. After some straightforward algebra and 
considering upto only first-order correction, the energy expression becomes, 
\begin{equation}
\epsilon_{n+1}(\eta) = \epsilon_{n+1}(0) + \eta \bigg[\frac{1}{6} - \frac{1}{(n+1)^2 \pi^2} \bigg] , \quad n=0,1,2,3, \cdots
\end{equation} 
 where $\epsilon_{n+1}(0)$ denotes zeroth-order PIB energy spectrum. Table~I reports estimated energy values 
for $n=0-4$, in three small representative $\eta$. Column labeled PT gives results corresponding 
to perturbation expression, Eq.~(21), while ``Exact" results refer to those obtained from the ``exact" 
analytical SCHO wave functions \cite{montgomery07}. As expected, PT energies deviate from actual values more 
and more as $\eta$ advances. Of late, an analogous study has been made by \cite{olendski15} for a particle confined 
in a well under the influence of an electric field. 

\begingroup     
\squeezetable
\begin{table}
\centering
\caption{$I_{x^{\prime}}$ and $I_{p^{\prime}}$ for first three states of SCHO potential, at six different $\eta$, 
namely, 0.001,~0.01,~0.1,~25,~50,~100. See text for details.} 
\begin{ruledtabular}
\begin{tabular}{c|cc|cc|cc}
$\eta$  & $I_{x^{\prime}}^{0}$ & $I_{p^{\prime}}^{0}$ & $I_{x^{\prime}}^{1}$ & $I_{p^{\prime}}^{1}$ & $I_{x^{\prime}}^{2}$ & $I_{p^{\prime}}^{2}$ \\
   \hline
0.001 & 9.869604  &0.522753  & 39.478417 & 1.125516 & 88.826439 & 1.243259 \\
0.01 & 9.869605 & 0.522767 & 39.478418  & 1.125436 & 88.826439 & 1.243276 \\
0.1 & 9.869651 & 0.521816  & 39.478464  & 1.124636 & 88.826430 & 1.243445 \\
25 & 11.678374  & 0.357351 & 41.880599  & 0.926130 & 89.290266  & 1.224696 \\
50 & 14.637965 & 0.275574 &  47.436915 & 0.776019 & 92.551459  & 1.139451 \\
100 & 20.063762 & 0.199485 & 60.996032 & 0.589550 & 106.634933  & 0.950170 \\
\end{tabular}
\end{ruledtabular}
\end{table} 
\endgroup 

Since wave function, energy, position expectation values in a SCHO were presented earlier (see, e.g., \cite{navarro80, 
campoy02, montgomery10, roy15}) in some detail, we do not discuss them in this work. Further, a thorough discussion on 
accuracy, convergence of ITP eigenvalues, eigenfunctions with respect to grid parameters, in context of 
confinement, has been published in \cite{roy15}; hence omitted here to avoid repetition. Thus, our primary focus is on 
information analysis. For that, at first we explore the influence of $\eta$, followed by the effect of $x_c$ on IE. At 
the outset, it may be mentioned that, for \emph{symmetric confinement}, some information theoretic ($S$ only) measures in $x$, $p$ and 
Wigner phase-space have been published \cite{laguna2014}. However, to the best of our knowledge, no such attempt is known for 
$I$, $E$ for a SCHO. In this sense, present work takes some inspiration from \cite{laguna2014} and extends it further.

\begingroup      
\squeezetable
\begin{table}
\centering
\caption{$E_{x^{\prime}}$ and $E_{p^{\prime}}$ for first three states of SCHO potential, at six different $\eta$, 
namely, 0.001,~0.01,~0.1,~25,~50,~100. See text for details.} 
\begin{ruledtabular}
\begin{tabular}{c|cc|cc|cc}
$\eta$  & $E_{x^{\prime}}^{0}$ & $E_{p^{\prime}}^{0}$ & $E_{x^{\prime}}^{1}$ & $E_{p^{\prime}}^{1}$ & $E_{x^{\prime}}^{2}$ & $E_{p^{\prime}}^{2}$ \\
   \hline
0.001 & 0.750007  & 0.186731 & 0.7500005  & 0.126261 & 0.7500001 & 0.115063 \\
0.01 & 0.750076 & 0.186723 & 0.750005  & 0.126259 & 0.750001 & 0.115059 \\
0.1 & 0.750769 & 0.186637 & 0.750048 & 0.126237 & 0.750010 & 0.115026 \\
25 & 0.930001 & 0.163488 & 0.782014   & 0.118493 & 0.7599928 & 0.105239 \\
50 & 1.060491 & 0.146313 & 0.837179 & 0.109491 & 0.778344 & 0.096475 \\
100 & 1.262383 & 0.125699 & 0.952923 & 0.095041 & 0.835886  & 0.0831471 \\
\end{tabular}
\end{ruledtabular}
\end{table} 
\endgroup

Let us consider the variation of IE as a function of $n$ (state index) for two special cases: (a) $\eta = 0$ 
which corresponds to a PIB model, (b) $\eta \rightarrow \infty $ leads to a QHO. 
IE measures in $x$-space, for a PIB are as follows, 
\begin{equation}
S_{x}^{n} = \log{4x_c}-1, \quad E_x^{n} = \frac{3x_c}{4}, \quad  I_x^{n} = \frac{n^2\pi^2}{x_{c}^{2}}, \ \ \ \  n=1,2,3, \cdots
\end{equation}
One notices that $S_x^{n}$ and $E_x^{n}$ remain stationary with $n$, but $I_x^{n}$ varies as $n^2$. More information about 
$S$ can be found in \citep{majernik97,sanchez97,laguna2014,olendski15}. The IE measures for QHO in $x,p$ space, on the other hand, 
were discussed in a number of works \cite{yanez94,majernik96} including a recent one from our laboratory \citep{mukherjee15}. 
It was found that $I_x^{n}$ changes with $\eta$ (linearly) and $n$ ($I_x=4(2n+1)\eta$), whereas $I_p^{n}$ varies inversely 
with $\eta$ 
and increases with $n$ ($I_p=\frac{2n+1}{\eta}$). Further, $S_x^{n}$ decreases with increase in $\eta$ but increases with 
$n$, whereas $S_p^{n}$ increases with both $\eta$ and $n$. And $E_x^{n}$, $E_p^n$ show opposite dependence of $S_x^n$, $S_p^n$
with both $\eta$ and $n$. Tables~II-IV illustrate these behaviors of $S, I, E$ respectively, for arbitrary $\eta$, which also 
includes the limiting case of $\eta \rightarrow 0$. Note that, now onwards (both in SCHO and ACHO), all information measures 
are suffixed with a a prime to denote the respective 
quantities for an arbitrary $\eta$, whereas unprimed variables are used for extreme cases (like PIB or QHO). All these 
calculations were performed by taking exact analytical wave functions of SCHO \cite{montgomery07} (see, Eqs.~(25) and 
(26) later). A detailed analysis of IE from these tables confirms a PIB behavior at small $\eta$. But it is not so obvious 
to conclude any kind of QHO behavior 
at large $\eta$ region. This requires a much larger $\eta$ than considered here and not approached; rather it is straightforward
to demonstrate this from an analysis of $x_c$ (Figs.~3, 4, as well as Table~VI later illustrate this).  

\begingroup      
\squeezetable
\begin{table}
\caption{$S_{x^{\prime}}$,~$I_{x^{\prime}}$ and $E_{x^{\prime}}$ for first three states of SCHO potential at small $\eta$,
namely, 0.001,~0.01,~0.1. PT implies Perturbation Theory. See text for details.} 
\begin{ruledtabular}
\begin{tabular}{c|cc|cc|cc} 
Quantity  & \multicolumn{2}{c}{$\eta=0.001$}  & \multicolumn{2}{c}{$\eta=0.01$}  & \multicolumn{2}{c}{$\eta=0.1$}   \\
\cline{2-7}
     & PT$^{\S}$  & Exact$^\dagger$  & PT$^{\S}$ &  Exact$^\dagger$ & PT$^{\S}$ &  Exact$^\dagger$  \\
\hline
$S_{x^\prime}^{0}$ & 0.386287 & 0.386386 & 0.386216 & 0.386293 & 0.385508 & 0.386294 \\
$S_{x^\prime}^{1}$ & 0.386246 & 0.386215 & 0.386289 & 0.386289 & 0.386246 & 0.386294\\
$S_{x^\prime}^{2}$ & 0.386294 & 0.385506 & 0.386294 & 0.386244 & 0.386294 & 0.386291\\
 \hline
$I_{x^{\prime}}^{0}$ & 9.869604 & 9.869604 & 9.869605 & 9.869605 & 9.869652 & 9.869651 \\
$I_{x^\prime}^{1}$ & 39.478417 & 39.478417 & 39.478418 & 39.478418 & 39.478462 & 39.478464\\
$I_{x^\prime}^{2}$ & 88.826439 & 88.826439& 88.826439 & 88.826439 & 88.826429 & 88.826430 \\
\hline
$E_{x^\prime}^{0}$ & 0.750008 & 0.750007 & 0.750077 & 0.750000 & 0.750771 & 0.750000 \\
$E_{x^\prime}^{1}$ & 0.750000 & 0.750076 & 0.750005 & 0.750005 & 0.750048 &  0.750001\\
$E_{x^\prime}^{2}$ & 0.750001  & 0.750769 & 0.750048 & 0.750048 & 0.750004 & 0.750010\\
\end{tabular}
\end{ruledtabular}
\begin{tabbing}
$^{\S}$1st-order perturbation theory.    \hspace{50pt} \=
$^\dagger$Calculated from exact analytical wave function, given in {\cite{montgomery07}}. 
\end{tabbing}
\end{table} 
\endgroup 

In order to probe further, at small $\eta$ limit, a perturbation treatment as in energy, can be employed. 
Considering up to first order, the perturbed wave functions can be expressed as,
\begin{equation}
\psi_n(\eta; x') = \psi_n^{(0)}(0;x') + \sum_{m \ne n}^{\infty} \frac{V_{mn}^{\eta}}{\epsilon_n^{0}-\epsilon_m^{0}} \psi_m^{(0)}(0;x')
\end{equation}
Here, we only retain only two consecutive corrected terms in the summation and 
$V_{mn}^{\eta} = \langle \psi_m^{(0)} |\frac{1}{2} \eta x'^2 | \psi_n^{(0)} \rangle$. After some tedious algebraic manipulation, 
one can write the following expressions for $I_{x'}$ and $E_{x'}$, 
\begin{eqnarray}
I_{x'}^0  & =  & \pi^2 \ \frac{a_0\pi^8 + b_0 \eta^2}{a_0 \pi^8+c_0\eta^2}; \ \ \  
I_{x'}^1   =   4\pi^2 \ \frac{a_1\pi^8 + b_1 \eta^2}{a_1 \pi^8+c_1\eta^2}; \ \  \ \ 
I_{x'}^2   =   9\pi^2 \ \frac{a_2\pi^8 + b_2 \eta^2}{a_2 \pi^8+c_2\eta^2} \\  
E_{x'}^0 & = & \frac{3}{8} \ \frac{a_3\pi^{16}+a_3 \pi^{12}\eta+ a_4\pi^8\eta^2-a_5 \pi^4 \eta^3
+a_6\eta^4}{(a_0\pi^8+c_0\eta^2)^2}, \nonumber  
\end{eqnarray}
where $a_0=5832, b_0=29837, c_0=3293; a_1=2985984, b_1=4253353, c_1=1055137; a_2=1024, b_2=689, c_2=801,
a_3=68024448, a_4=144190368, a_5=7085880, a_6=21851723$.

Table~V gives a cross section of $S, I$ and $E$ in $x$ space for three low-lying states at three specific $\eta$. The four 
quantities, given in Eq.~(24) are estimated from the perturbation expressions, whereas all others including
Shannon entropies are numerically evaluated by taking approximate wave function of Eq.~(23). For sake of easy comparison, 
we also quote respective exact results from Tables~II-IV. The two results show quite good agreement in 
small-$\eta$ region and disagreements tend to set in as $\eta$ assumes larger values.  

\begin{figure}                         
\begin{minipage}[c]{0.25\textwidth}\centering
\includegraphics[scale=0.45]{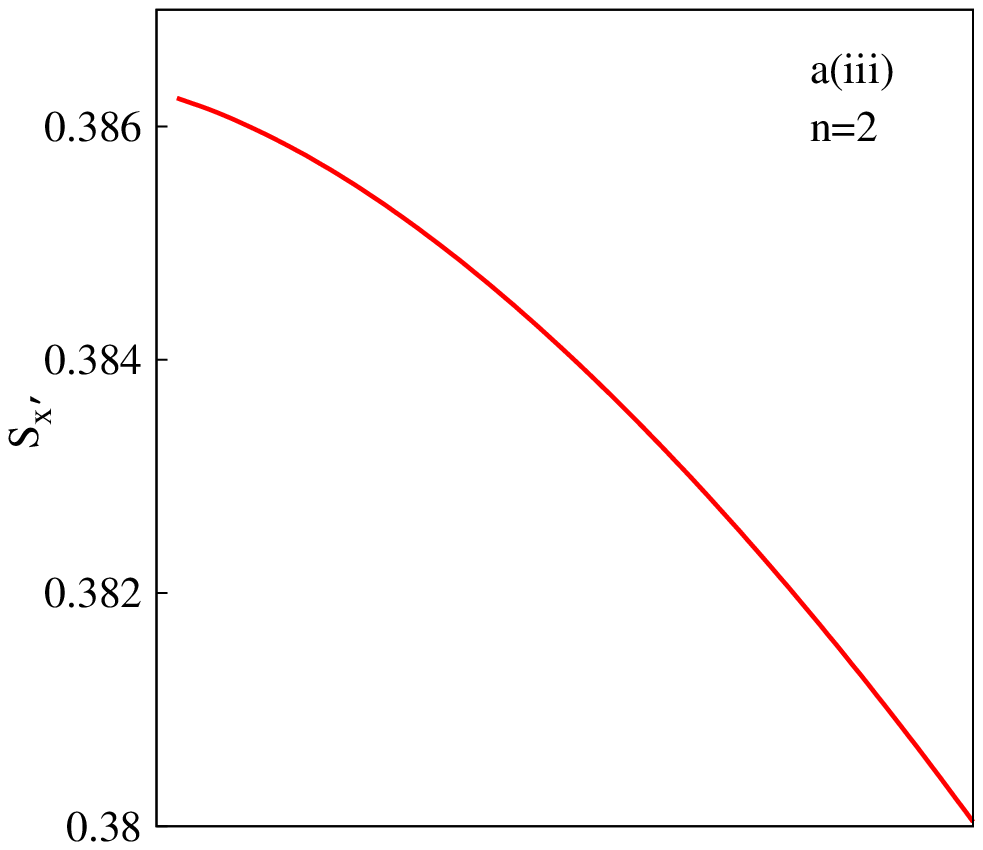}
\end{minipage}%
\hspace{0.45in}
\begin{minipage}[c]{0.25\textwidth}\centering
\includegraphics[scale=0.45]{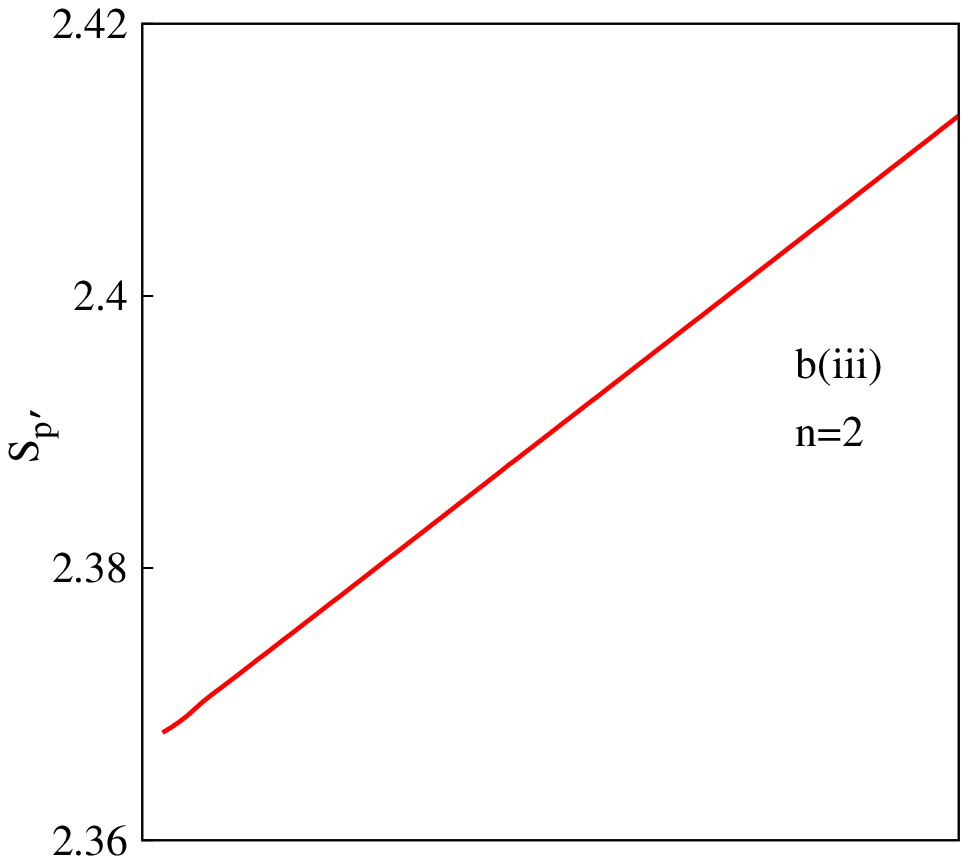}
\end{minipage}%
\hspace{0.45in}
\vspace{0.15in}
\begin{minipage}[c]{0.25\textwidth}\centering
\includegraphics[scale=0.45]{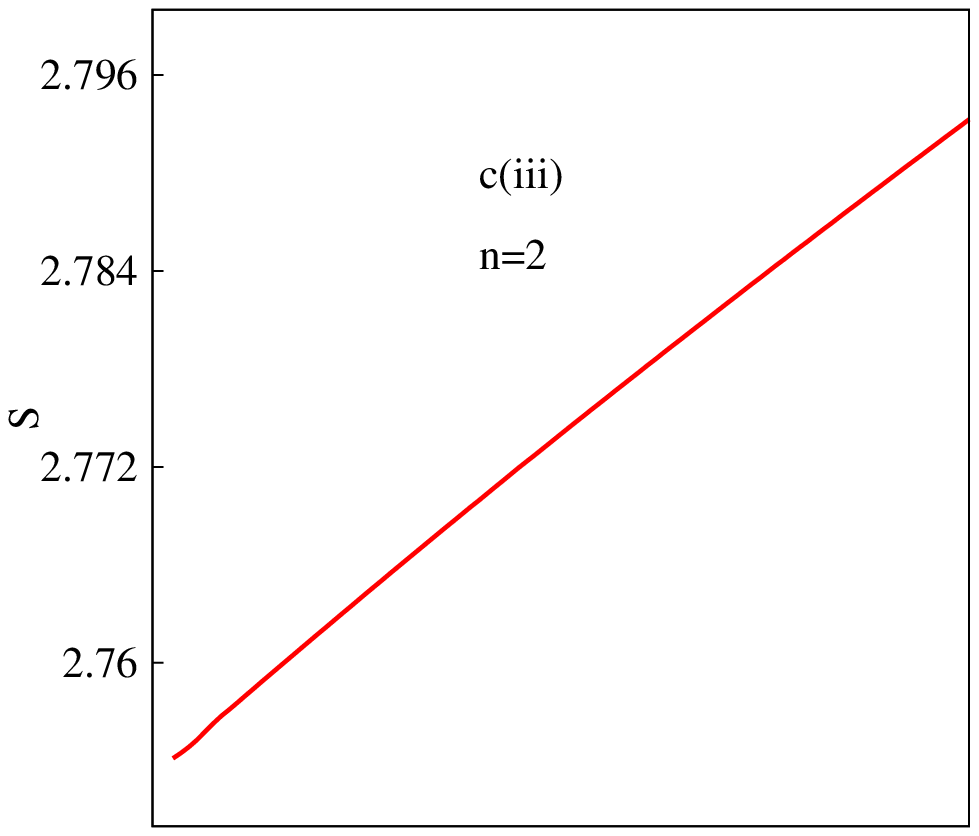}
\end{minipage}%
\hspace{0.45in}
\begin{minipage}[c]{0.25\textwidth}\centering
\includegraphics[scale=0.45]{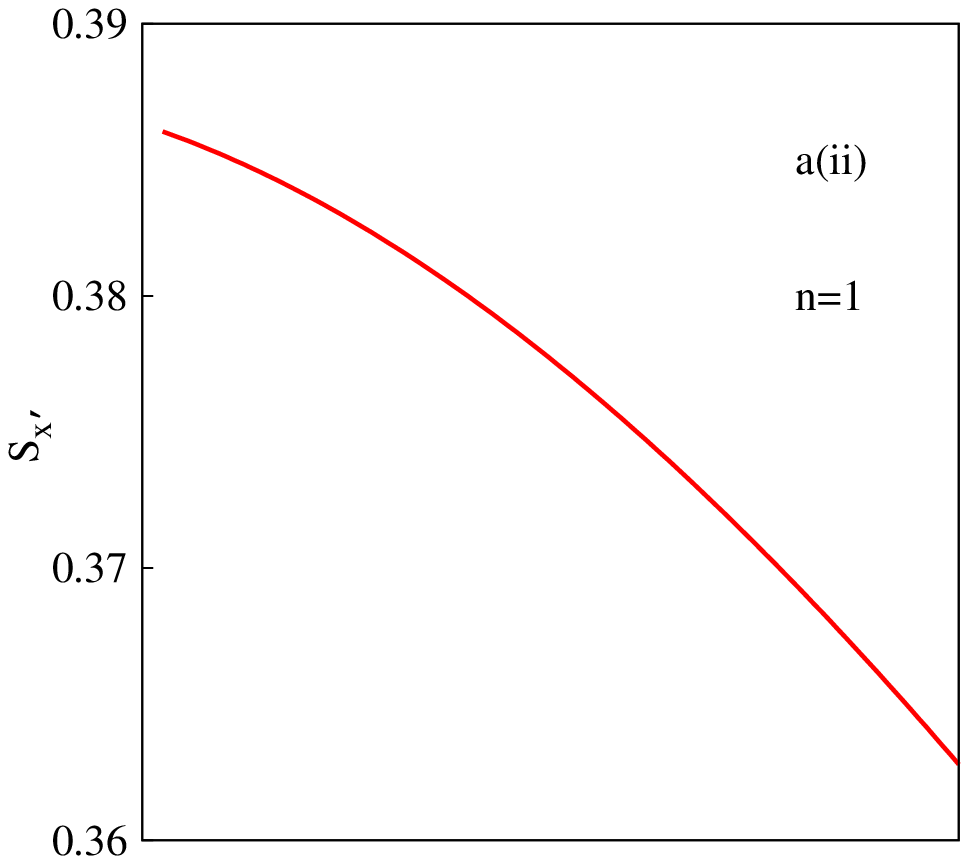}
\end{minipage}%
\hspace{0.45in}
\begin{minipage}[c]{0.25\textwidth}\centering
\includegraphics[scale=0.45]{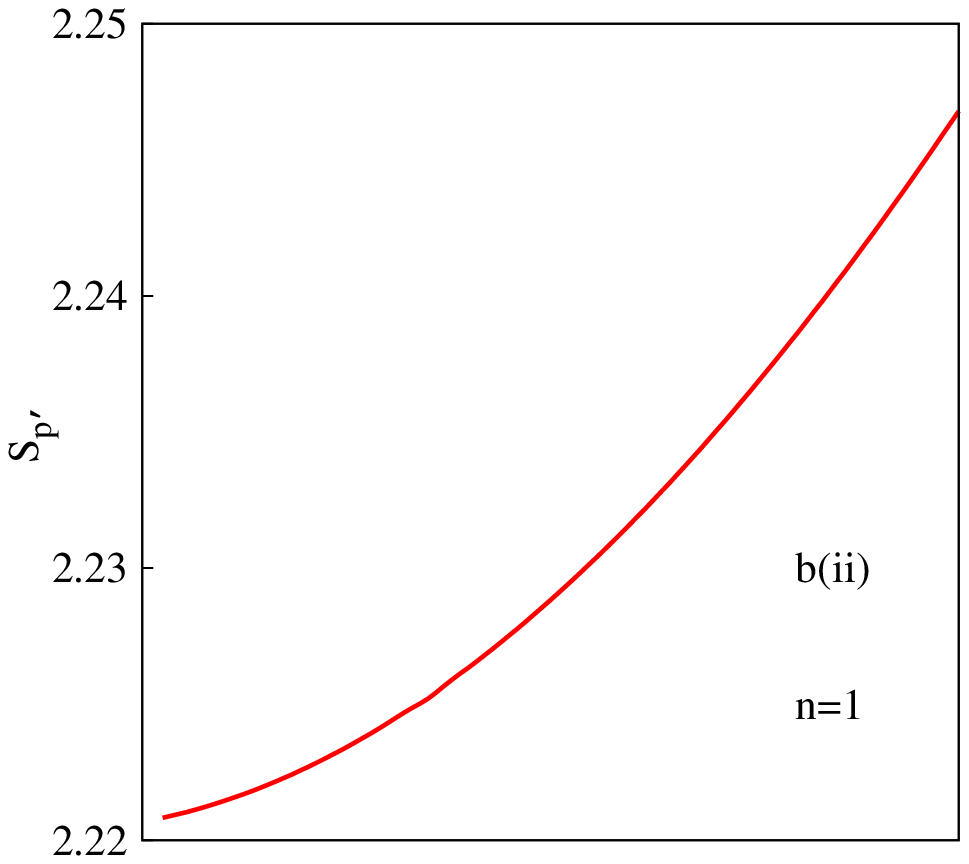}
\end{minipage}%
\hspace{0.45in}
\vspace{0.15in}
\begin{minipage}[c]{0.25\textwidth}\centering
\includegraphics[scale=0.45]{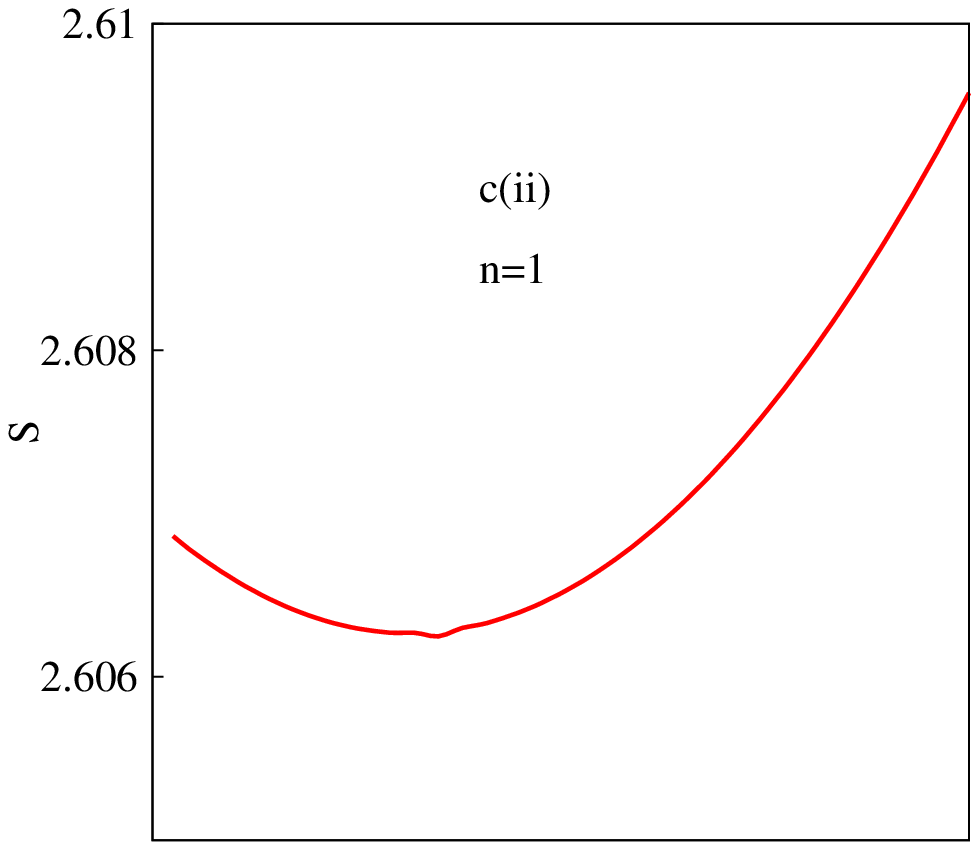}
\end{minipage}%
\hspace{0.45in}
\begin{minipage}[c]{0.25\textwidth}\centering
\includegraphics[scale=0.48]{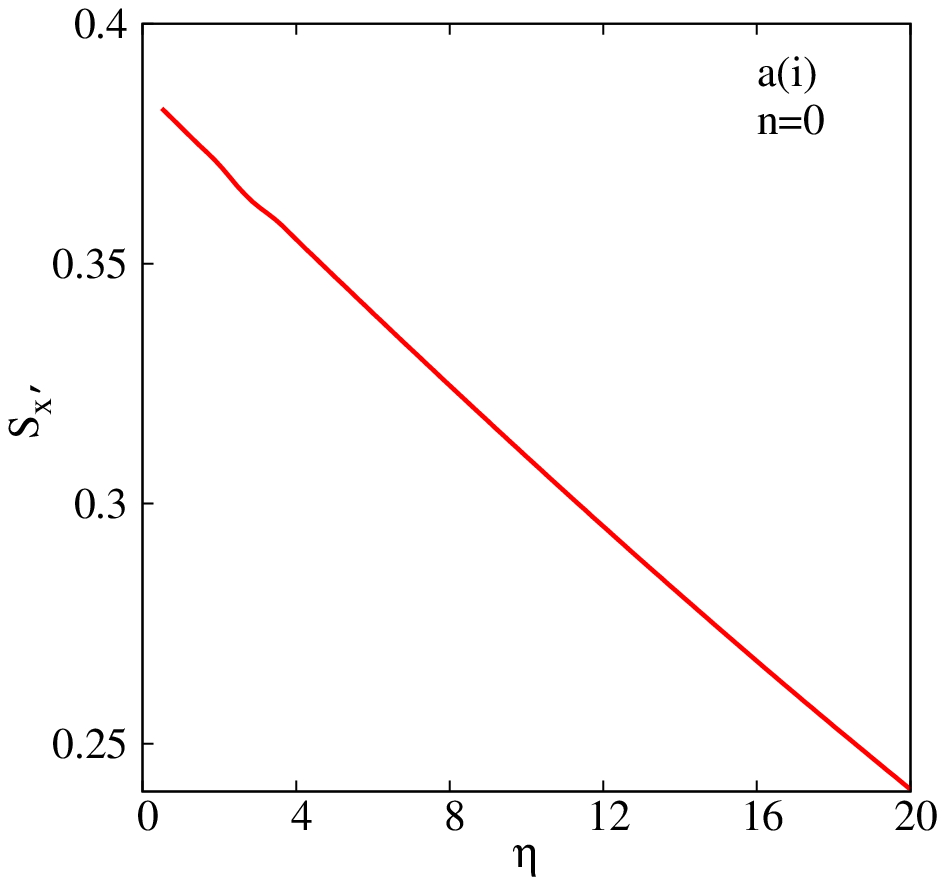}
\end{minipage}%
\hspace{0.45in}
\begin{minipage}[c]{0.25\textwidth}\centering
\includegraphics[scale=0.48]{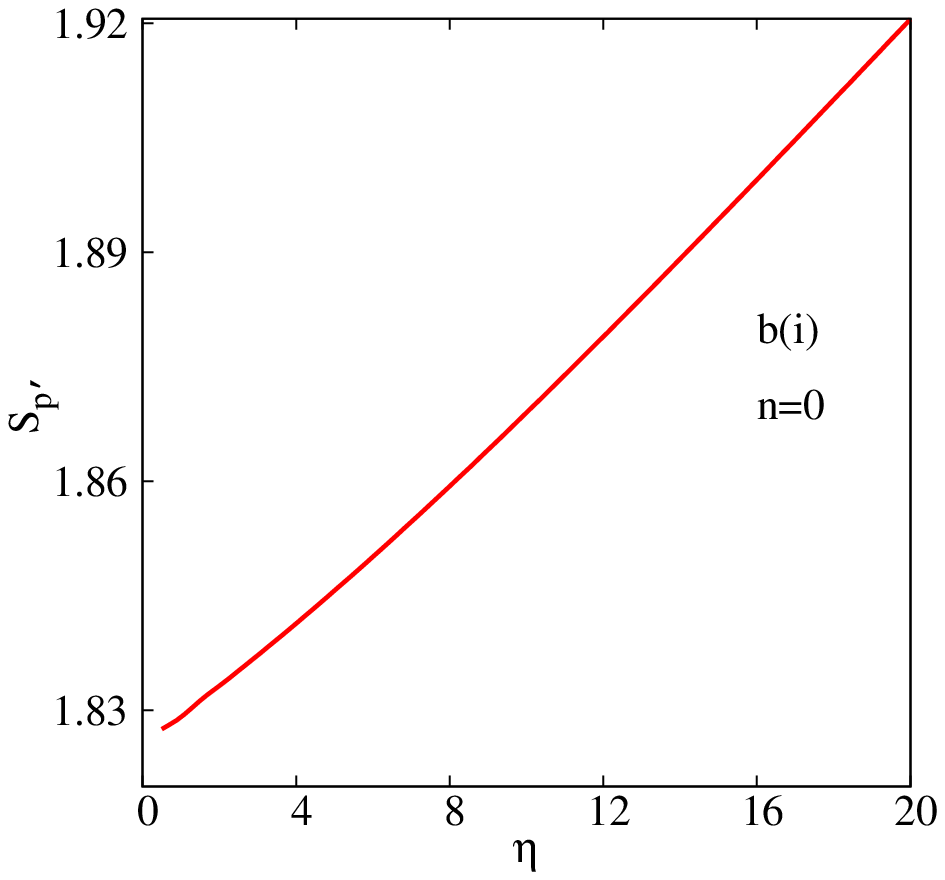}
\end{minipage}%
\hspace{0.45in}
\begin{minipage}[c]{0.25\textwidth}\centering
\includegraphics[scale=0.48]{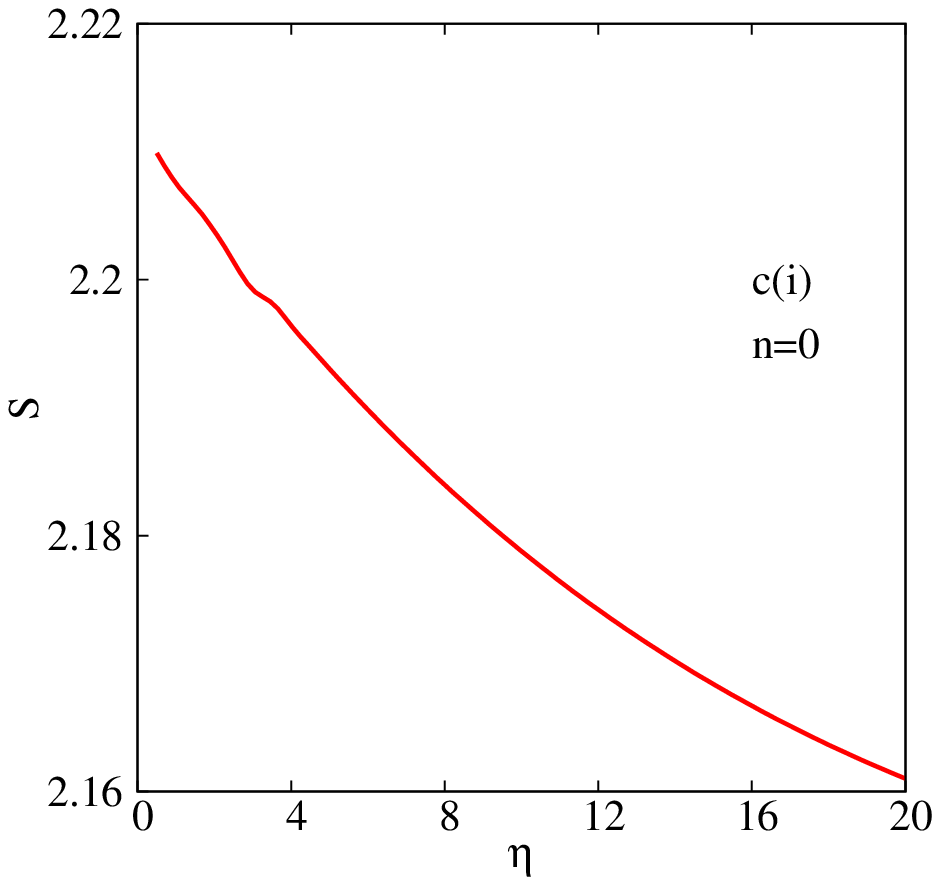}
\end{minipage}%
\caption{Plot of $S_{x'}$, $S_{p'}$, $S$ of first three states of SCHO potential, as function of $\eta$, in left (a), middle (b), 
right (c) columns; (i)--(iii) represent $n=0-2$ states. See text for details.}
\end{figure}

The general variation of IE with respect to $\eta$ is now presented graphically. For this, all three information measures
$I, S, E$ were computed for $n=0-4$ states. Moderate ranges of $\eta$ were considered in all cases. From this set, here we present 
in Fig.~(2), only $S_{x'},S_{p'}$ and $S$ in three lowest states. Higher states as well as $I,E$ plots are given in 
supplementary material. In all cases, a gain in $\eta$ leads to a consistent decrease in $S_{x'}$, which clearly 
indicates the localization of particle in $x$ space. On the other hand, $S_{p'}$ shows a reverse trend with rise in 
$\eta$, signifying a de-localization of particle in $p$ space. Panel c(i) of Fig.~(2) shows that, $S$ for $n=0$ 
decreases with $\eta$ steadily, whereas, from the remaining two rightmost panels, behavior in excited states remains somehow
less clear cut, because of the composite behavior of $S_{x'}$, $S_{p'}$. Interestingly, in $S$ plots, $n=1$ possesses a 
minimum and $n=3,4$ states in Fig.~(S1) tend to show a steady progress throughout. Similar plots of $I$ and $E$, covering 
same range of $\eta$, are produced in Figs.~(S2) and (S3) respectively, for $n=0-4$ states. In $n=0-2$, $I_{x'}$ increases with 
$\eta$, whereas for $n >2$, it decreases. One sees that $I_{p'}$ decreases with $\eta$ in first two states, and then at $n=2$, 
there appears a dome-shaped structure with a maximum. Thereafter for $n >2$, it completely reverses the trend. 
For all states, $E_{x'}$ increases monotonically while $E_{p'}$ decreases monotonically, with $\eta$. And finally, except
for $n=0$, $E$ decreases with $\eta$. Hence, it can be concluded that $S$ and $E$ could be possibly used to explain the 
localization and de-localization behavior in a SCHO. 
 
\begingroup      
\squeezetable      
\begin{table}
\centering
\caption{Calculated $S_x^n, E_x^n$ of a SCHO at three chosen $x_c$'s; 0.25, 2 and 5, for 
five lowest states ($n \! = \! 0-4$). Reference results are obtained by taking \emph{exact} analytical wave functions, 
reported in \cite{montgomery07}. PR implies Present Result. See text for more details.} 
\begin{ruledtabular}
\begin{tabular}{|l|ll|ll|ll|} 
   & \multicolumn{2}{c}{$x_c=0.25$}   & \multicolumn{2}{c}{$x_c=2$}    & \multicolumn{2}{c}{$x_c=5$}  \\
\hline
      Property & PR & Reference & PR &  Reference & PR & Reference \\
\hline
$S_{x}^{0}$ & $-$1.0000307  & $-$1.0000307 & 0.9603491 & 0.9603491 & 1.0767572 & 1.0767573  \\
$S_{x}^{1}$ & $-$1.0000019  & $-$1.0000019 & 1.0625421 & 1.0625421 & 1.3427276 & 1.3427277  \\
$S_{x}^{2}$ & $-$1.0000003  & $-$1.0000003 & 1.0749785 & 1.0749785 & 1.4986081 & 1.4986082  \\
$S_{x}^{3}$ & $-$1.0000001  & $-$1.0000001 & 1.0776239 & 1.0776239 & 1.6097006 & 1.6097006  \\
$S_{x}^{4}$ & $-$0.9999999  & $-$0.9999999 & 1.0786018 & 1.0786018 & 1.6964625 & 1.6964627  \\
\hline
$E_{x}^{0}$ & 3.0001200   & 3.0001200  & 0.4347289 & 0.4347291  & 0.3989422 & 0.3989422  \\
$E_{x}^{1}$ & 3.0000070   & 3.0000075  & 0.3832620 & 0.3832622  & 0.2992065 & 0.2992067  \\
$E_{x}^{2}$ & 3.0000010   & 3.0000014  & 0.3775775 & 0.3775772  & 0.2555729 & 0.2555724  \\
$E_{x}^{3}$ & 3.0000005   & 3.0000005  & 0.3761693 & 0.3761693  & 0.2290809 & 0.2290810  \\
$E_{x}^{4}$ & 3.0000001   & 3.0000001  & 0.3755618 & 0.3755620  & 0.2106059 & 0.2106061  \\
\end{tabular}
\end{ruledtabular}
\end{table} 
\endgroup

Up to now, we were concerned with $\eta$ variations. But since from Eq.~(20), $\eta \propto k x_c^4$, these results 
include combined effects from both $k$ and $x_c$. In order to get a clear picture of confinement, these two effects need to be 
separated. This motivates our forthcoming analysis with respect to $x_c$ on IE, keeping $k=1$. For all $x_c$ variation
throughout the article, unprimed variables are used in IE suffixes. 
At first Table~VI presents sample results for $S_x$, $E_x$ for lowest five energy states of SCHO potential 
at three chosen $x_c$, \emph{viz.,} 0.25, 2 and 5 covering small, intermediate and large regions. Now onwards, all IE
results on SCHO were performed following the ITP method as prescribed in Sec.~(II). Several
test calculations were done on various number of grid points $N$, to check convergence. Generally, quality of 
results improves as $N$ increases and after some value, these remain practically unchanged. Similar conclusions also 
hold for $I_x$ and hence not repeated. While one can think of ways 
to improve such results further by employing higher precision wave function on finer grid, present results are sufficiently 
accurate for the purpose at hand. Thus there is no necessity to pursue such accurate calculations, as we are 
primarily interested in the qualitative trend of IE. For each of these 
quantities we also report the respective reference results, obtained by taking \emph{exact} analytical 
eigenfunctions \cite{montgomery07}.

\begin{figure}                         
\begin{minipage}[c]{0.45\textwidth}\centering
\includegraphics[scale=0.80]{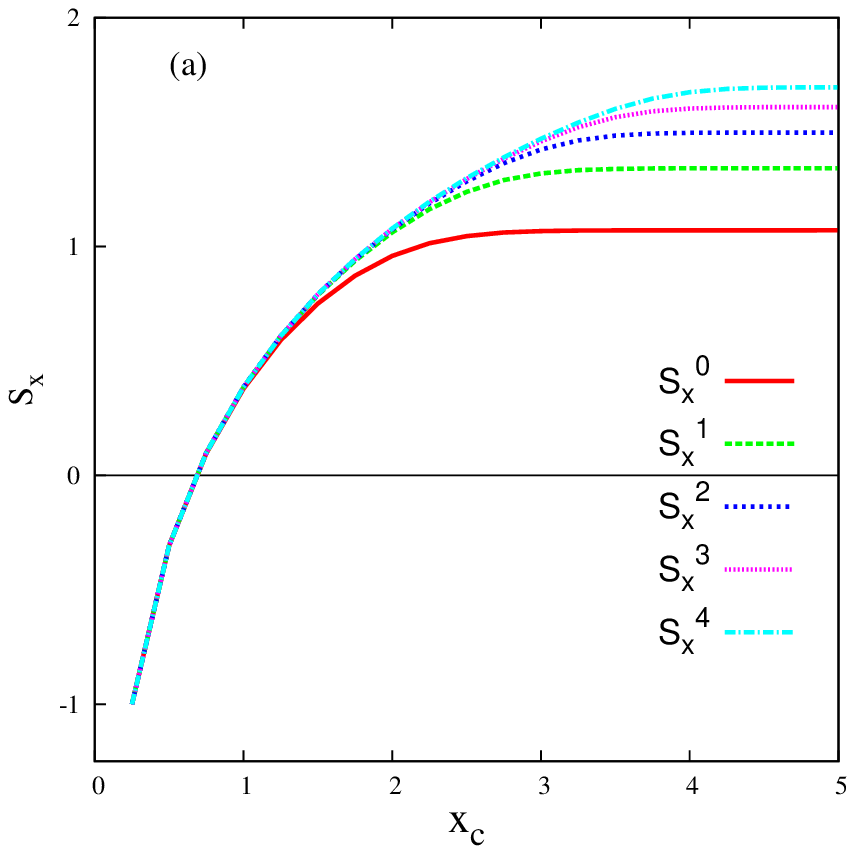}
\end{minipage}%
\hspace{0.1in}
\begin{minipage}[c]{0.45\textwidth}\centering
\includegraphics[scale=0.80]{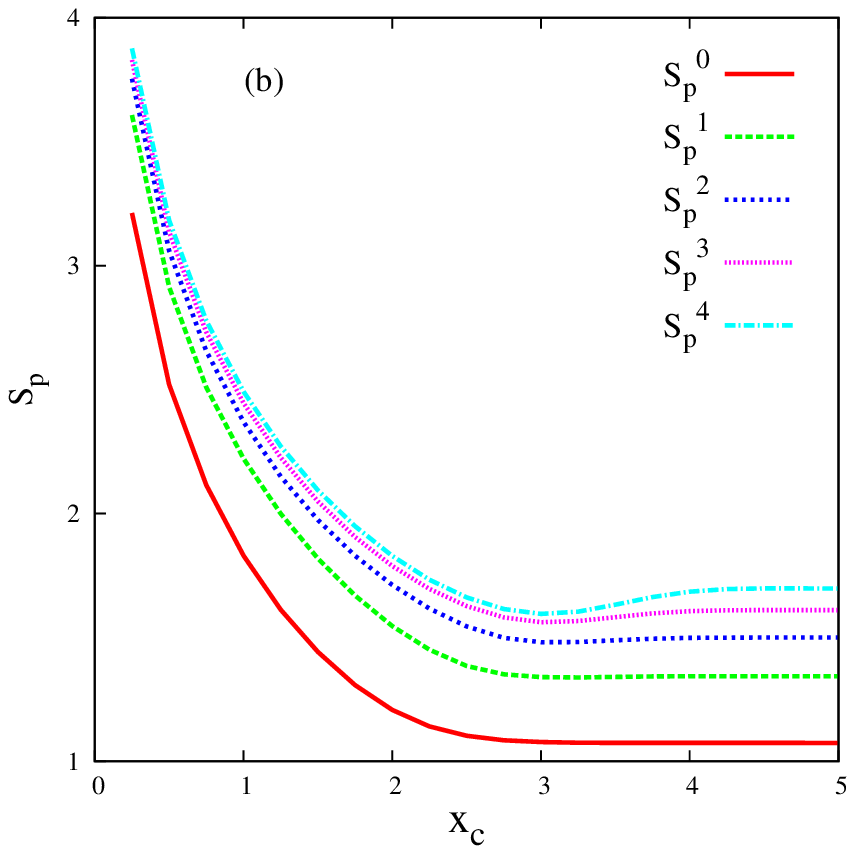}
\end{minipage}%
\caption{Plot of (a) $S_x$ (b) $S_p$, of SCHO potential, as function of $x_c$, for first five states.}
\end{figure} 

For even and odd states, these are given as:
\begin{equation}
\psi_e(x) = e^{-\frac{x^2}{2}}~_1F_1\left(\frac{1}{4}-\frac{\epsilon_n}{2},\frac{1}{2},x^2\right), \ \ \ \ \ \
\psi_o(x) = xe^{-\frac{x^2}{2}}~_1F_1\left(\frac{3}{4}-\frac{\epsilon_n}{2},\frac{3}{2},x^2\right).
\end{equation}
Here, $\epsilon_n$ and $_1F_1\left(a,b,x\right)$ denote eigenvalues and Kummer confluent hypergeometric
function. Energies are computed by putting $x=x_c$ and numerically solving the following equations:
\begin{equation}
_1F_1\left(\frac{1}{4}-\frac{\epsilon_n}{2},\frac{1}{2},x_c^2\right)=0, \ \ \ \ \ \
_1F_1\left(\frac{3}{4}-\frac{\epsilon_n}{2},\frac{3}{2},x_c^2\right)=0
\end{equation}
for even and odd case. In all cases, we notice that present results are practically coincident with those 
from reference. Similar conclusions hold for $I_x$.  

\begin{figure}             
\centering
\begin{minipage}[c]{0.45\textwidth}\centering
\includegraphics[scale=0.74]{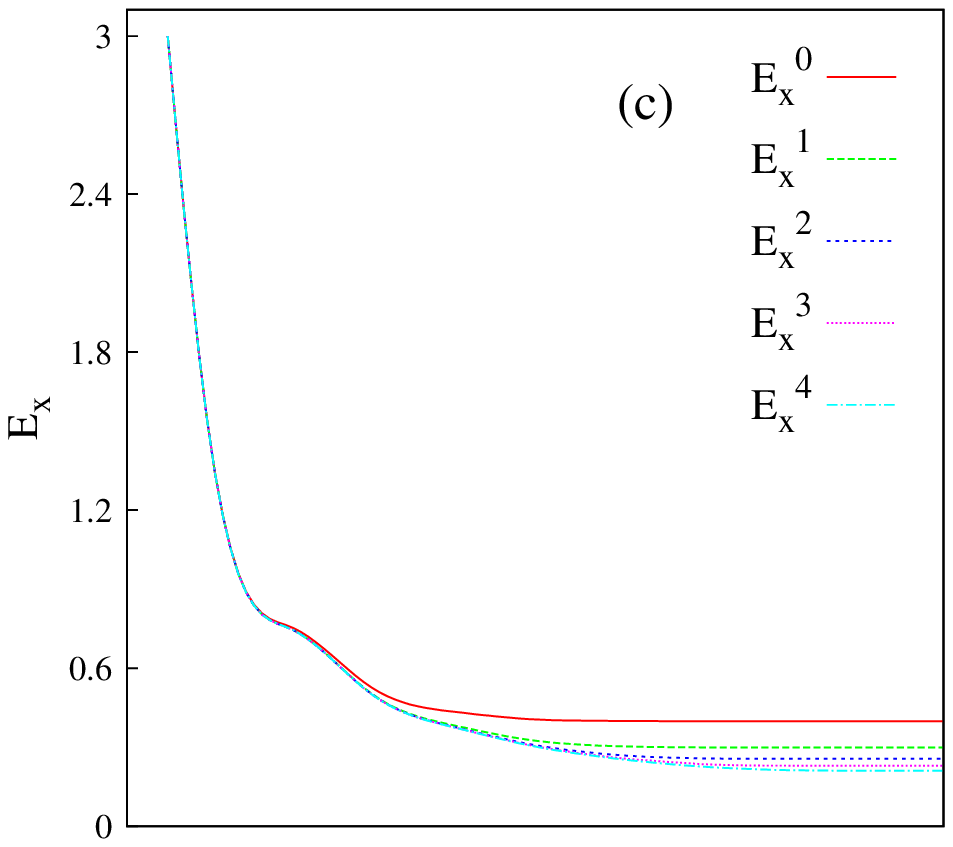}
\end{minipage}\hspace{0.2in}
\begin{minipage}[c]{0.45\textwidth}\centering
\includegraphics[scale=0.74]{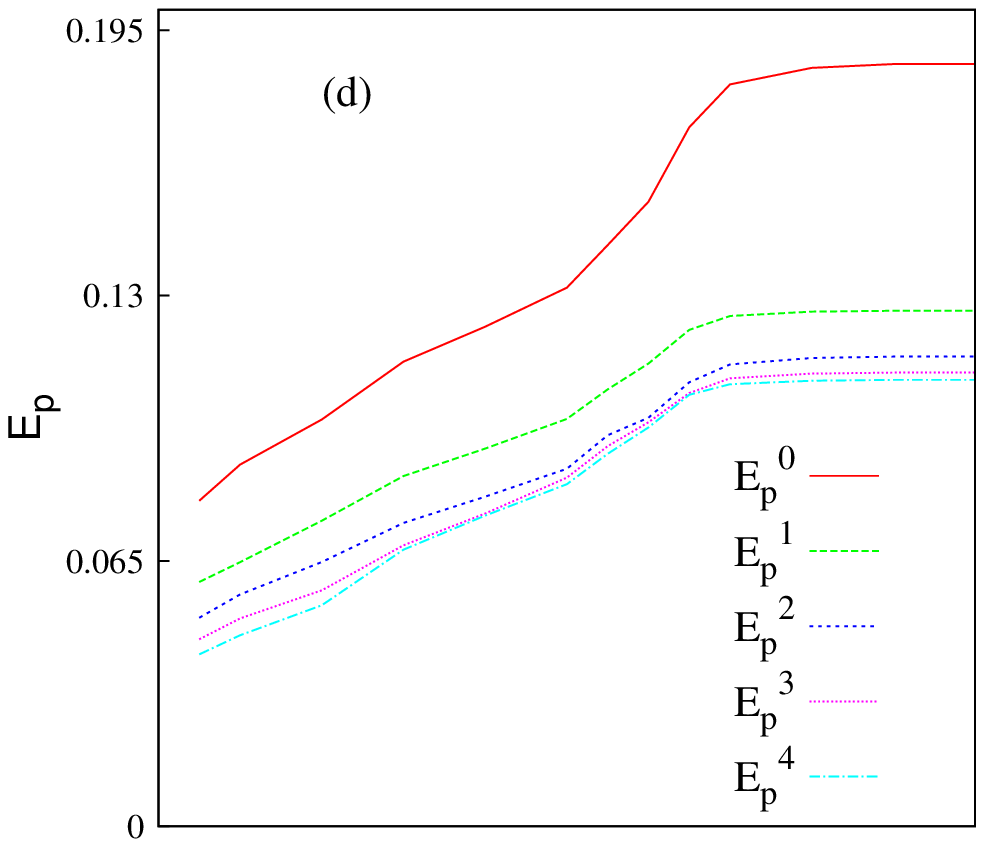}
\end{minipage}\hspace{0.2in}
\\[15pt]
\begin{minipage}[c]{0.47\textwidth}\centering
\includegraphics[scale=0.80]{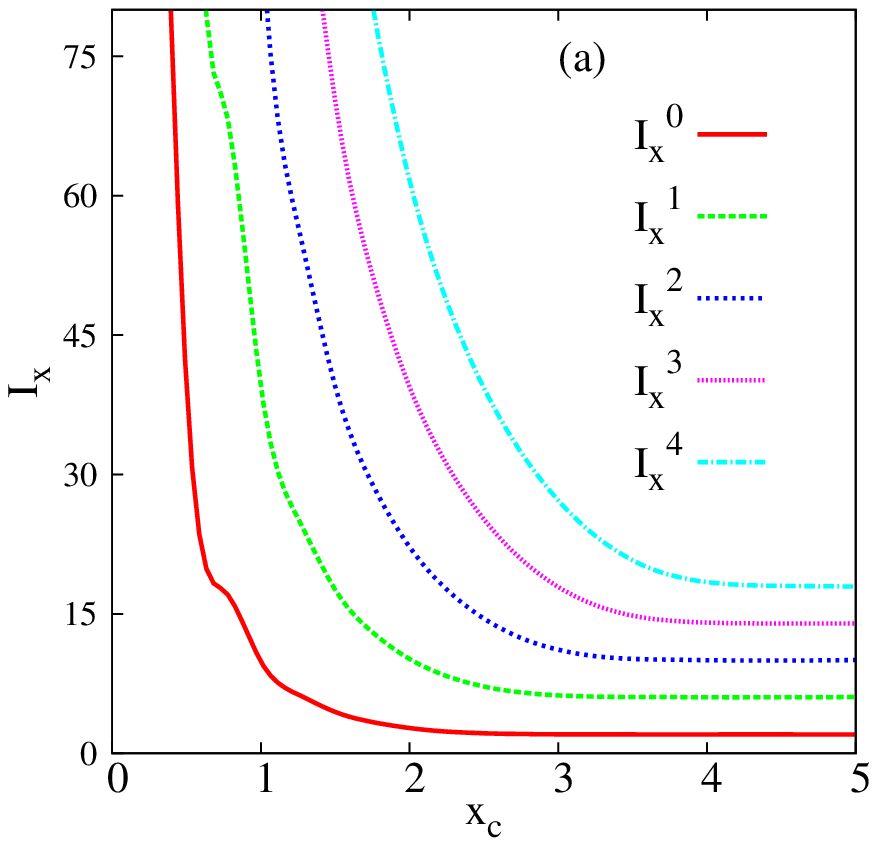}
\end{minipage}\hspace{0.2in}
\begin{minipage}[c]{0.47\textwidth}\centering
\includegraphics[scale=0.80]{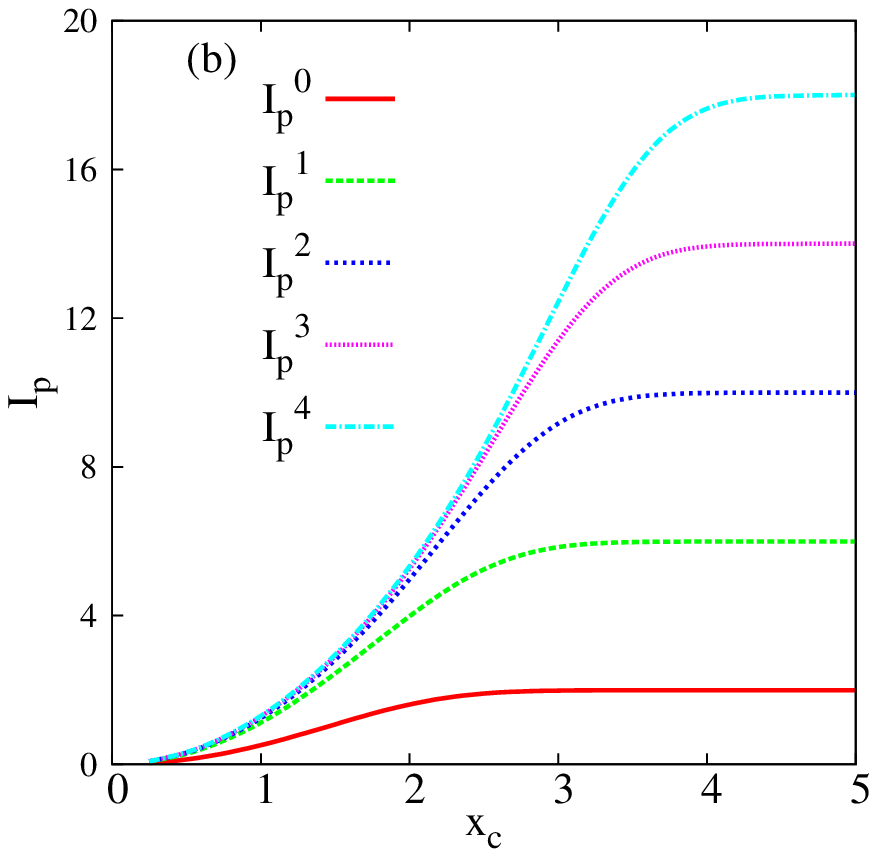}
\end{minipage}
\caption{Plot of $I_x$,~$I_p$, $E_x$,~$E_p$ for first five states of SCHO potential as a function of $x_c$, in panels 
(a), (b), (c) and (d). See text for details}.
\end{figure}

Once the authenticity of ITP method is established, now we proceed for a detailed analysis on information measures 
with respect to $x_c$. 
First, Fig.~(3) shows plots of $S_{x}$, $S_{p}$ versus $x_c$, for five low-lying states of SCHO. In panel (a), 
$S_{x}$ increases with box length (indicating delocalization of particle) and converges to a constant value ($S_{x}$ 
of QHO) at sufficiently large box length. At small $x_c$ region, $S_x$ changes insignificantly with state index $n$,
resembling the behavior of a PIB problem (where $S_x$ remains stationary with $n$). Next, panel (b), 
suggests that, $S_{p}$ decreases with increase in box length and finally merges to corresponding QHO value. It is 
interesting to note that, there appears a minimum (which becomes progressively more prominent as $n$ increases)
in $S_p$ for all excited states. Appearance of such minimum may be attributed to the competing effect in $p$ space. 
Actually, three possibilities could be envisaged ($l_{x}$, $l_{p}$ are box lengths in $x$, $p$ space):
\begin{enumerate}[(a)]
\item
When $l_{x} \rightarrow 0$ then $l_{p} \rightarrow \infty$.
\item
When $l_{x}$ finite then $l_{p}$ is also finite. But an increase in $l_{x}$ leads to a decrease in $l_{p}$.
\item
When $l_{x} \rightarrow \infty$ then also $l_{p} \rightarrow \infty$. 
\end{enumerate}
From above plots it is clear that, initially with increase in $l_{x}$, particle gets localized in $p$ space 
($S_{p}$ decreases), but when potential starts behaving like QHO, de-localization occurs. 
Thus, existence of minimum in $S_{p}$ is due to balance of two conjugate forces. These variations of 
$S_x$, $S_p$, total $S$ (not presented here) with $x_c$ are in harmony with the findings of 
\cite{laguna2014}. 

\begin{figure}                         
\begin{minipage}[c]{0.25\textwidth}\centering
\includegraphics[scale=0.47]{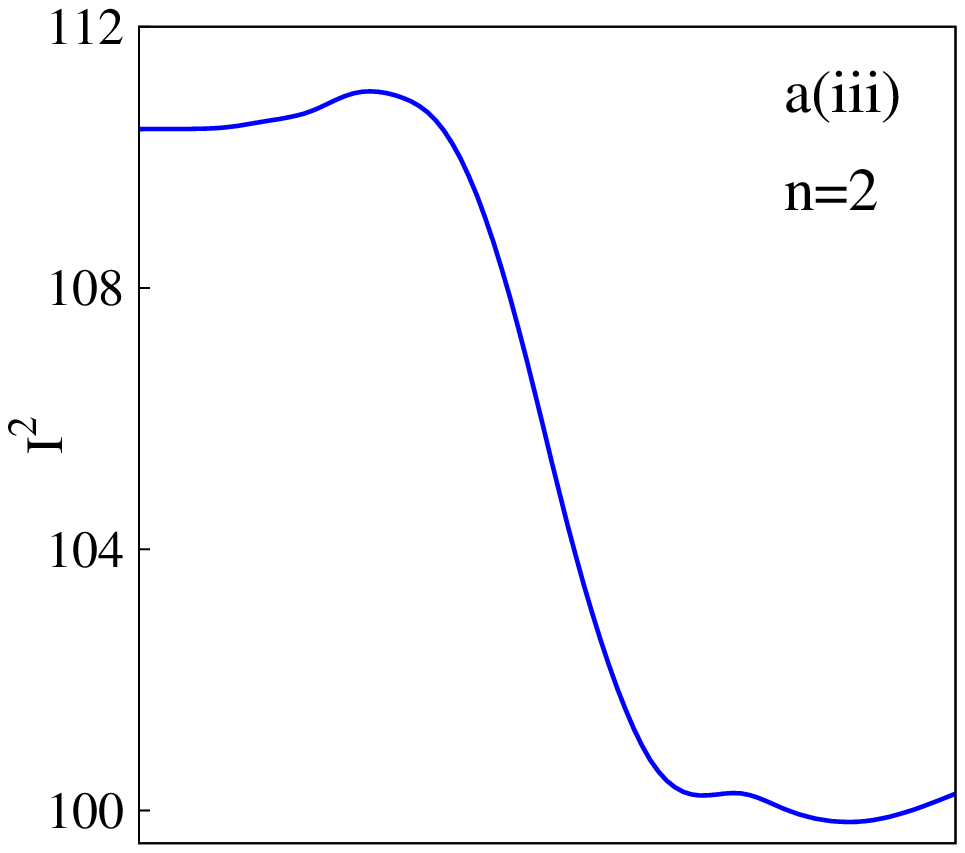}
\end{minipage}%
\hspace{0.5in}
\begin{minipage}[c]{0.25\textwidth}\centering
\includegraphics[scale=0.47]{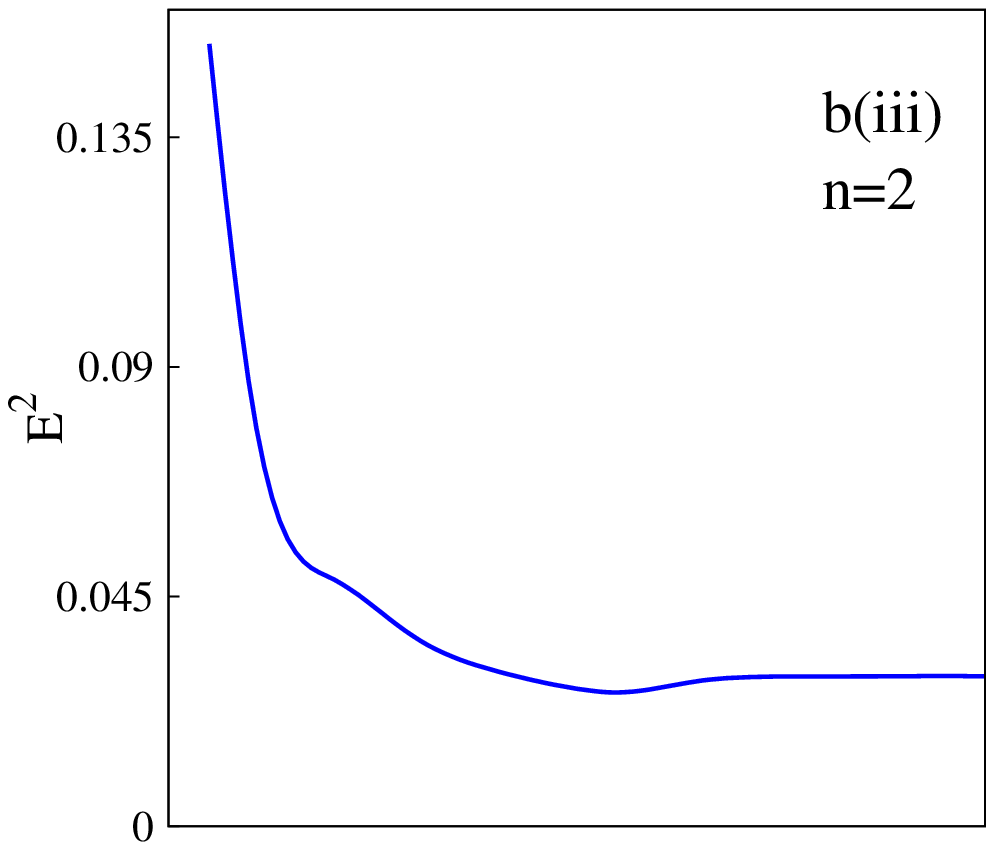}
\end{minipage}
\hspace{1in}
\vspace{0.1in}
\begin{minipage}[c]{0.25\textwidth}\centering
\includegraphics[scale=0.47]{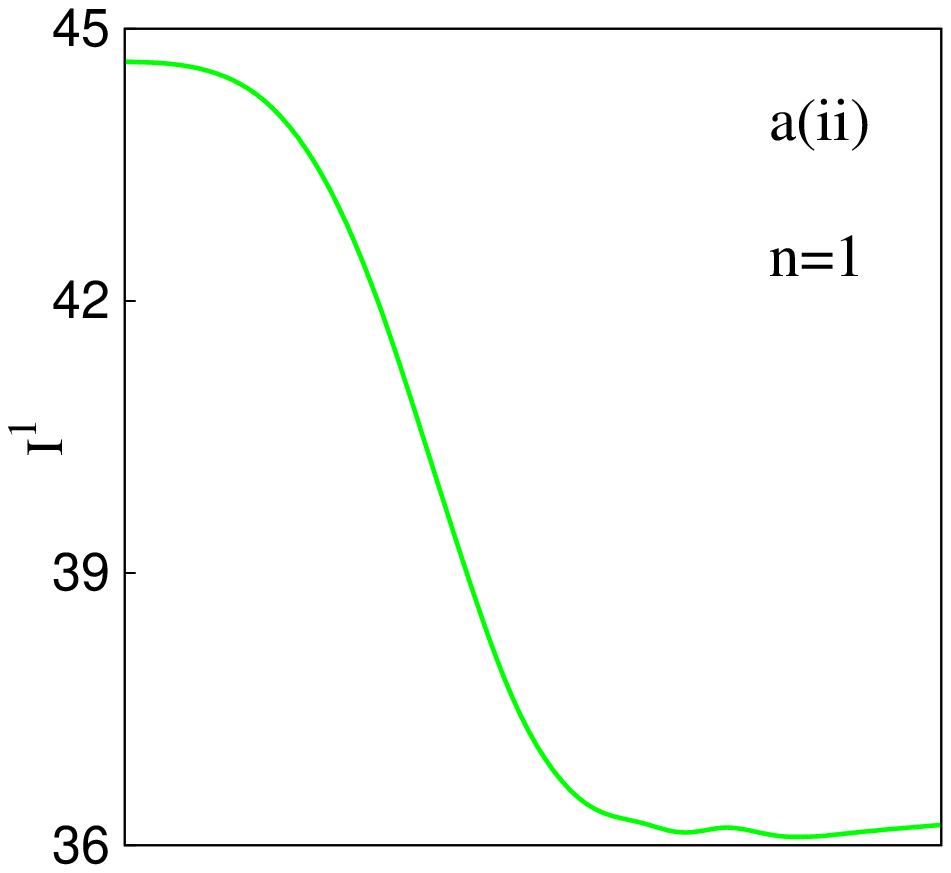}
\end{minipage}
\hspace{0.5in}
\begin{minipage}[c]{0.25\textwidth}\centering
\includegraphics[scale=0.47]{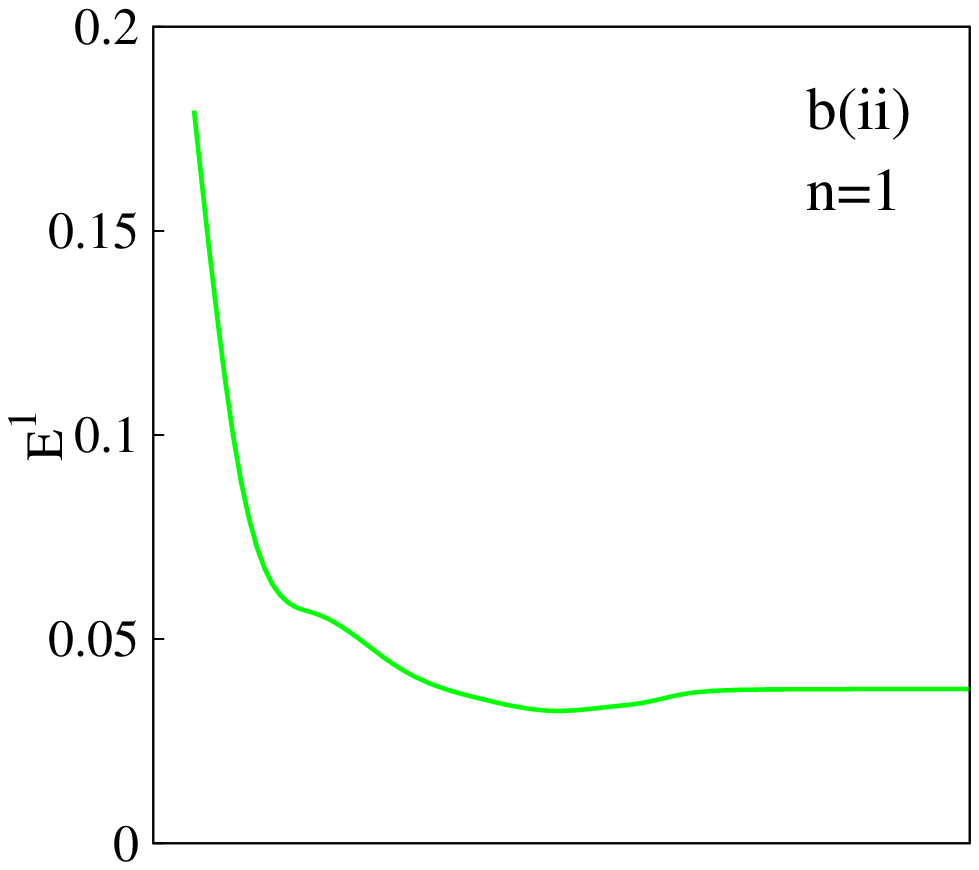}
\end{minipage}
\hspace{1in}
\vspace{0.1in}
\begin{minipage}[c]{0.28\textwidth}\centering
\includegraphics[scale=0.52]{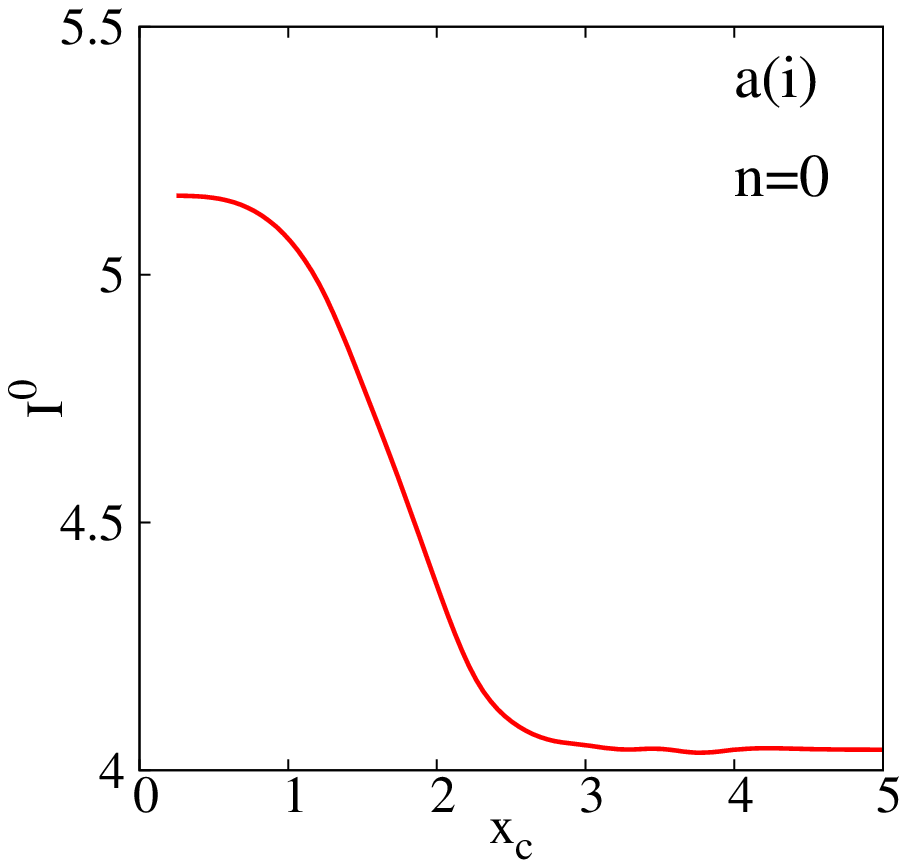}
\end{minipage}
\hspace{0.3in}
\begin{minipage}[c]{0.28\textwidth}\centering
\includegraphics[scale=0.52]{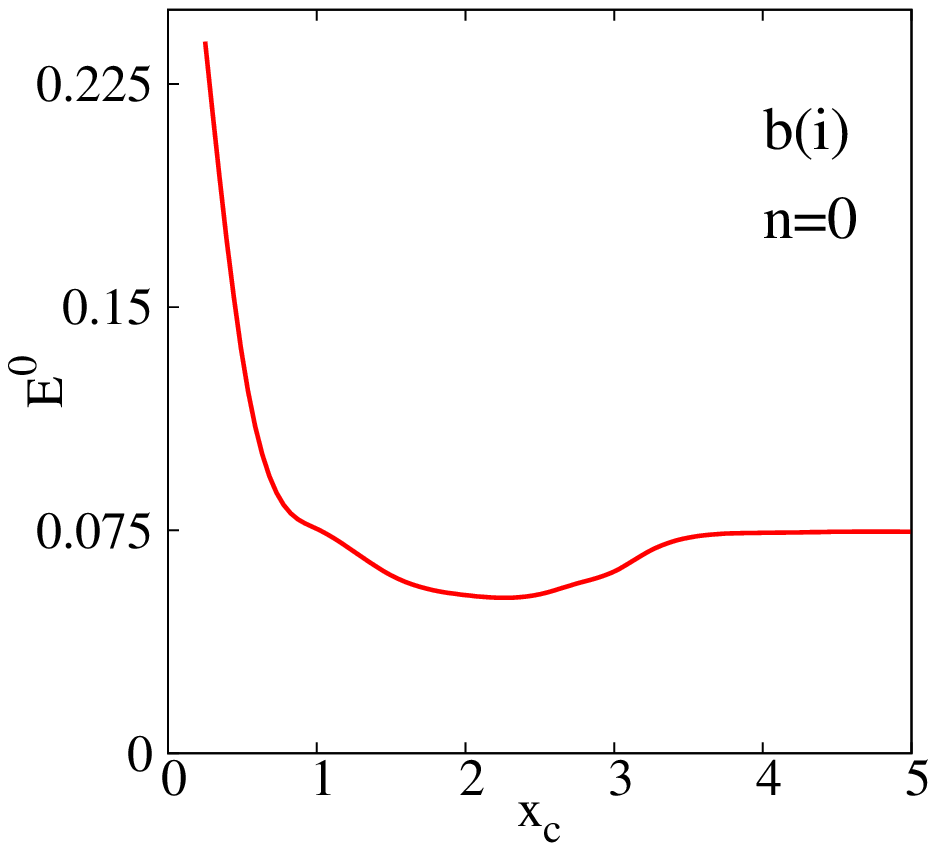}
\end{minipage}%
\caption{Plots of $I$, $E$ for first three energy states of SCHO potential as function of $x_c$. Panels
(a), (b) represent $I$ and $E$. For more details, see text.}
\end{figure}

Next we move on to $I$ and $E$, in a SCHO potential, which have not been studied before. For this, first we analyze 
$x$ space localization, then $p$ space de-localization and finally composite space localization-de-localization by 
computing $I_x,~E_x$; $I_p,~E_p$; $I,~E$. These calculation will consolidate the conclusions obtained from the study 
of $S_x$,~$S_p$, $S$ delineated earlier. Behavior of $I_{x}$, $I_{p}$ 
of first five stationary-states are demonstrated in two bottom panels (a), (b) of Fig.~(4). One sees that
initially $I_{x}$ falls sharply, extent of which is maximum for lowest state and progressively decreases with $n$. 
Then it assumes a constant value beyond a certain $x_c$. All the states remain well separated; no mixing occurs 
amongst them. On the other hand, in panel (b), $I_{p}$ strongly increases with $x_{c}$ at first; the extent 
increases with $n$; lowest state producing lowest. Finally, for all states, it again reaches a state-dependent constant 
value as in $I_x$. It is relevant to mention that, behavior of $I_x$, $I_p$ under confinement matches with those 
of $\Delta p$, $\Delta x$ \cite{laguna2014}. Next, from panel (c), we discern that, $E_x$ for all states remain 
very close to each other at smaller $x_c$; then as $x_c$ increases, $E_x$ for individual states branch out and decreases 
indicating localization. At the end, it assumes some constant value for all $n$. Interestingly, up to about 
$x_c \approx 1$, $E_x^n$ does not depend 
on $n$ quantum number. Lastly in (d), $E_p$s tend to increase in the beginning for all states at low $x_c$ region 
(indicating localization), eventually becoming flat after some threshold $x_c$.  

\begin{figure}                         
\begin{minipage}[c]{0.3\textwidth}\centering
\includegraphics[scale=0.65]{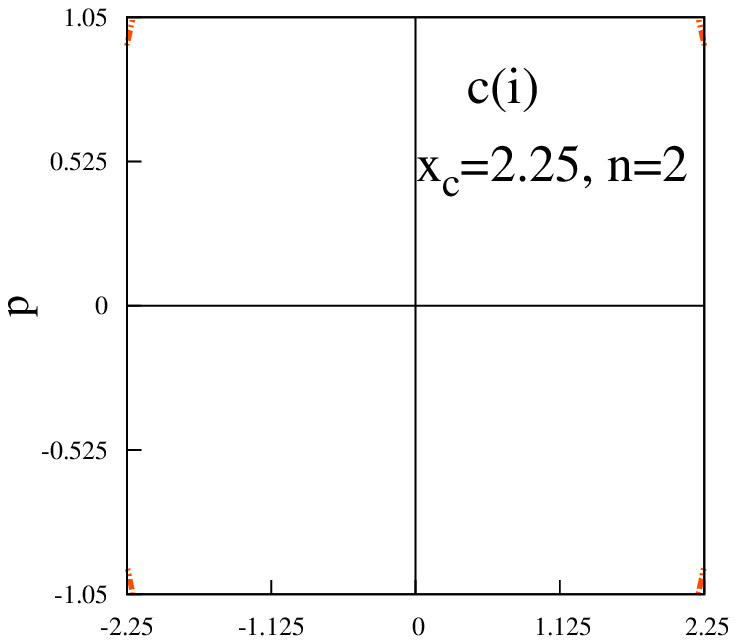}
\end{minipage}
\hspace{0.05in}
\begin{minipage}[c]{0.3\textwidth}\centering
\includegraphics[scale=0.65]{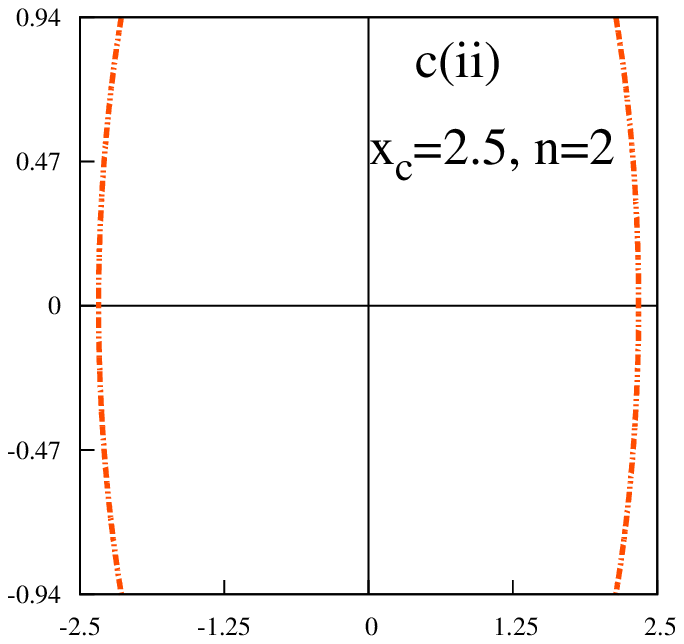}
\end{minipage}
\hspace{0.05in}
\begin{minipage}[c]{0.3\textwidth}\centering
\includegraphics[scale=0.65]{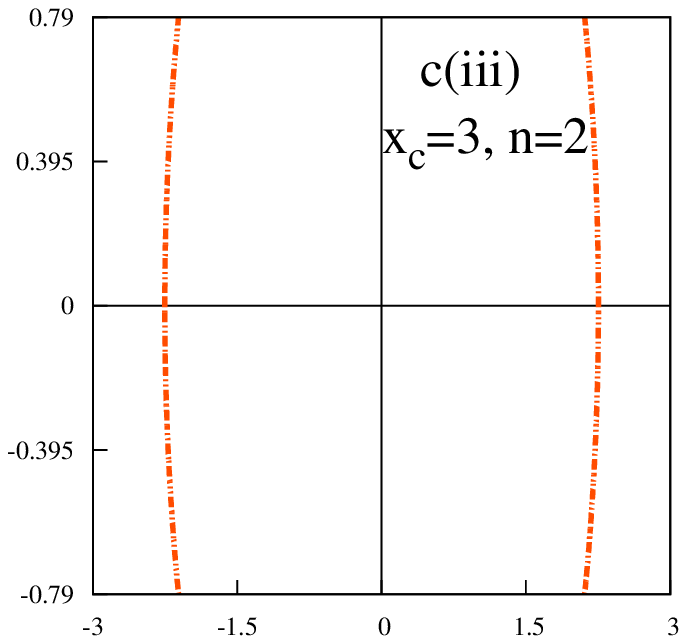}
\end{minipage}
\hspace{0.05in}
\vspace{0.1in}
\begin{minipage}[c]{0.3\textwidth}\centering
\includegraphics[scale=0.65]{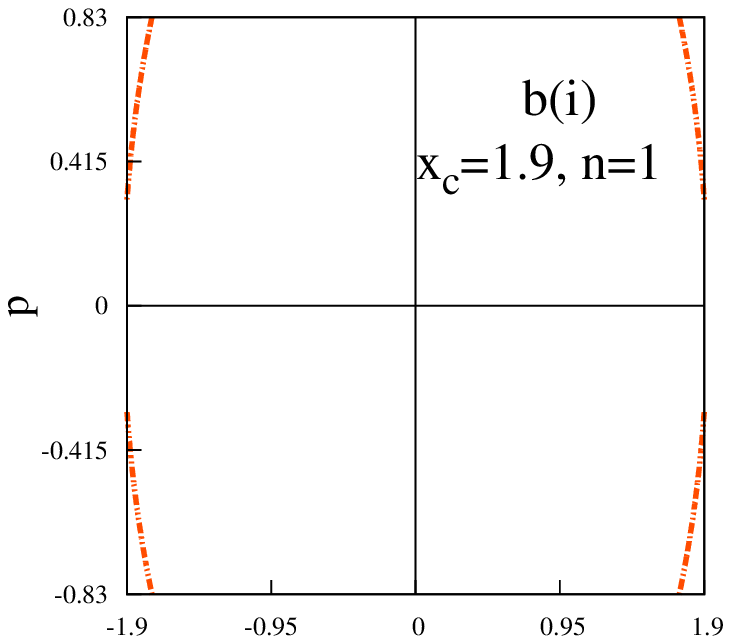}
\end{minipage}%
\hspace{0.05in}
\begin{minipage}[c]{0.3\textwidth}\centering
\includegraphics[scale=0.65]{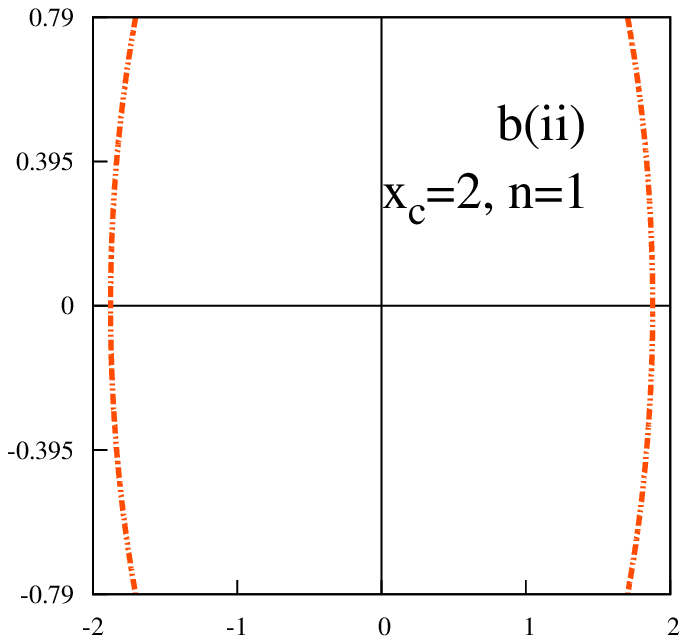}
\end{minipage}%
\hspace{0.05in}
\begin{minipage}[c]{0.3\textwidth}\centering
\includegraphics[scale=0.65]{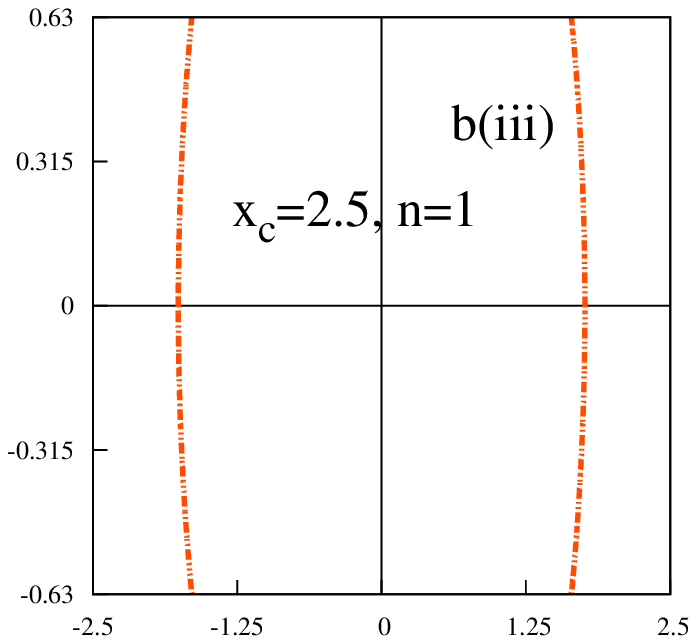}
\end{minipage}%
\hspace{0.05in}
\begin{minipage}[c]{0.3\textwidth}\centering
\includegraphics[scale=0.65]{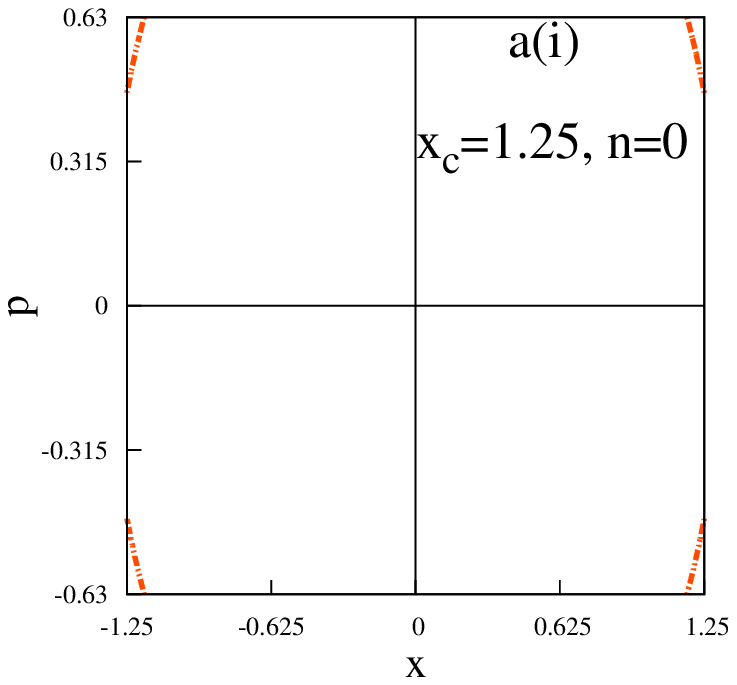}
\end{minipage}%
\hspace{0.05in}
\begin{minipage}[c]{0.3\textwidth}\centering
\includegraphics[scale=0.65]{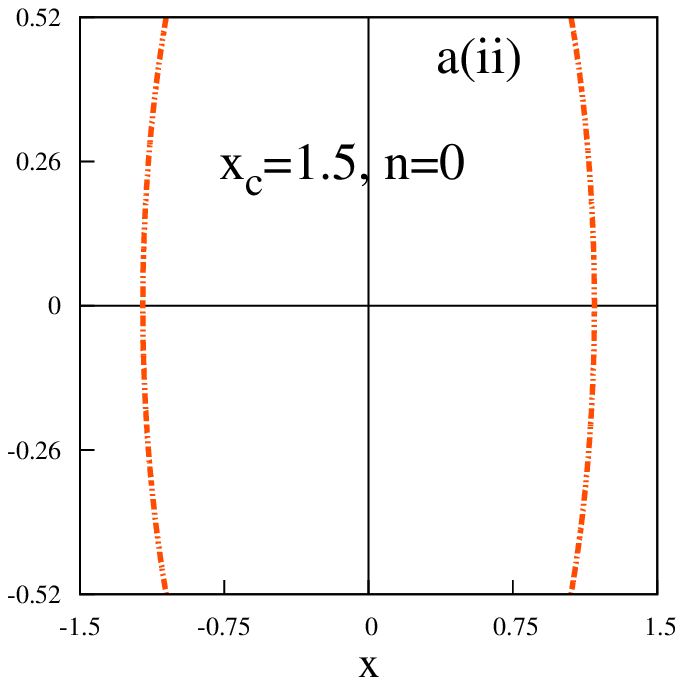}
\end{minipage}%
\hspace{0.05in}
\begin{minipage}[c]{0.3\textwidth}\centering
\includegraphics[scale=0.65]{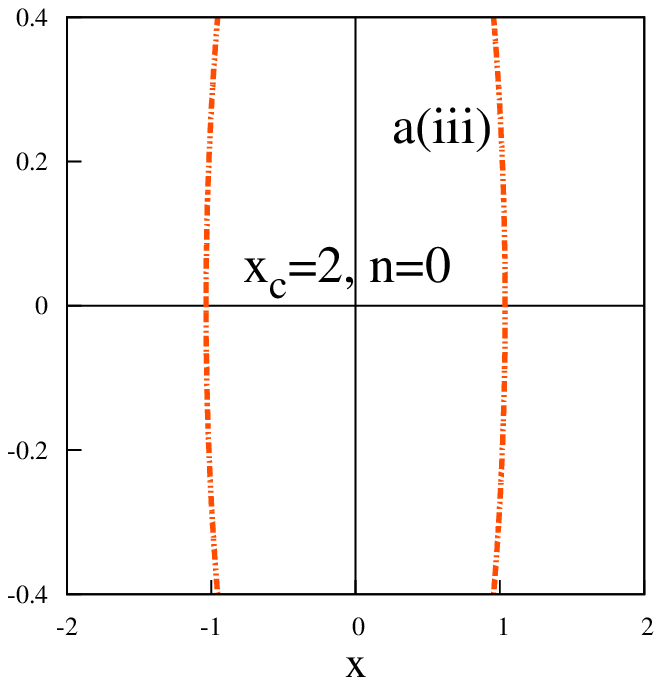}
\end{minipage}%
\caption{Semi-classical phase space contour plots for first three states of SCHO. 
For each state, three $x_c$s are chosen. See text for details.}
\end{figure}

Now, results on $I$, $E$ are given in left and right columns of Fig.~(5) and Fig.~(S4). Total $I$, for three states 
having to $n=0-2$ and remaining two excited states corresponding to $n=3,4$ are depicted, from bottom to top in 
a(i)--a(iii) of Fig.~(5) and a(i)--a(ii) of Fig.~(S4) respectively. For $n=2,3,4$, one notices maximum in $I$ plots, 
signifying a competition in localization-delocalization. Large $I$ values in small $x_c$ region correspond to PIB model, 
while smaller values in large $x_c$ signify those for a free QHO. Middle region of these plots indicates effect of 
confinement, implying that SCHO may be viewed as an intermediate between these two limiting cases. Further, it is
verified that behavior of $I$ resembles closely that of $\Delta x \Delta p$ for a SCHO \cite{laguna2014}. For 
highest states, however, we notice a reversal in these limiting $I$ values. Right-hand side panels b(i)--b(iii) of  
Fig.~(5) and b(i)--b(ii) of Fig.~(S4) portray $E$ plots for same five states, which in all cases, consistently 
decrease with increase in $x_c$. 

\begin{figure}                         
\begin{minipage}[c]{0.25\textwidth}\centering
\includegraphics[scale=0.46]{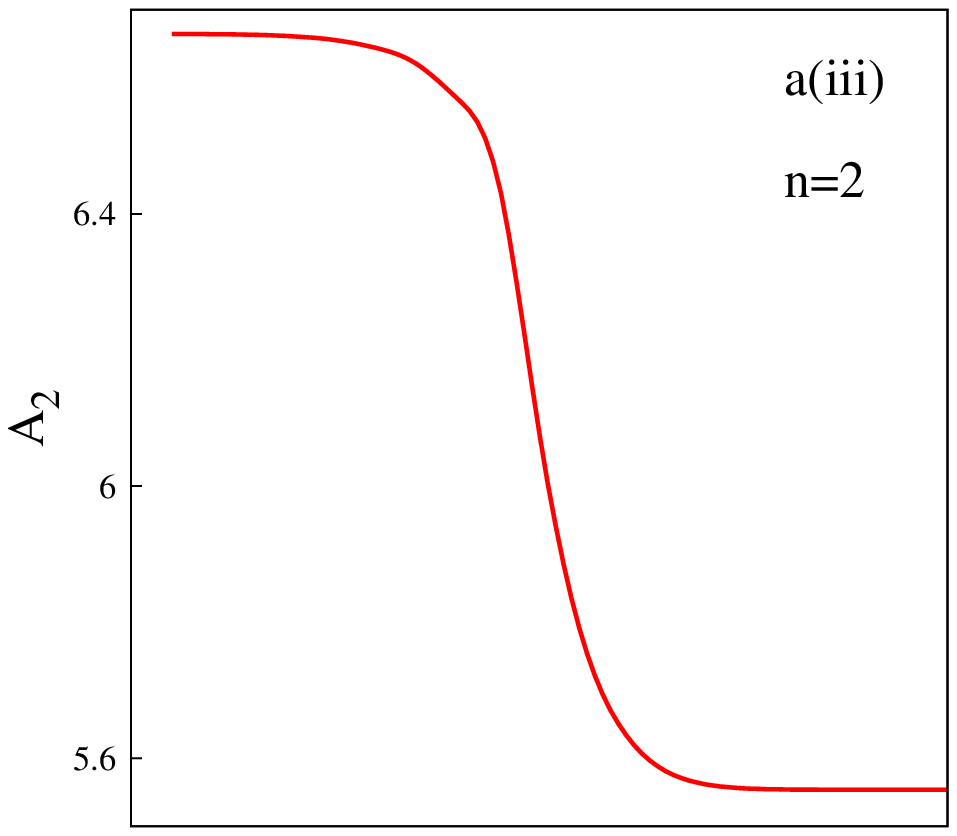}
\end{minipage}%
\hspace{0.8in}
\begin{minipage}[c]{0.25\textwidth}\centering
\includegraphics[scale=0.46]{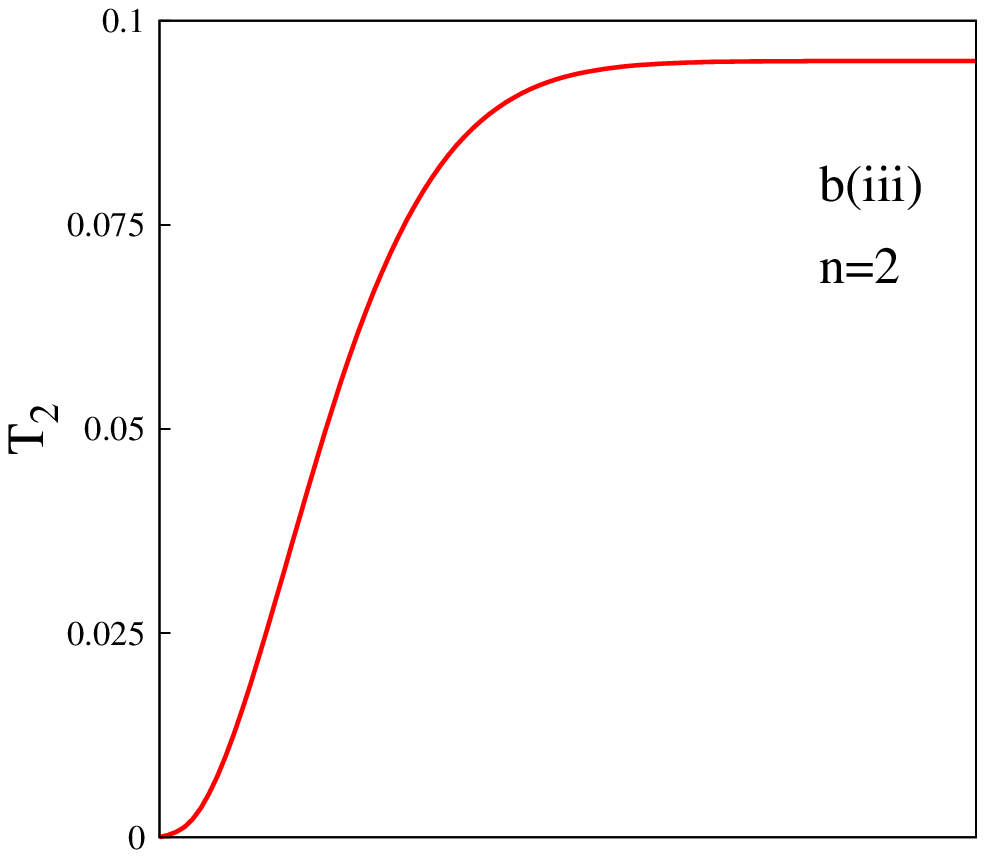}
\end{minipage}
\hspace{0.8in}
\vspace{0.15in}
\begin{minipage}[c]{0.25\textwidth}\centering
\includegraphics[scale=0.46]{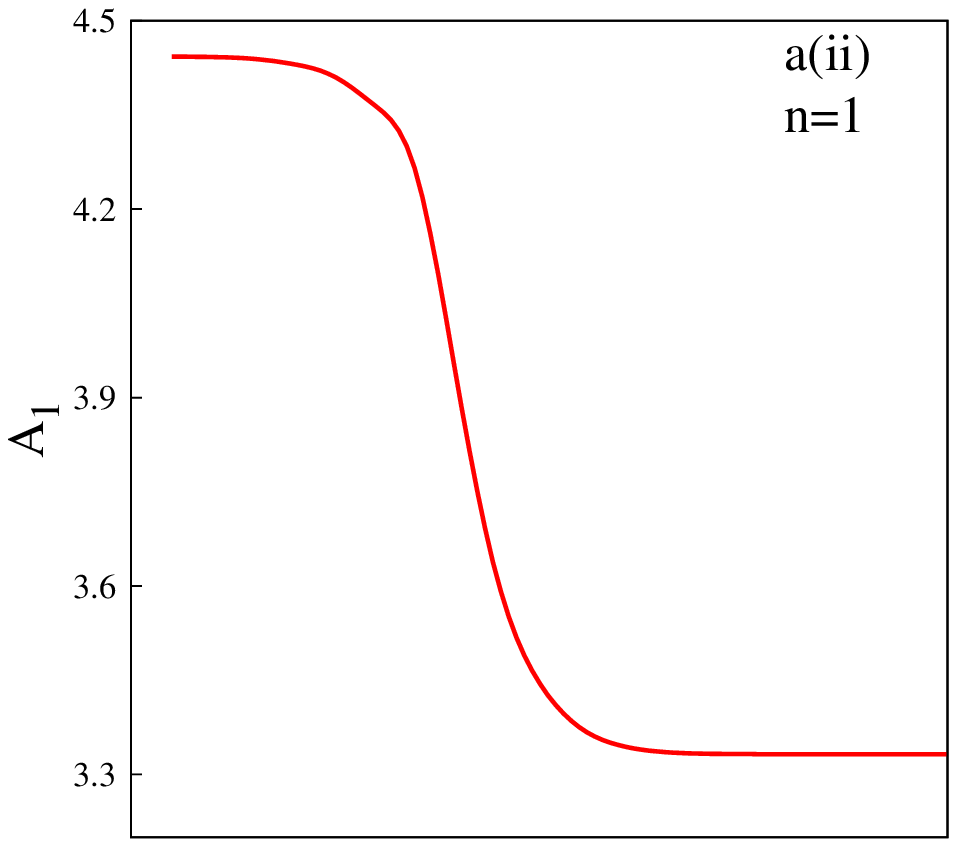}
\end{minipage}
\hspace{0.8in}
\begin{minipage}[c]{0.25\textwidth}\centering
\includegraphics[scale=0.46]{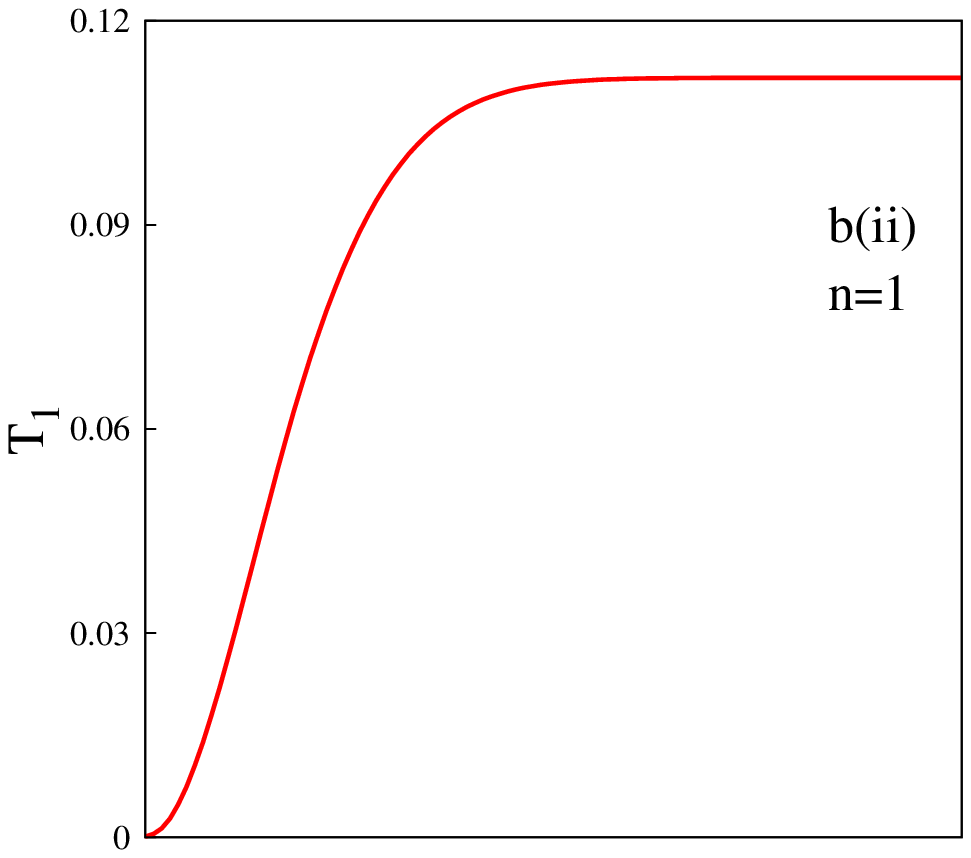}
\end{minipage}
\hspace{0.8in}
\begin{minipage}[c]{0.27\textwidth}\centering
\includegraphics[scale=0.5]{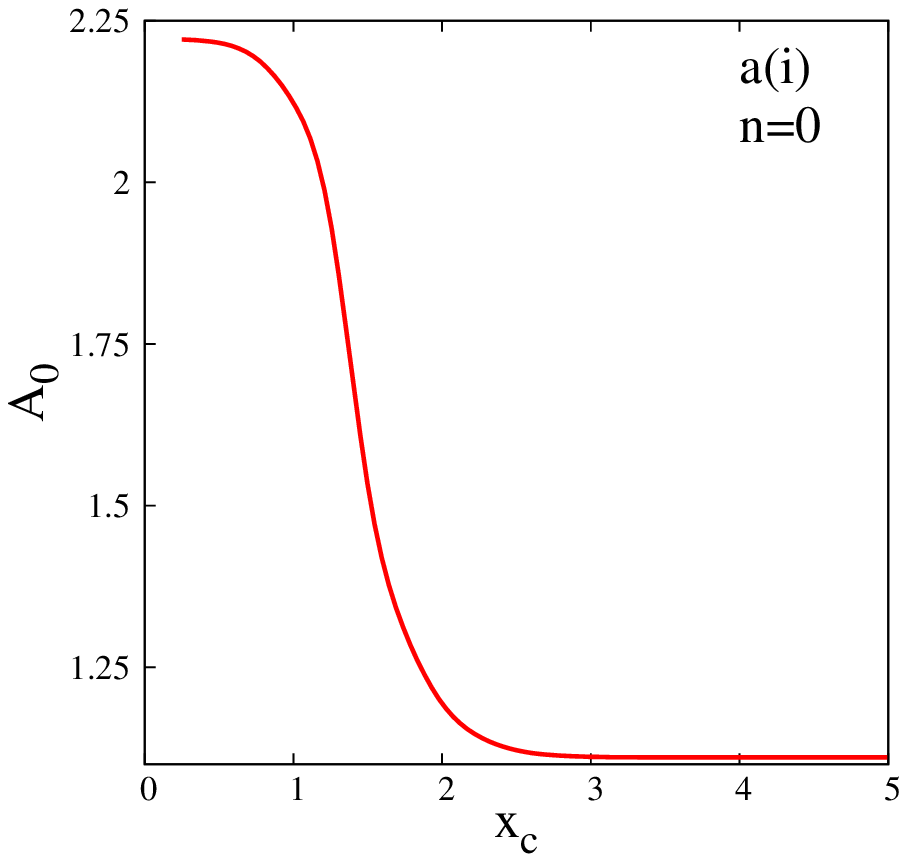}
\end{minipage}
\hspace{0.8in}
\begin{minipage}[c]{0.27\textwidth}\centering
\includegraphics[scale=0.5]{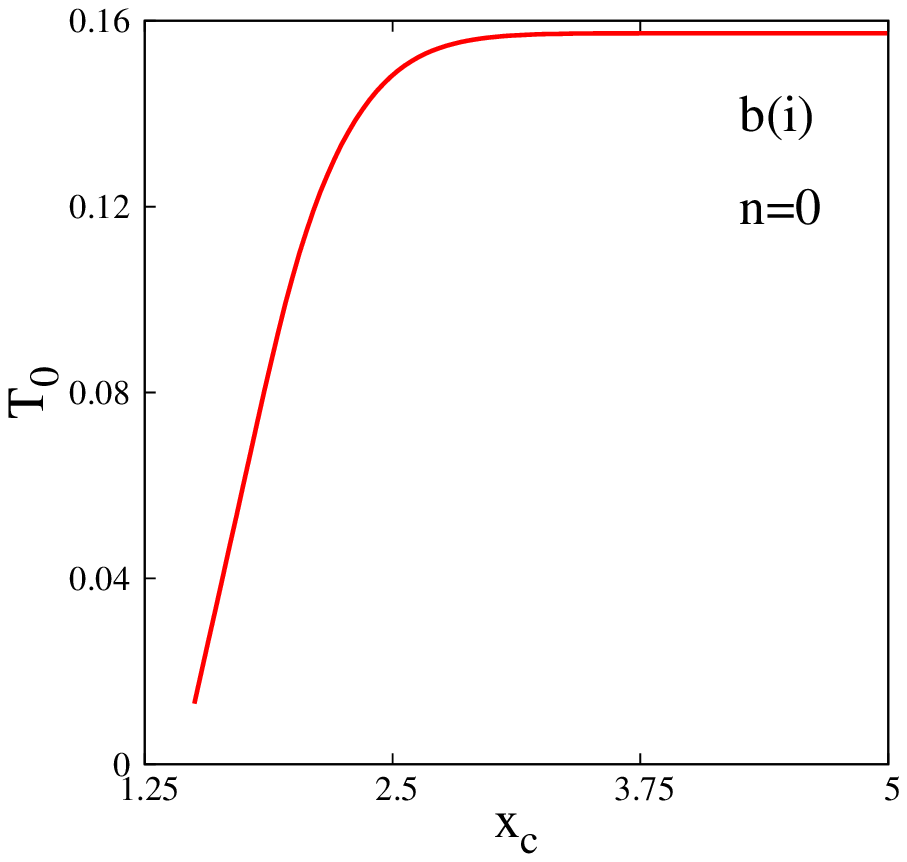}
\end{minipage}%
\caption{Phase-space area ($A_n$) and probability ($T_n$) of finding the particle outside classical region with 
changes in $x_c$ for $n=0-2$ of SCHO, in panels (a) and (b). For details, see text.}
\end{figure}

In order to get further insight, next we present phase space for same lowest five states of SCHO 
potential for three distinct sets of $x_c$ (not necessarily same for all $n$). These are 
displayed in Fig.~(6) ($n=0-2$) and Fig.~(S5) ($n=3,4$). For $n=0$, phase space is examined at $x_c=1.25, 1.5, 2$ in 
panels a(i)--a(iii) of Fig.~(6). At the first 
$x_c$, it indicates a PIB-like behavior; at remaining two $x_c$, however, its shape changes. Appearance of a line 
in phase space in a(i) of Fig.~(6) (corresponding to $x_c = 1.25$) indicates the onset of tunneling. Similarly, the first 
excited state phase spaces are plotted for $x_c= 1.9,2$ and 2.5 respectively in segments b(i)--b(iii) of Fig.~(6).   
In this occasion too, shape of phase space changes with increase in $x_c$. From b(i), we discern that, 
tunneling begins approximately at $x_c=1.9$. Analogous conclusions can be drawn from phase spaces of $n=2-4$,
where tunneling starts at $x_c \approx 2.25, 2.75 \ \mathrm{and} \ 3$. Actually, this study only exhibits the PIB 
behavior in small $x_c$ region, but somehow insufficient to explain the large $x_c$ nature of SCHO potential. 

\begingroup      
\squeezetable    
\begin{table}
\centering
\caption{Comparison of energies of first six states of ACHO potential at four distinct $d_m$. Length of the box is 
fixed at 2. PR implies Present Result. See text for details.} 
\begin{ruledtabular}
\begin{tabular}{|l|ll|ll|} 
$d_m$ & $\epsilon_0$ (PR) &  $\epsilon_0$ (Ref.) & $\epsilon_1$ (PR) &  $\epsilon_1$(Ref.) \\
\hline
0.36 & 2.7177633960054 & 2.59691966$^\dag$, 2.7177633960054$^\ddag$         
     & 10.283146010610 & 10.28314602$^\dag$, 10.283146010610$^\ddag$         \\
1.92 & 6.0383021056781 & 6.03830195$^\dag$, 6.0383021056781$^\ddag$             
     & 13.901445986629 & 13.90144582$^\dag$, 13.901445986629$^\ddag$         \\
5.00 & 26.065225076406 & 26.065225076406$^{\S}$ & 35.462261039378 & 35.462261039378$^{\S}$ \\
10.0 & 97.474035270680 & 97.474035270680$^{\S}$ & 110.51944554927 & 110.51944554927$^{\S}$ \\
   \hline
    & $\epsilon_2$ (PR) &  $\epsilon_2$ (Ref.) & $\epsilon_3$ (PR) &  $\epsilon_3$(Ref.) \\
    \hline
0.36 & 22.648848755052 & 22.64884877$^\dag$, 22.648848755052$^\ddag$           
     & 39.929984298830 & 32.92998431$^\dag$, 39.929984298830$^\ddag$         \\
1.92 & 26.249310409373 & 26.24931024$^\dag$, 26.249310409373$^\ddag$          
     & 43.513981920357 & 43.51398176$^\dag$, 43.513981920357$^\ddag$          \\
5.00 & 47.817024422796 & 47.817024422796$^{\S}$  & 64.900200447511 & 64.900200447511$^{\S}$  \\
10.0 & 123.593144939095& 123.593144939095$^{\S}$ & 140.555432078323  & 140.555432078323$^{\S}$  \\
    \hline
    & $\epsilon_4$ (PR) &  $\epsilon_4$ (Ref.) & $\epsilon_5$ (PR) &  $\epsilon_5$(Ref.) \\
    \hline
0.36 & 62.140768627508 & 62.140768627508$^\ddag$ & 89.284409553063  & 89.284409553063$^\ddag$ \\
1.92 & 65.715672311936 & 65.715672311936$^\ddag$ & 92.854029622882  & 92.854029622882$^\ddag$ \\
5.00 & 87.137790461503 & 87.137790461503$^{\S}$ & 114.244486402564 & 114.244486402564$^{\S}$  \\
10.0 & 162.519960161732 & 162.519960161732$^{\S}$ & 189.515389275133 & 189.515389275133$^{\S}$  \\
\end{tabular}
\begin{tabbing}
$^\dag$Power-series solution \cite{campoy02}. \hspace{35pt} \=  
$^\ddag$ITP result \cite{roy15}. \hspace{35pt} \=
$^{\S}$ITP result, newly done here, for this table.
\end{tabbing}
\end{ruledtabular}
\end{table} 
\endgroup  

Additional information on confinement could be gathered from a semi-classical interpretation. For this we computed
phase-space area $A_n$ (left side) as well as probability $T_n$ (right side) of finding the particle outside classical 
region for $n=0-4$; these are depicted in panels a(i)--a(iii), b(i)--b(iii) of Fig.~(7) for $n=0-2$, and 
in a(i)--a(ii), b(i)--b(ii) of Fig.~(S6) for $n=3,4$. For all $n$,
we observe a swift jump of $A_n$ from a higher to lower value at a certain characteristic $x_c$, marking the 
onset of tunneling, as encountered earlier. For $n=0-4$, this occurs roughly when $x_c$ falls in the range of 
1.25--1.5, 1.75--2, 2.25--2, 2.75--3 and 3--3.25. Here tunneling begins at a point when particle starts 
sensing the presence of harmonic potential. Another fact is that $A_n$ values converge to 1.11, 3.33, 5.55, 7.77 and 
9.99 for $n=0-4$. Appearance of such convergence in $A_n$ implies that at large $x_c$, SCHO 
potential behaves much like a QHO potential. Next, we present $T_n$  
for same lowest five states in right panels b(i)--b(iii) of Fig.~(7) ($n=0-2$) and b(i)--b(ii) of Fig.~(S6) ($n=3,4$). 
For all $n$, $T_n$ increases steadily reaching a particular value characteristic for that given state. These values for 
$n=0-4$ are roughly
0.16, 0.11, 0.095, 0.085 and 0.078. This again explains that at large $x_c$, SCHO potential behaves like 
a QHO system. Here $T_n$ decreases with increase in $n$, as $A_n$ increases with $n$.   
In summary, small $x_c$ region can be understood by the nature of phase-space, whereas large $x_c$ domain can be 
interpreted in terms of $A_n$ and $T_n$. 

\begin{figure}                         
\begin{minipage}[c]{0.3\textwidth}\centering
\includegraphics[scale=0.5]{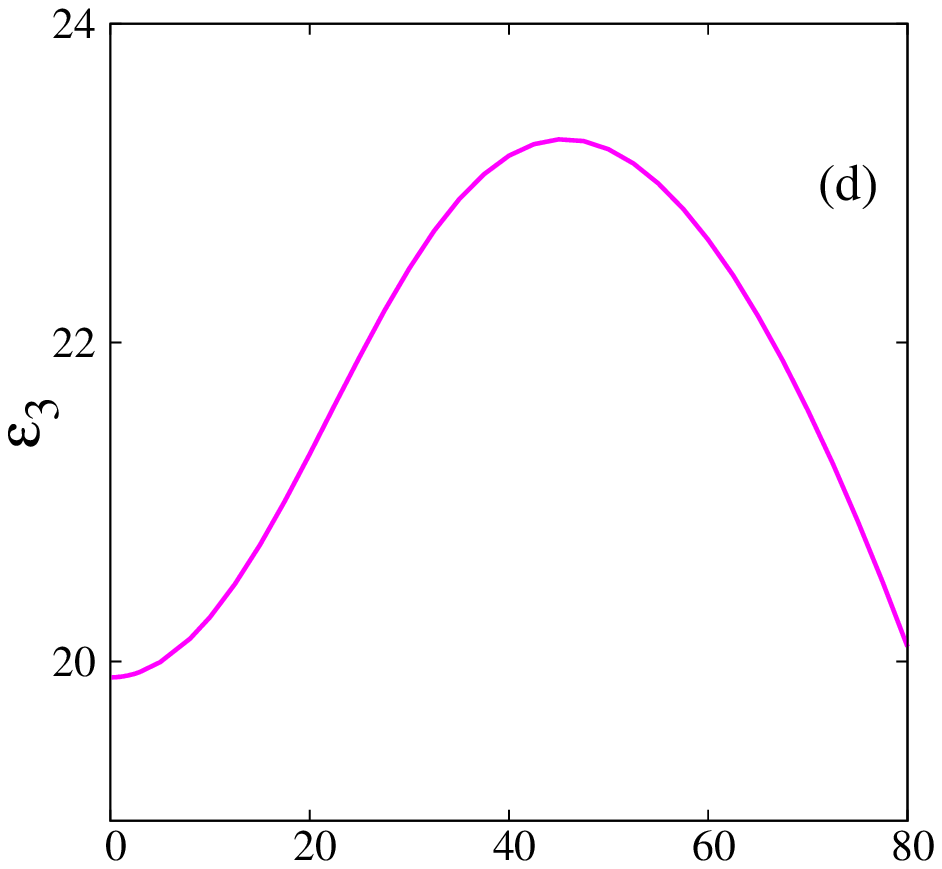}
\end{minipage}%
\hspace{0.15in}
\begin{minipage}[c]{0.3\textwidth}\centering
\includegraphics[scale=0.5]{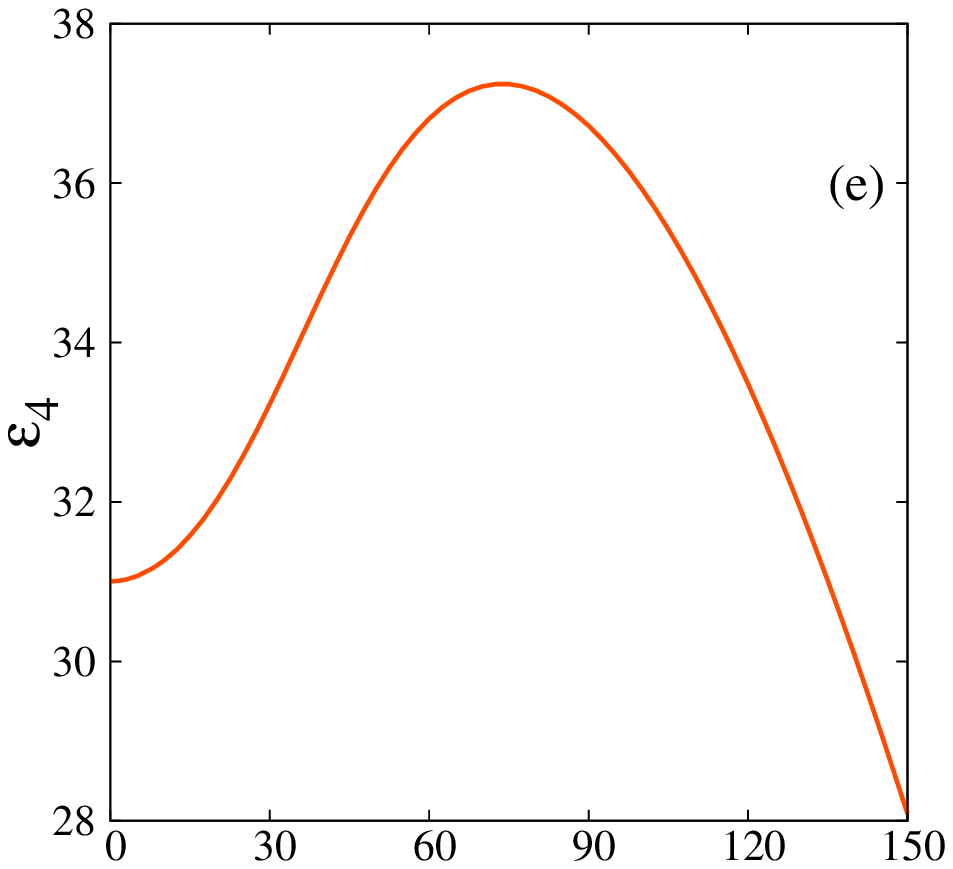}
\end{minipage}%
\hspace{0.15in}
\vspace{0.2in}
\begin{minipage}[c]{0.3\textwidth}\centering
\includegraphics[scale=0.5]{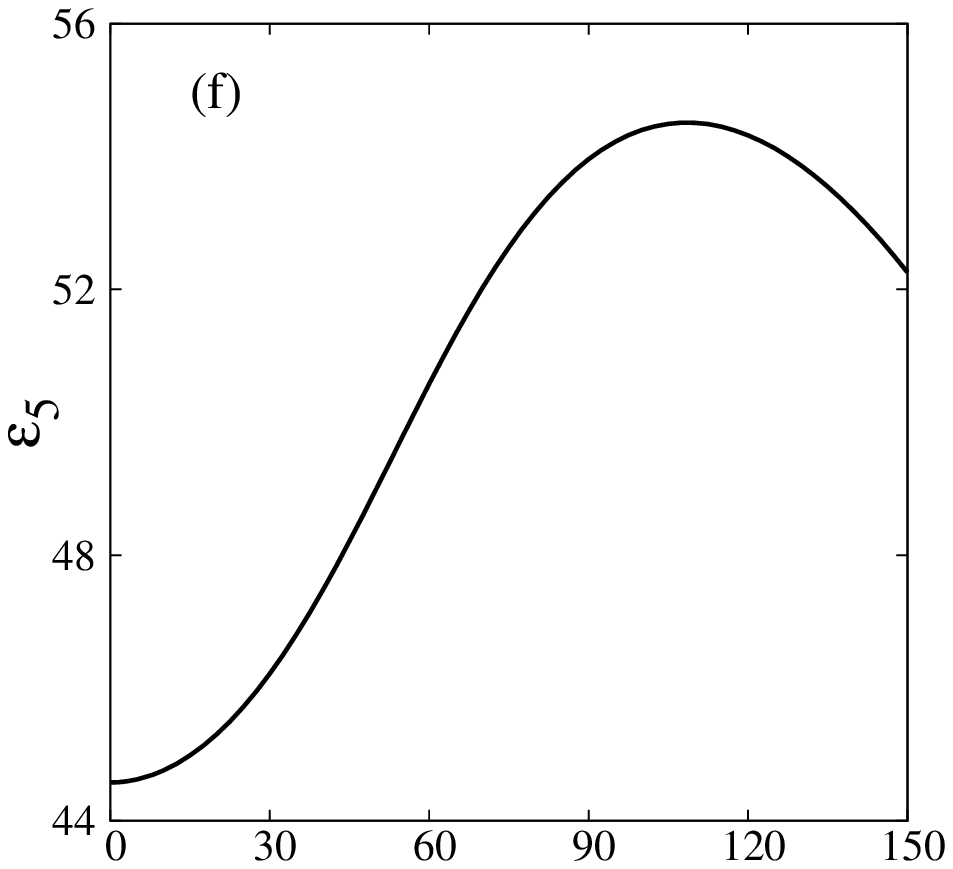}
\end{minipage}%
\hspace{0.01in}
\begin{minipage}[c]{0.31\textwidth}\centering
\includegraphics[scale=0.52]{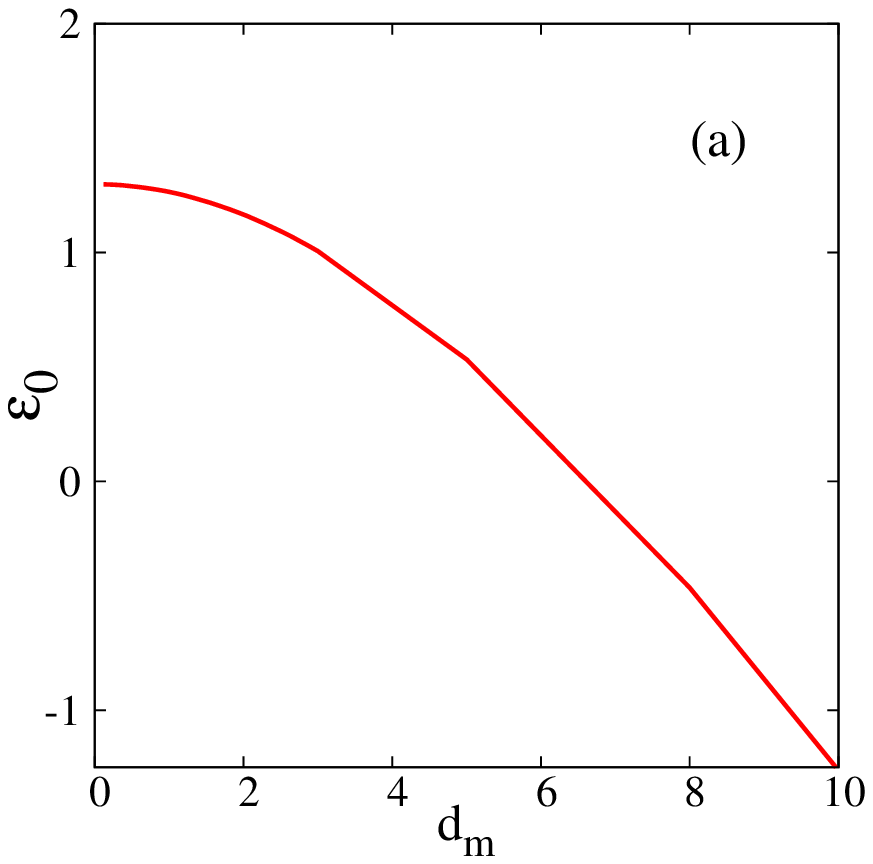}
\end{minipage}
\hspace{0.1in}
\begin{minipage}[c]{0.31\textwidth}\centering
\includegraphics[scale=0.52]{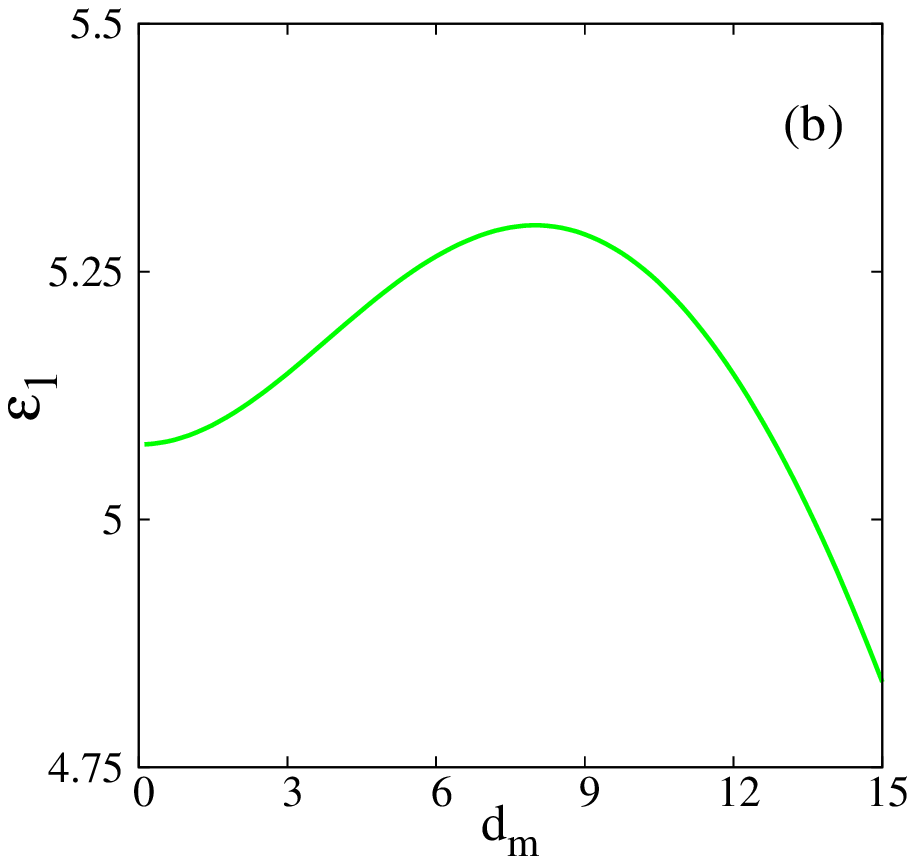}
\end{minipage}%
\hspace{0.01in}
\begin{minipage}[c]{0.31\textwidth}\centering
\includegraphics[scale=0.52]{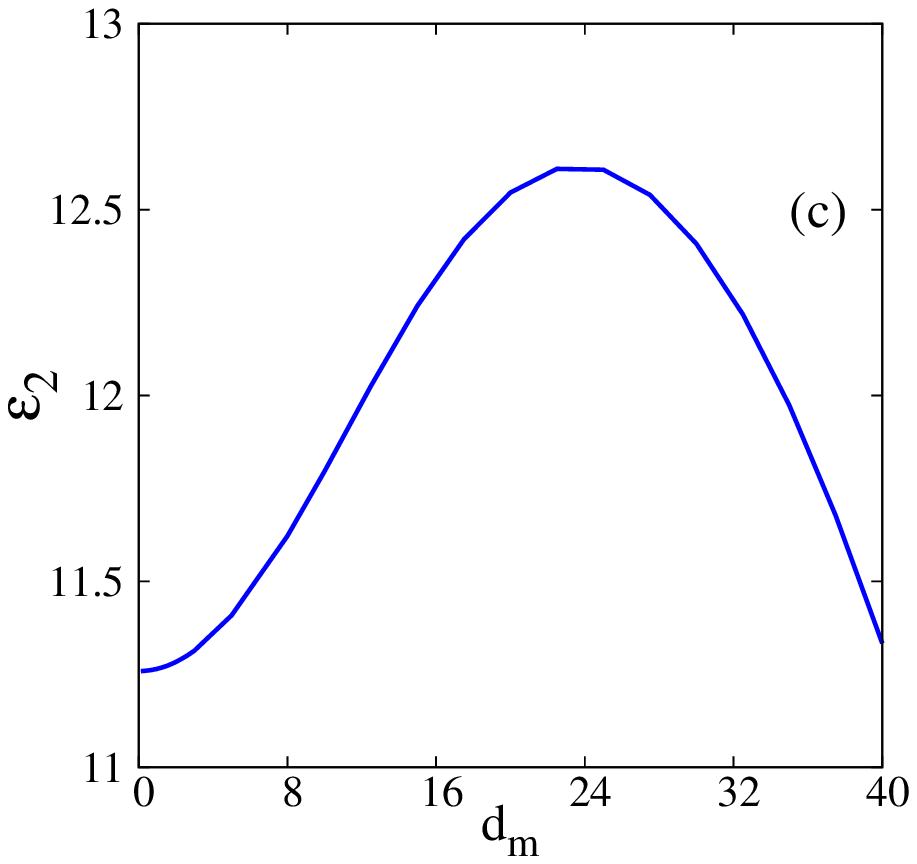}
\end{minipage}%
\caption{Variation of $\epsilon_n$ ($n \! = \! 0-5$) of ACHO potential as function of $d_m$. See text for details.}
\end{figure}

\subsection{Asymmetric confinement}
Now, we move on to a discussion on ACHO case. At first we make the ACHO Hamiltonian into dimensionless form, represented as
\begin{equation}
-\frac{\hbar^2}{2m} \frac{d^2 \psi}{dx^2} + \frac{1}{2} k(x-d_m)^2\psi + V\theta(x^2-x_{c}^{2}) \psi = \epsilon \psi, \\
\end{equation}
From this equation, one reads that, 
\begin{align*}
 \epsilon \equiv  \epsilon \bigg(\frac{\hbar^2}{m}, k, d_m, x_c\bigg) \quad \mathrm{and} 
 \quad \psi \equiv \psi \bigg(\frac{\hbar^2}{m}, k, d_m, x_c, x\bigg).
\end{align*}
As in the SCHO case, again incorporation of $x=\lambda x'$ and $x_c = \lambda $ into Eq.~(27), leads to a modified dimensionless 
Hamiltonian, expressed as, 
\begin{equation}
-\frac{1}{2} \ \frac{d^2 \psi (x')}{dx'^2} + \frac{1}{2} \ \frac{m}{\hbar^2} k x_c^{4} 
\left( x'-\frac{d_m}{\lambda} \right)^2 \psi(x') + 
V\theta(x'^2-1) \psi (x')= \frac{m x_c^{2}}{\hbar^2} \epsilon \psi (x')
\end{equation}
where $x'$ is a dimensionless variable. This gives the following set of equations,
\begin{align}
\begin{rcases}
 \epsilon \bigg(\frac{\hbar^2}{m}, k, d_m, x_c\bigg)  =  
 \frac{\hbar^2}{mx_c^{2}} \epsilon \bigg(1, \frac{mkx_c^{4}}{\hbar^2}, d_m, 1 \bigg) \\ 
 \psi \bigg(\frac{\hbar^2}{m}, k, d_m, x_c, x\bigg)  =  \frac{1}{\sqrt{\lambda}}  \psi \bigg(1, \eta, d_m, 1, x' \bigg)  \\
 \eta  =  \frac{mkx_c^{4}}{\hbar^2}.  
\end{rcases}
\end{align}

\begin{figure}                         
\begin{minipage}[c]{0.3\textwidth}\centering
\includegraphics[scale=0.5]{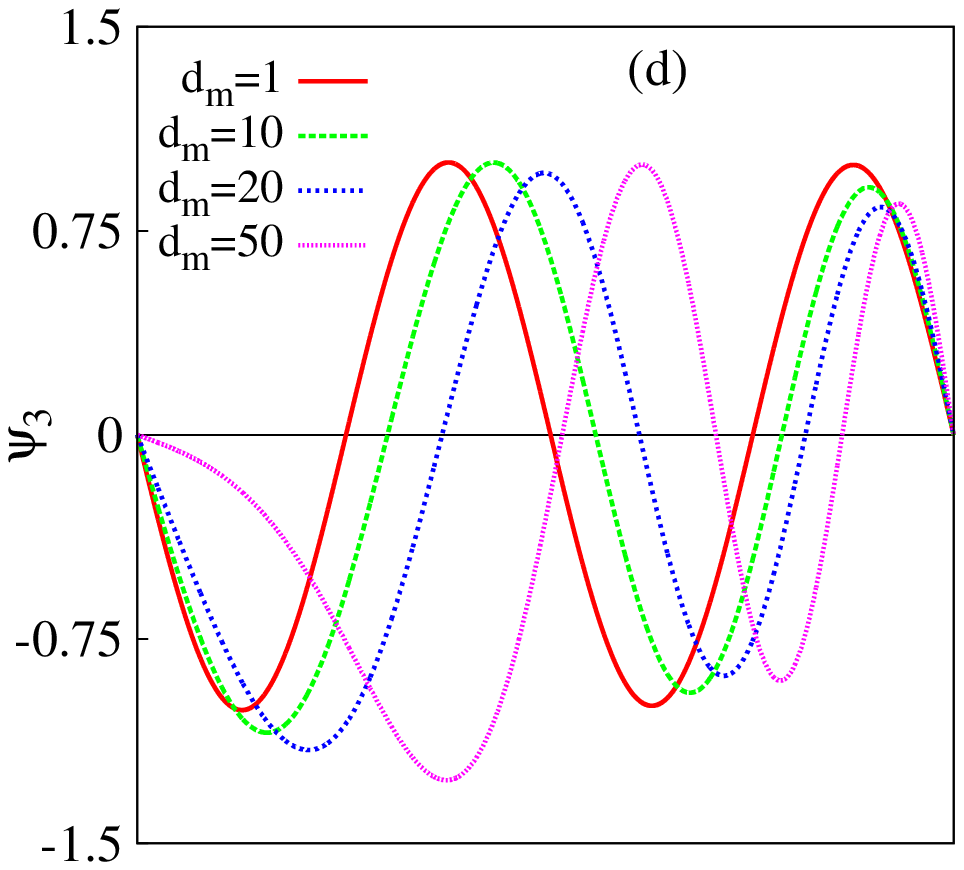}
\end{minipage}%
\hspace{0.1in}
\begin{minipage}[c]{0.3\textwidth}\centering
\includegraphics[scale=0.5]{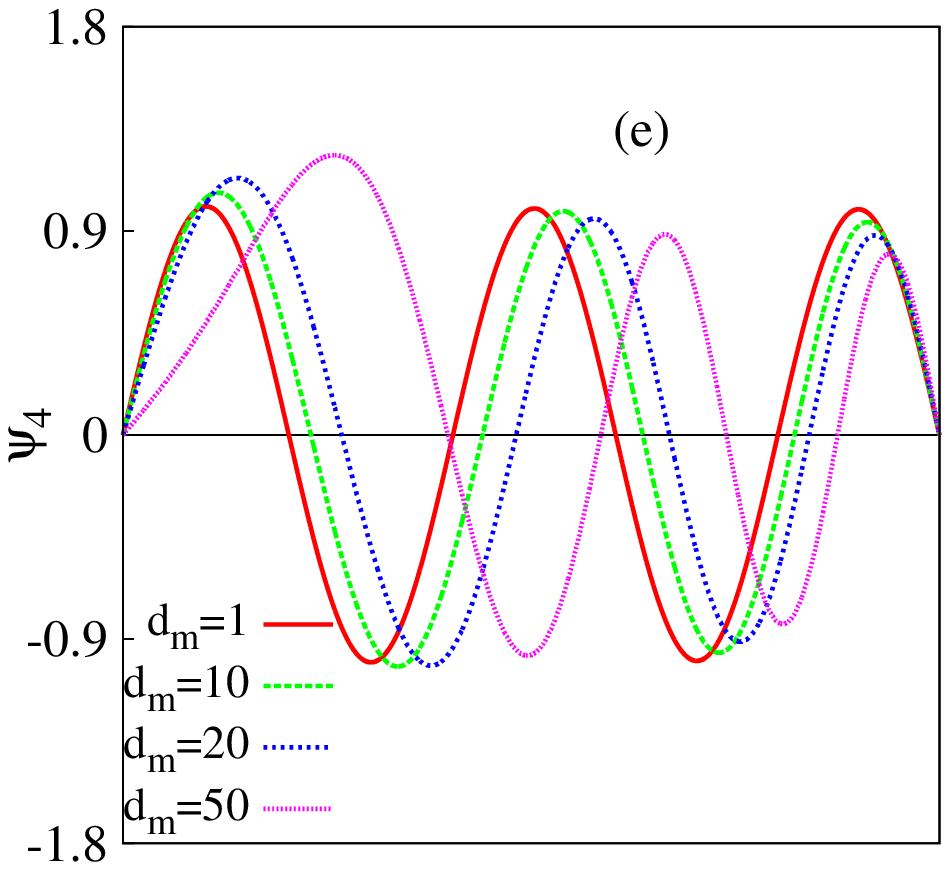}
\end{minipage}%
\hspace{0.1in}
\vspace{0.1in}
\begin{minipage}[c]{0.3\textwidth}\centering
\includegraphics[scale=0.5]{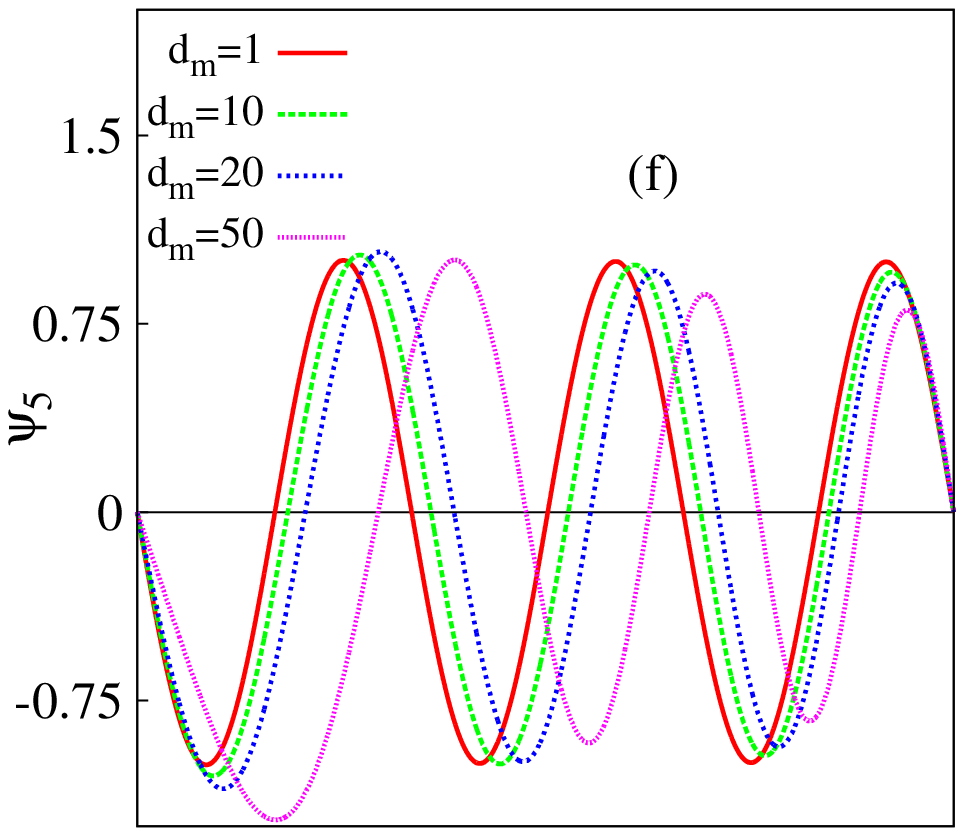}
\end{minipage}%
\hspace{0.1in}
\begin{minipage}[c]{0.3\textwidth}\centering
\includegraphics[scale=0.52]{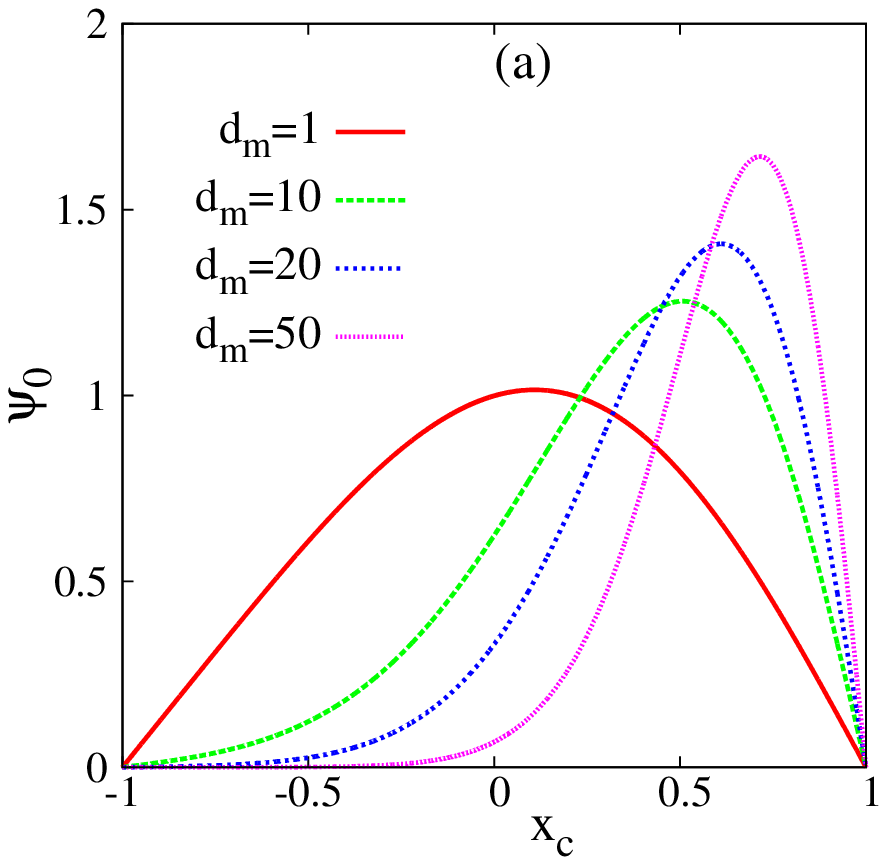}
\end{minipage}%
\hspace{0.1in}
\begin{minipage}[c]{0.3\textwidth}\centering
\includegraphics[scale=0.52]{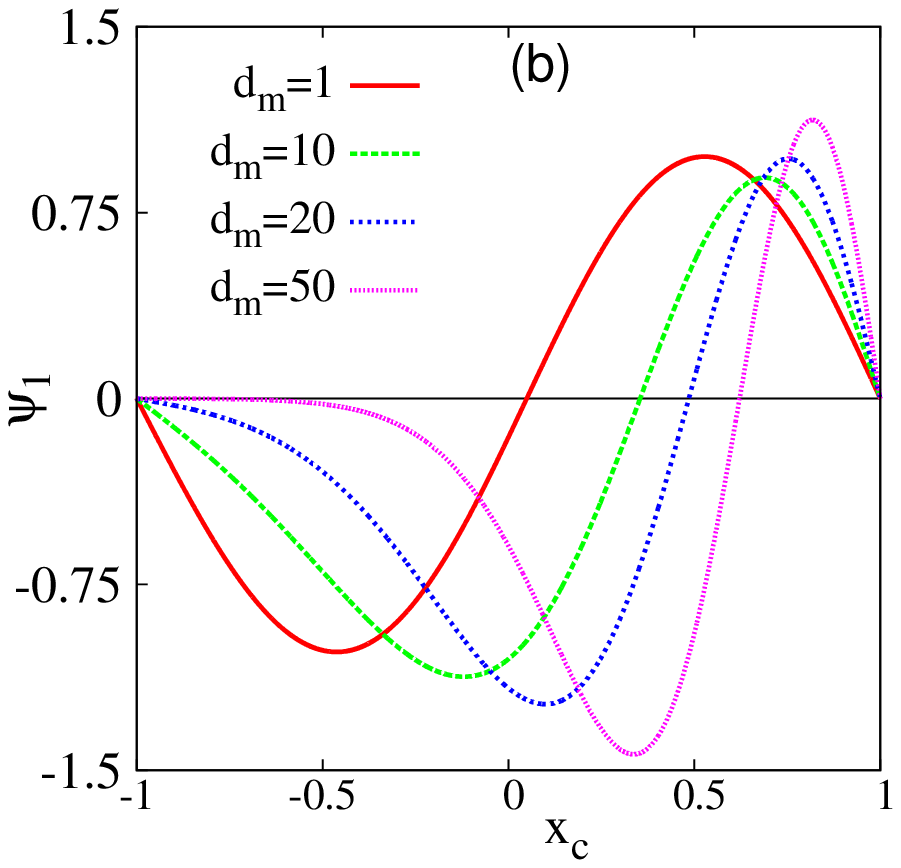}
\end{minipage}%
\hspace{0.1in}
\begin{minipage}[c]{0.3\textwidth}\centering
\includegraphics[scale=0.52]{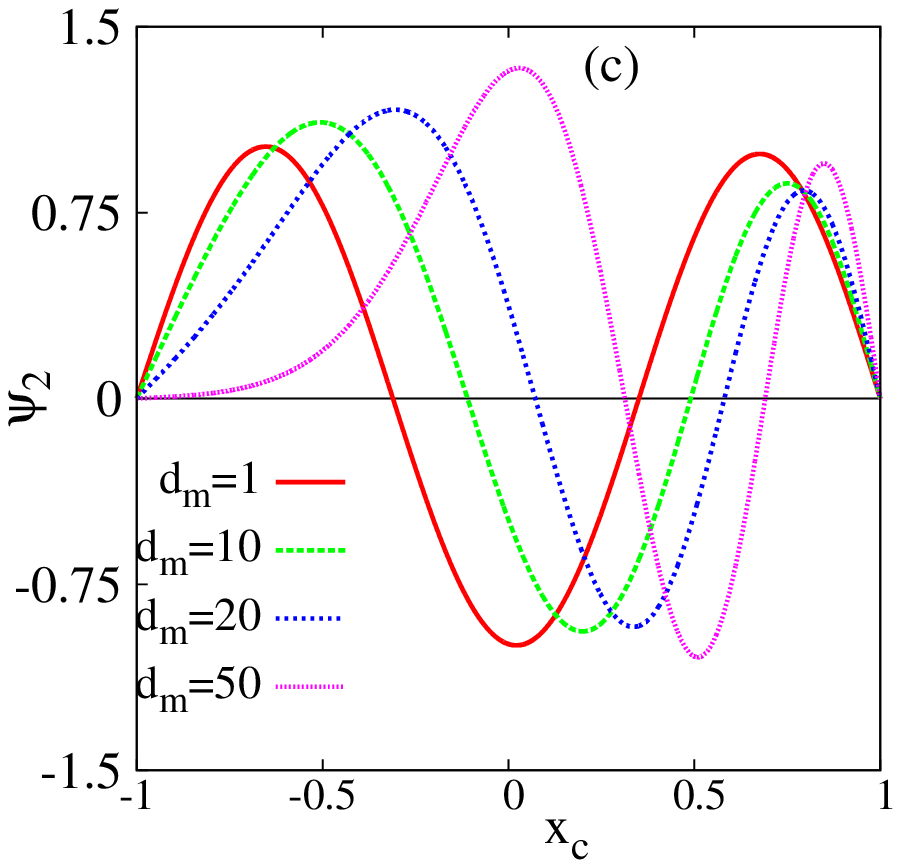}
\end{minipage}%
\caption{Wave functions of lowest six states of ACHO potential at four specific values of $d_m$, namely 1,
10, 20, 50. Panels (a)--(f) represent $n \! = \! 0-5$ states. See text for details.}
\end{figure}  

So far, no IE analysis has been done as yet for this system. Our current 
presentation is based on results of $\epsilon_n$; $S_x, S_p, S$; $I_x, I_p, I$; $E_x, E_p, E$ as 
functions of $d_m$ and $\eta$, for some low-lying states. For all forthcoming calculations box length has been kept 
constant at 2 and boundaries are placed at $-$1 and 1. At first, let us present the variation of classical 
turning point ($L_{cl}$) with $d_m$, in Fig.~(S7) for first six $n$ values. It is easily concluded that, an increase in 
$d_m$ leads to localization in $x$ space. Thus it is expected that, there will be delocalization in $p$ space. 
Note that, ACHO reduces to SCHO, if $d_m=0$. 

\begin{figure}                         
\begin{minipage}[c]{0.23\textwidth}\centering
\includegraphics[scale=0.45]{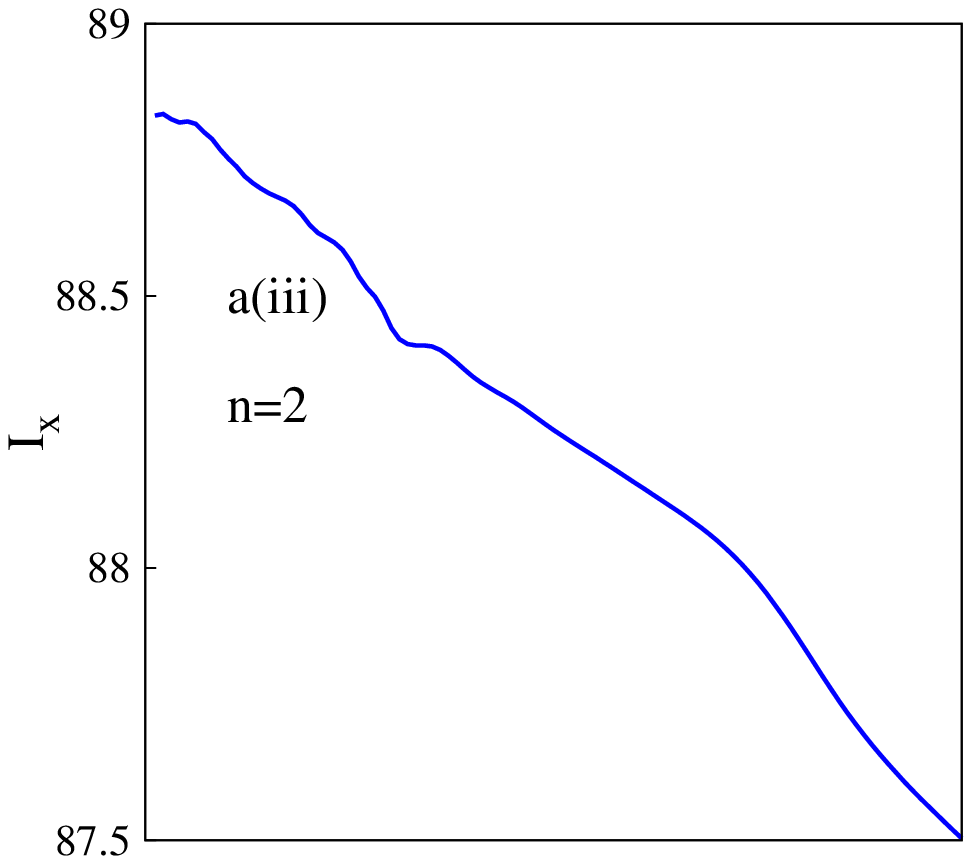}
\end{minipage}
\hspace{0.45in}
\begin{minipage}[c]{0.23\textwidth}\centering
\includegraphics[scale=0.45]{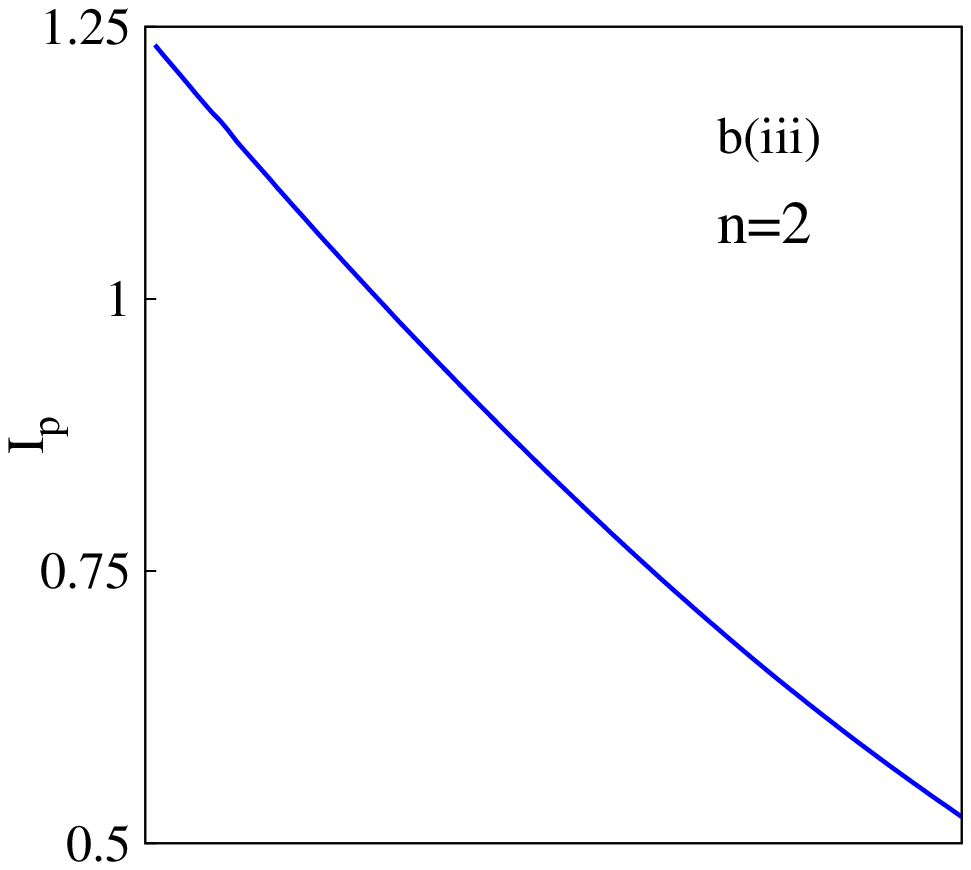}
\end{minipage}
\hspace{0.45in}
\vspace{0.15in}
\begin{minipage}[c]{0.23\textwidth}\centering
\includegraphics[scale=0.45]{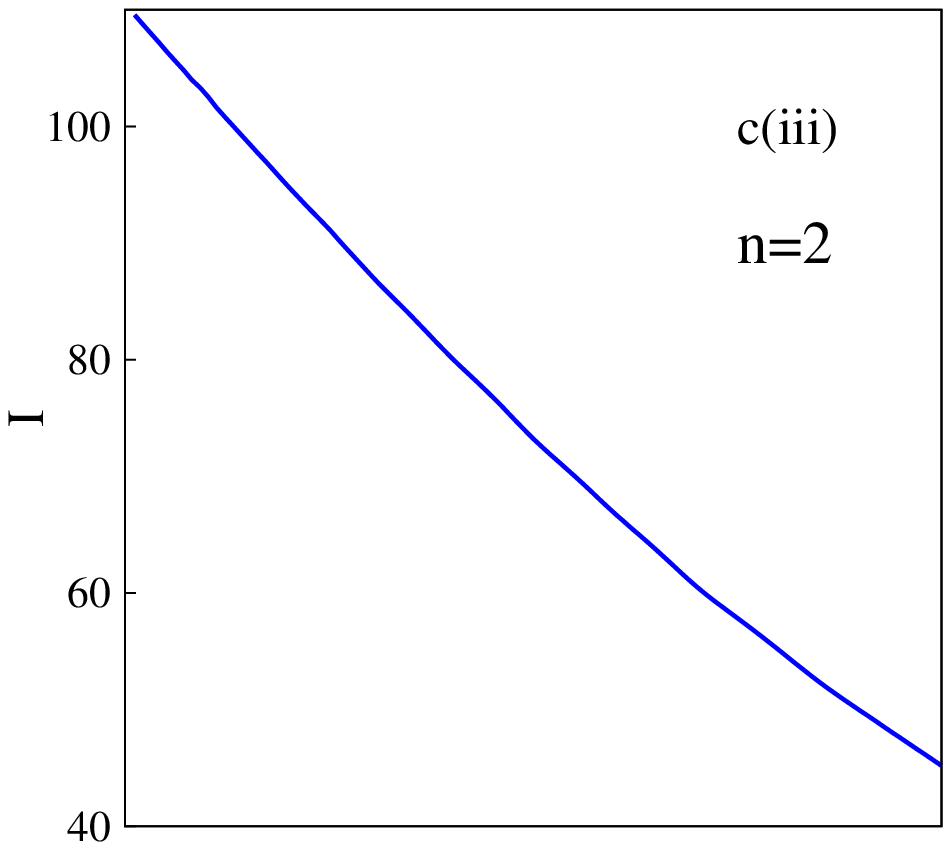}
\end{minipage}
\hspace{0.45in}
\begin{minipage}[c]{0.23\textwidth}\centering
\includegraphics[scale=0.45]{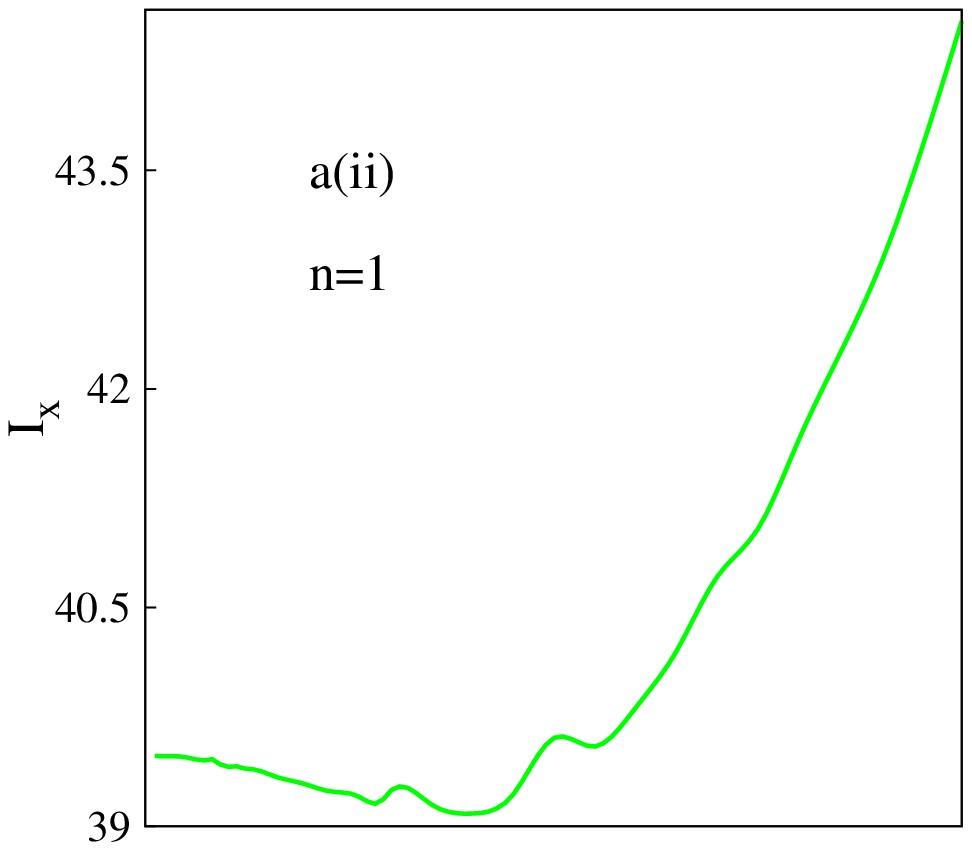}
\end{minipage}
\hspace{0.45in}
\begin{minipage}[c]{0.23\textwidth}\centering
\includegraphics[scale=0.45]{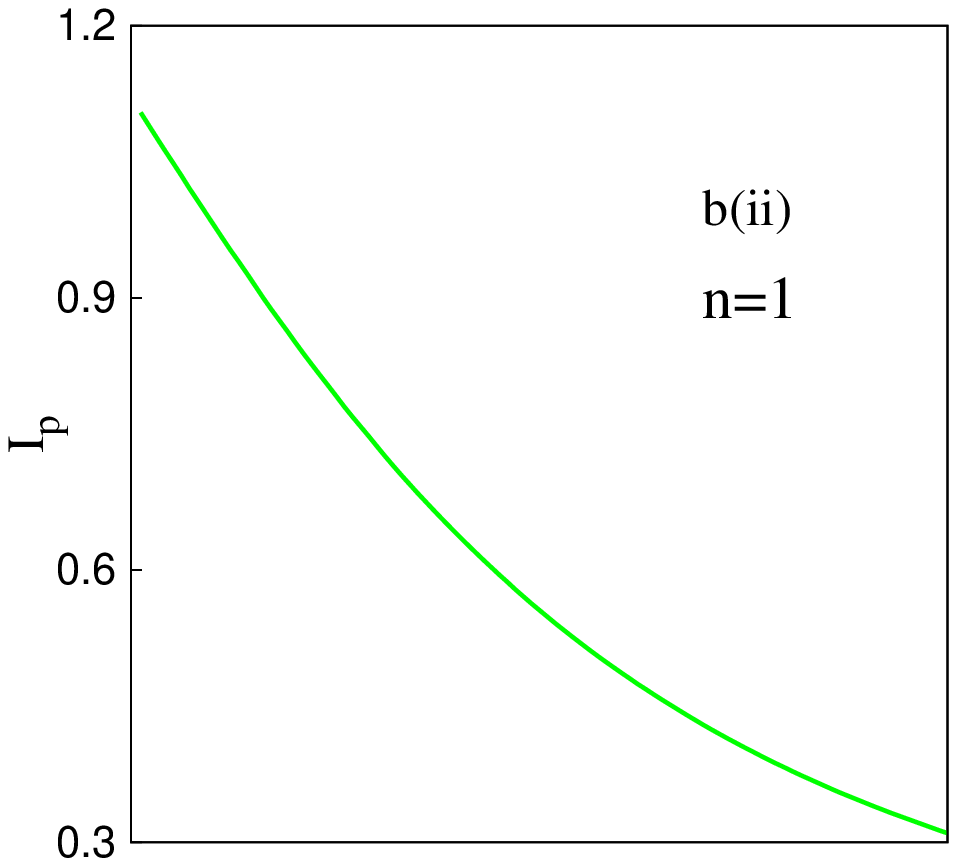}
\end{minipage}
\hspace{0.45in}
\vspace{0.15in}
\begin{minipage}[c]{0.23\textwidth}\centering
\includegraphics[scale=0.45]{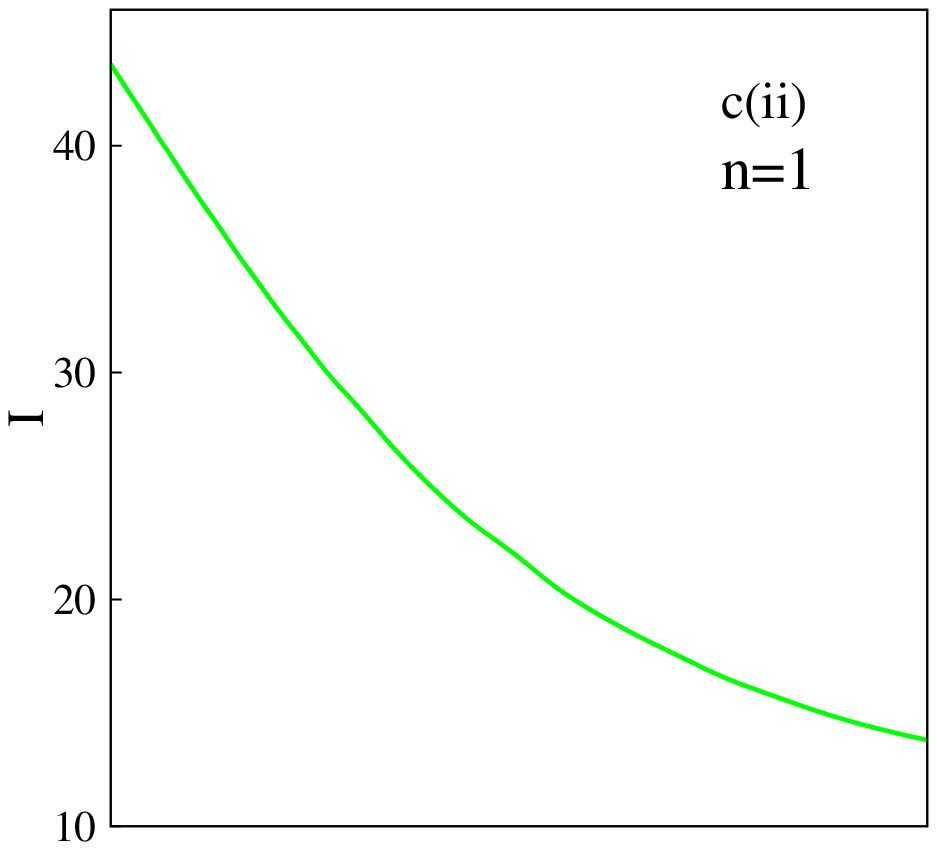}
\end{minipage}
\hspace{0.4in}
\begin{minipage}[c]{0.24\textwidth}\centering
\includegraphics[scale=0.5]{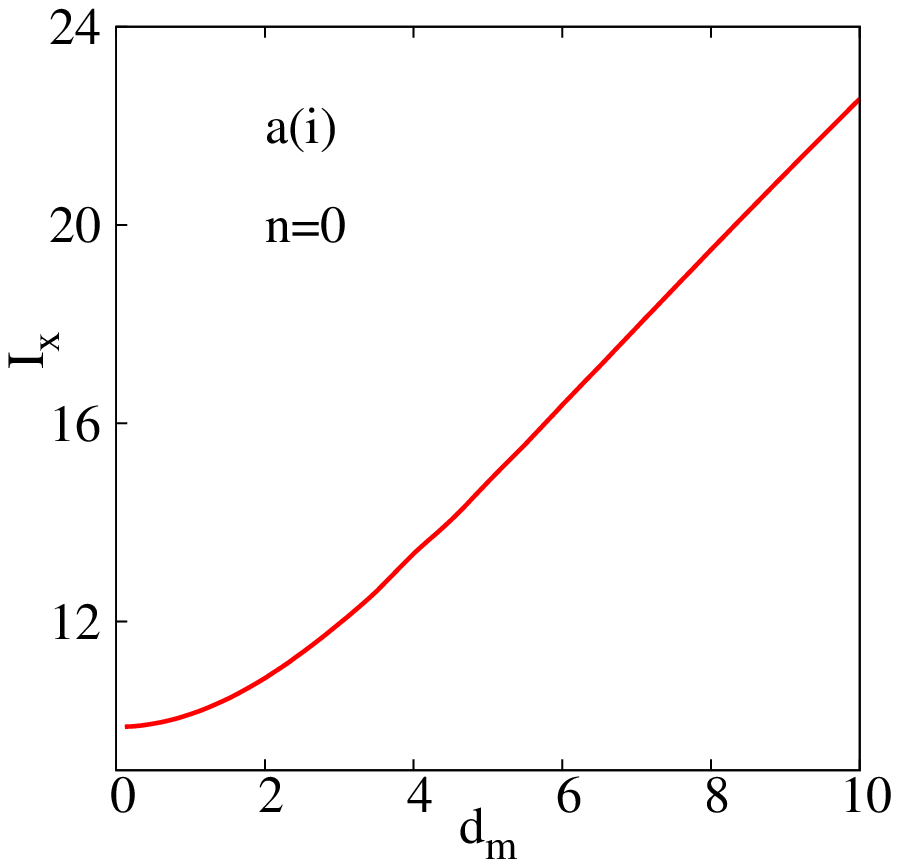}
\end{minipage}
\hspace{0.3in}
\begin{minipage}[c]{0.24\textwidth}\centering
\includegraphics[scale=0.5]{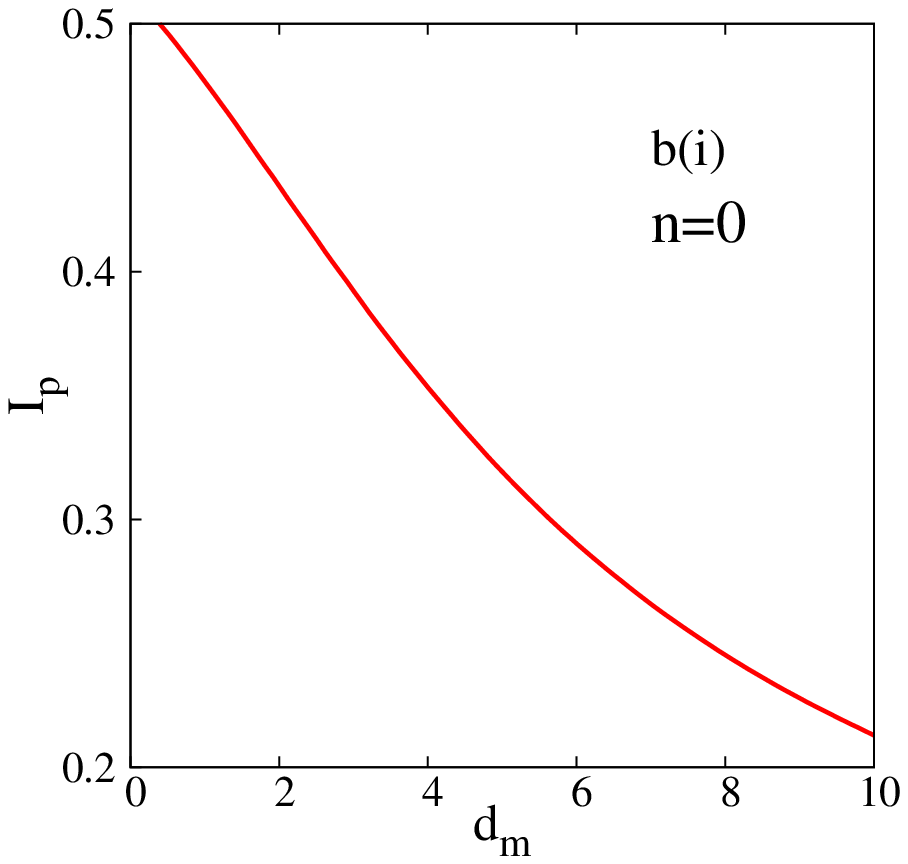}
\end{minipage}
\hspace{0.35in}
\begin{minipage}[c]{0.24\textwidth}\centering
\includegraphics[scale=0.5]{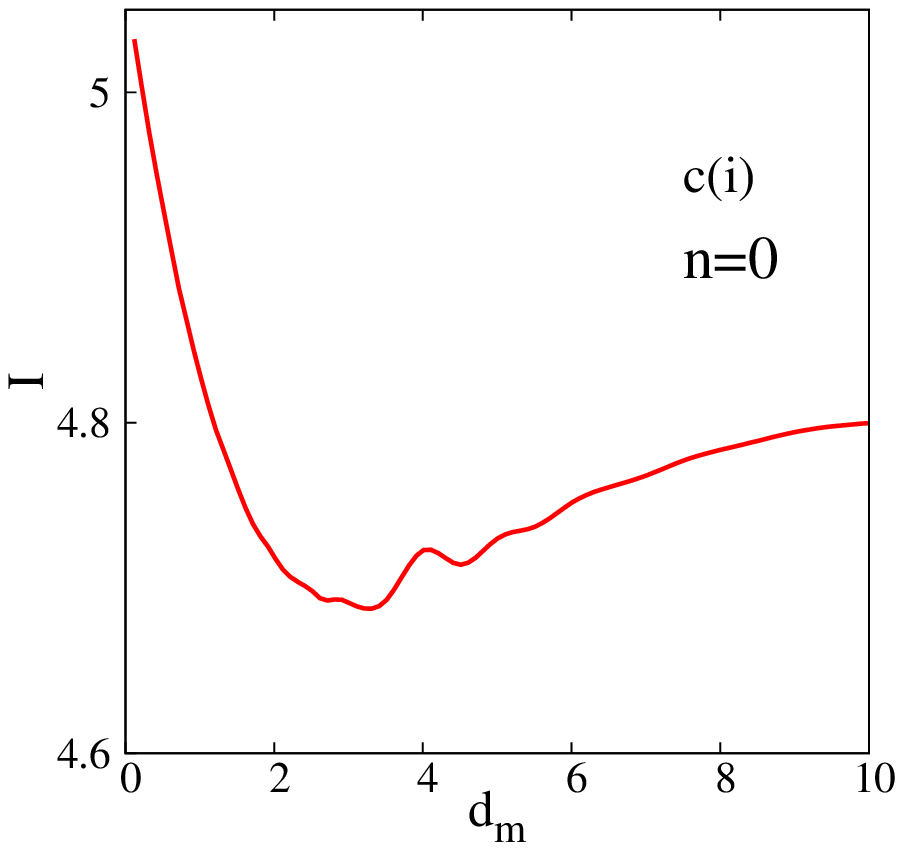}
\end{minipage}
\caption{Plot of $I_x$, $I_p$, $I$ of first five states of ACHO potential, as function of $d_m$, in left (a), middle (b), 
right (c) columns; (i)--(iii) represent $n=0-2$ states. See text for details.}
\end{figure}

Before proceeding for a discussion on IE in ACHO, it is appropriate to make some comments regarding its 
eigenvalues and eigenfunctions. Good-quality results for ACHO have been rather scarce; two notable ones are 
power-series solution \cite{campoy02} and ITP method \cite{roy15}. In this section, we take this opportunity to 
present another simple method for ACHO potential, which, as our future discussion confirms, produces quite accurate 
results. Although it is aside the main objective of this work, 
nevertheless quite relevant and worthwhile mentioning. In this so-called \emph{variation induced exact diagonalization}
procedure \cite{griffiths2004}, an energy functional is minimized using a SCHO basis set. Recently this method has been
used successfully in symmetric and asymmetric double-well potentials \cite{mukherjee15, mukherjee16}, where a QHO 
basis was utilized. This complete basis set corresponds to exact solution of SCHO potential, written as below 
($\alpha=\sqrt{\frac{1}{8}}$; while e,o signify even, odd states):
\begin{equation}
\psi_e  = \! _1F_1\left(\frac{1}{4}-\frac{\epsilon_e}{4\sqrt{2}\alpha},\frac{1}{2},2\sqrt{2}\alpha x^2\right) 
e^{-\sqrt{2}\alpha x^2}, \ \  
\psi_o  = x \ _1F_1\left[\frac{3}{4}-\frac{\epsilon_o}{4\sqrt{2}\alpha},\frac{3}{2},2\sqrt{2}\alpha x^2\right] 
e^{-\sqrt{2}\alpha x^2}. 
\end{equation}

\begin{figure}                         
\begin{minipage}[c]{0.25\textwidth}\centering
\includegraphics[scale=0.45]{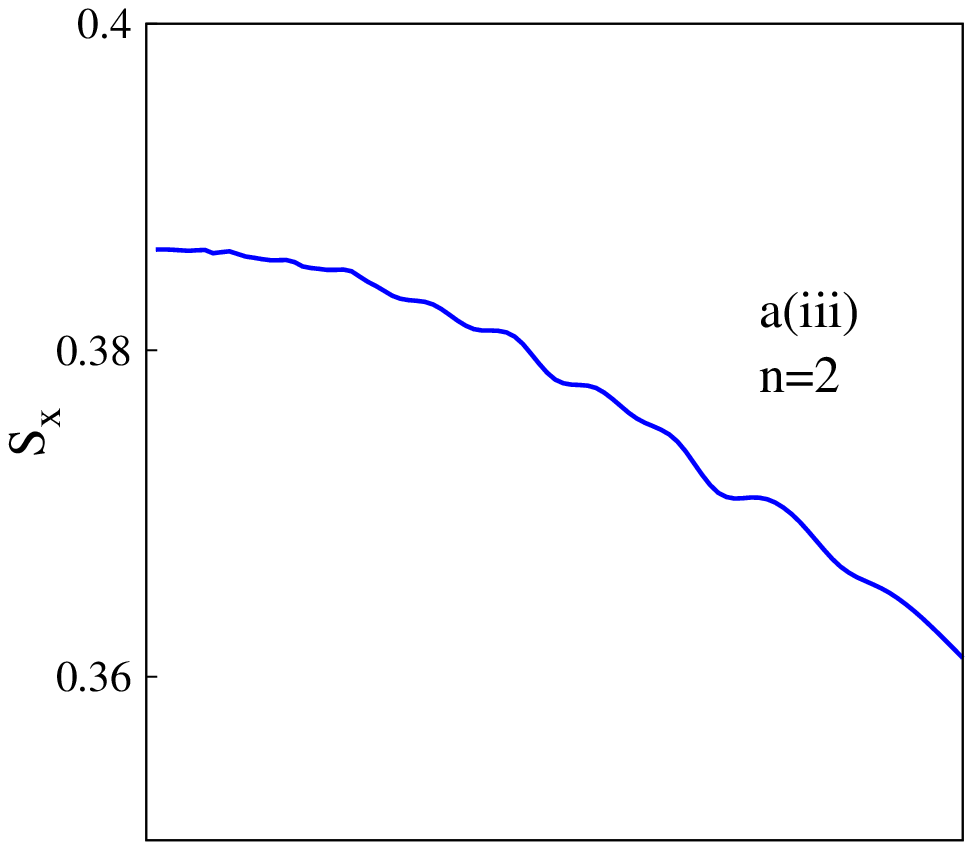}
\end{minipage}%
\hspace{0.45in}
\begin{minipage}[c]{0.25\textwidth}\centering
\includegraphics[scale=0.45]{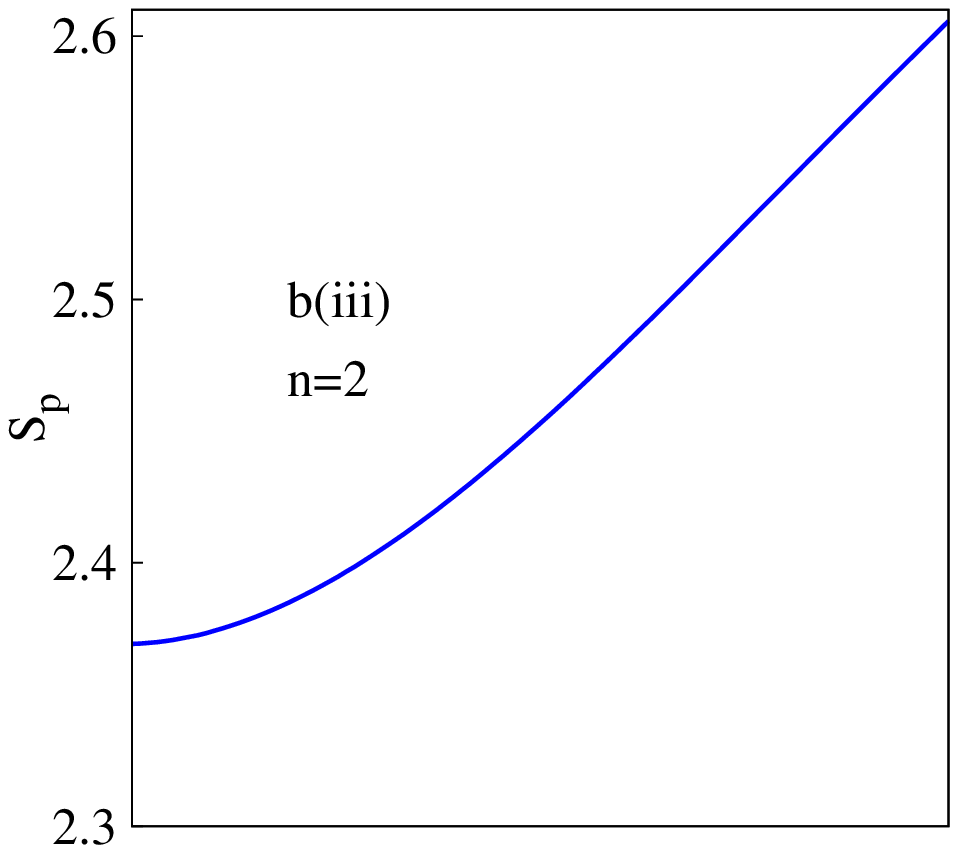}
\end{minipage}%
\hspace{0.45in}
\vspace{0.15in}
\begin{minipage}[c]{0.25\textwidth}\centering
\includegraphics[scale=0.45]{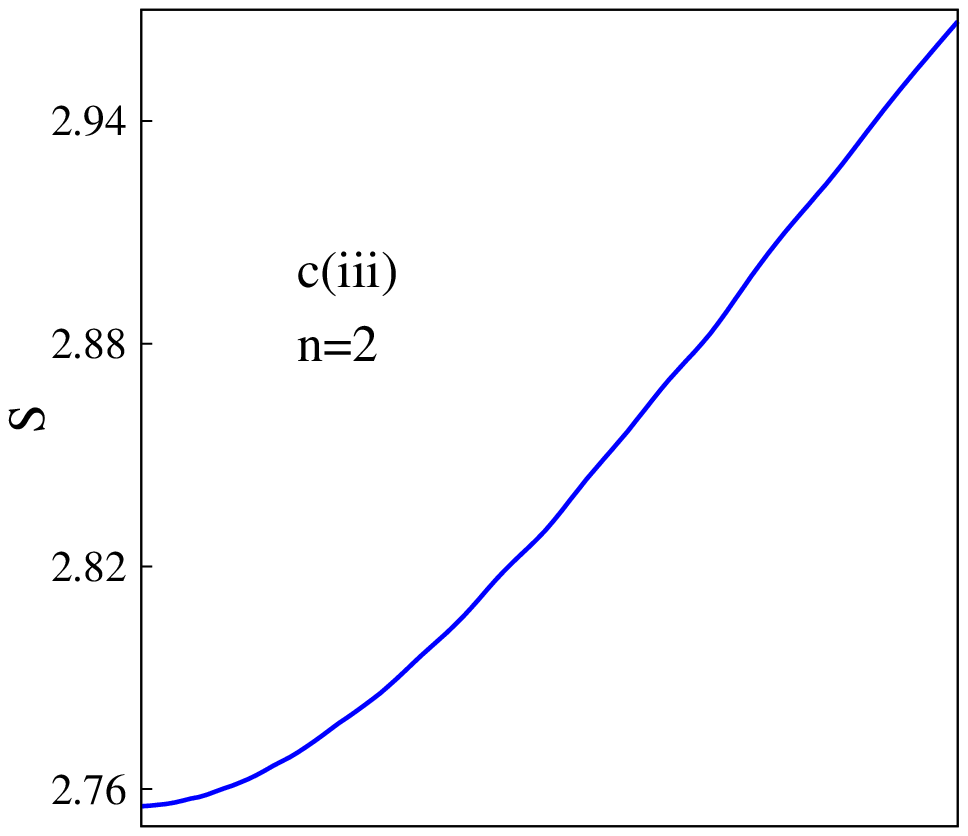}
\end{minipage}%
\hspace{0.45in}
\begin{minipage}[c]{0.25\textwidth}\centering
\includegraphics[scale=0.45]{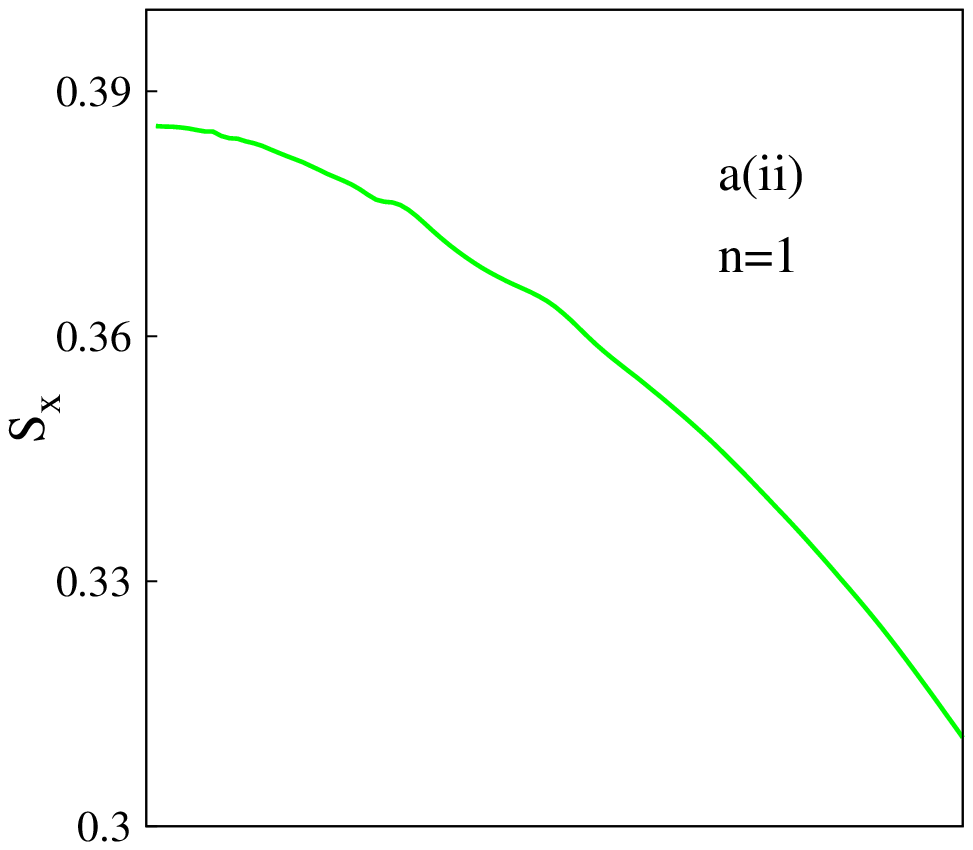}
\end{minipage}%
\hspace{0.45in}
\begin{minipage}[c]{0.25\textwidth}\centering
\includegraphics[scale=0.45]{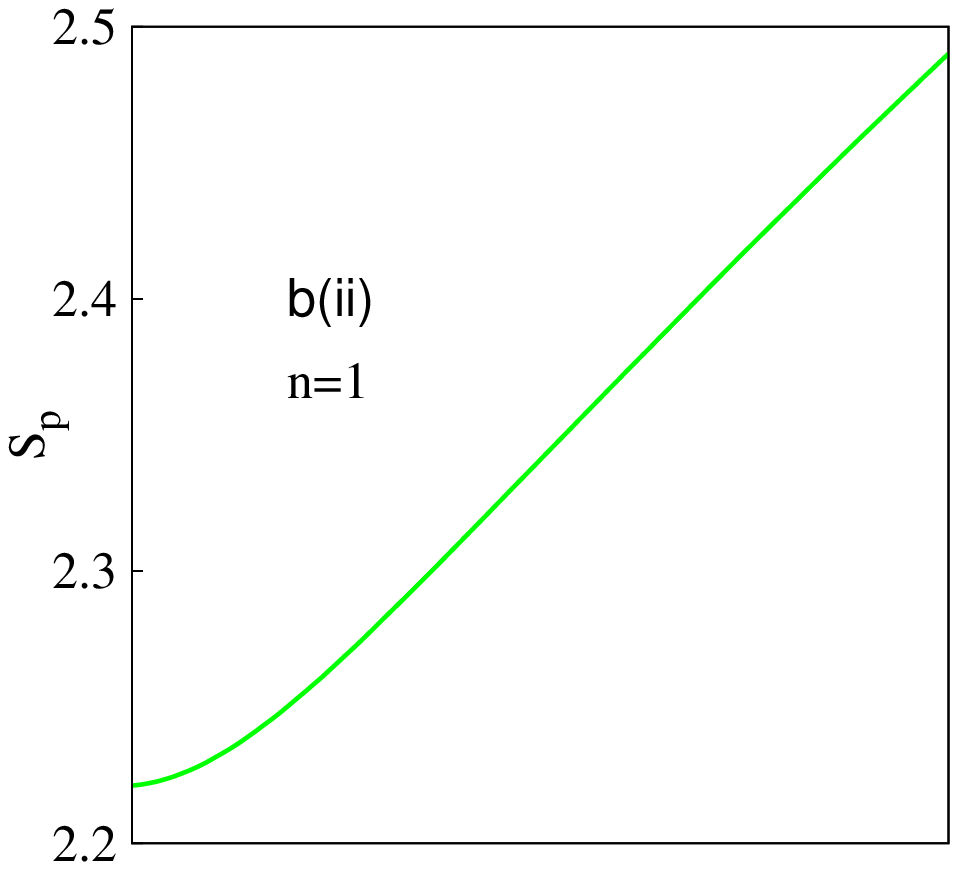}
\end{minipage}%
\hspace{0.45in}
\vspace{0.15in}
\begin{minipage}[c]{0.25\textwidth}\centering
\includegraphics[scale=0.45]{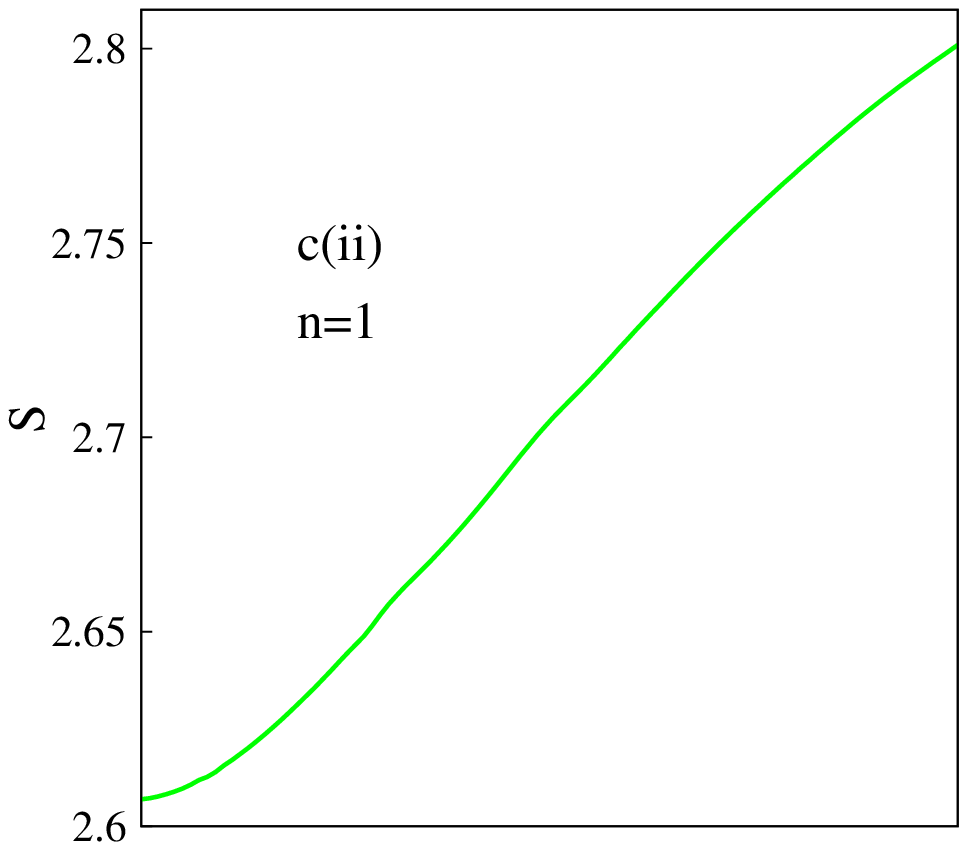}
\end{minipage}%
\hspace{0.45in}
\begin{minipage}[c]{0.25\textwidth}\centering
\includegraphics[scale=0.48]{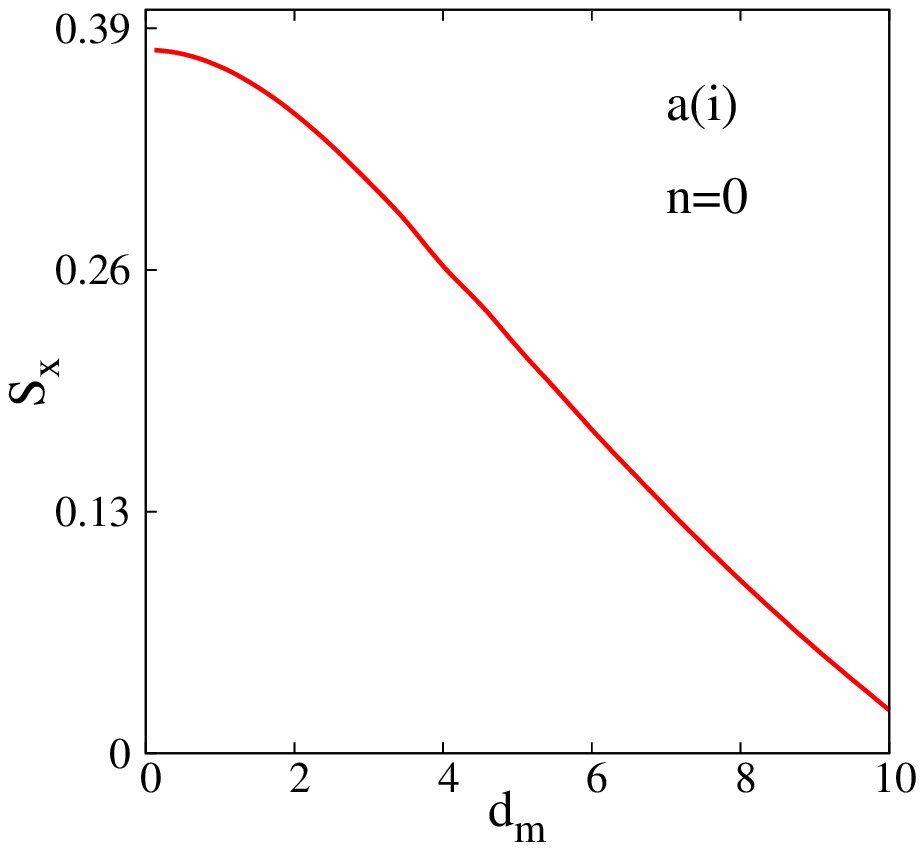}
\end{minipage}%
\hspace{0.45in}
\begin{minipage}[c]{0.25\textwidth}\centering
\includegraphics[scale=0.48]{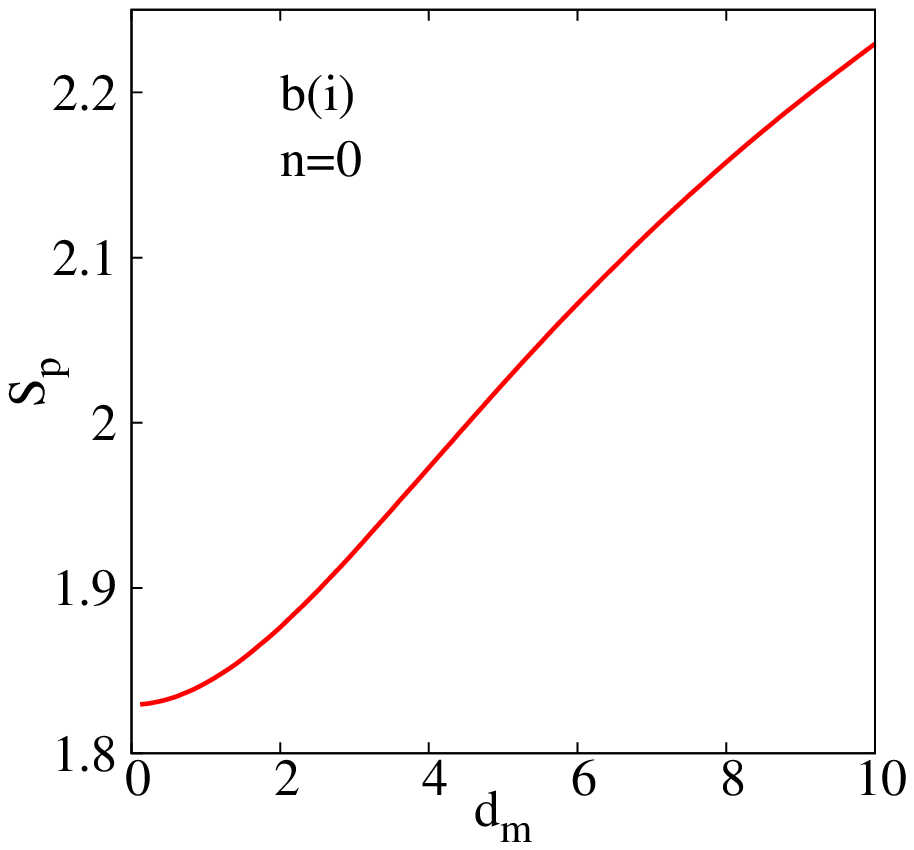}
\end{minipage}%
\hspace{0.45in}
\begin{minipage}[c]{0.25\textwidth}\centering
\includegraphics[scale=0.48]{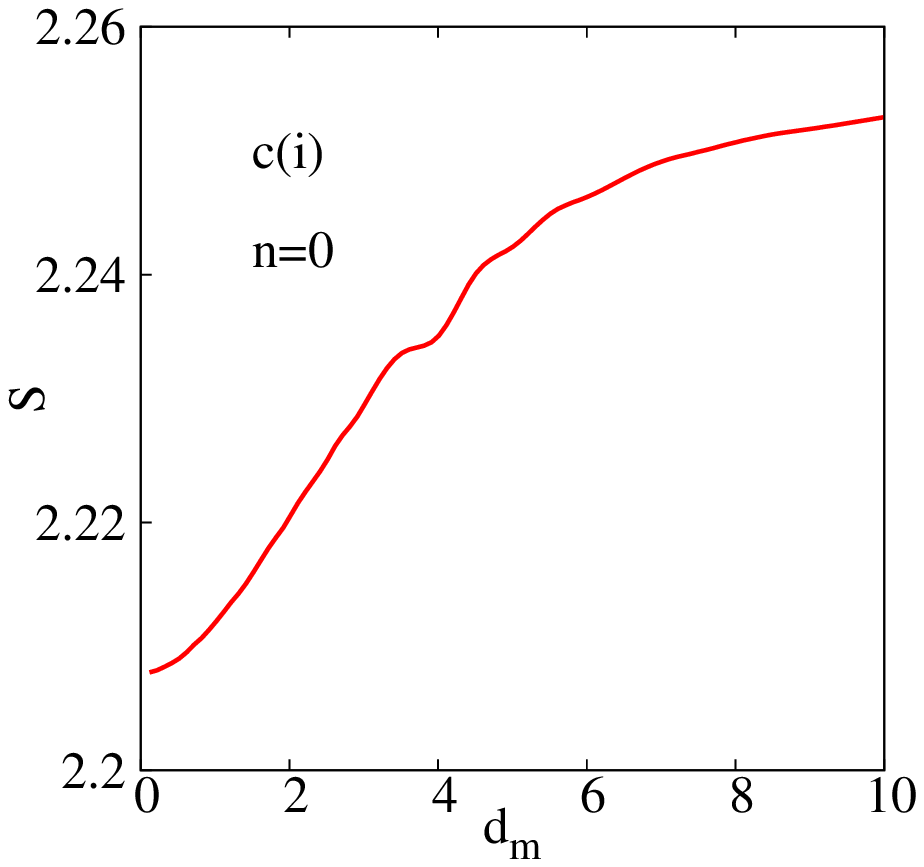}
\end{minipage}%
\caption{Plot of $S_x$, $S_p$, $S$ of first three states of ACHO potential, as function of $d_m$, in left (a), middle (b), 
right (c) columns; (i)--(iii) represent $n=0-2$ states. See text for details.}
\end{figure}

The integrals involved in Hamiltonian matrix are easy to evaluate by using these functions. Presence of a single 
non-linear parameter $\alpha$ permits us to adopt a coupled variation strategy, which is advantageous than purely  
linear variation. Further it is easy to ensure convergence of results with respect to basis dimension $N$. Thus it 
offers a secular equation at each $\alpha$. The kinetic energy part becomes infinitely large when $\alpha 
\rightarrow \infty$, whereas, on other extreme ($\alpha \rightarrow 0$), potential energy part behaves in a 
similar fashion. This qualitative analysis through uncertainty principle, ensures the existence of such basis. 
This fulfills one of the requirements of a satisfactory basis set and therefore justifies its adoption. 
Diagonalization of $H_{mn} = \langle m | \hat{H}| n \rangle$ leads to accurate eigenvalues and eigenfunctions, which
is accomplished through MATHEMATICA package. In principle, to obtain exact solution, one needs to employ a \emph{complete} 
basis; however for practical purposes a truncated basis of finite dimension is invoked. $N=50$ appears sufficient; with 
further increase in basis, result improves. A cross-section of energies
generated from above scheme is produced in Table~VII, for lowest six states of ACHO for four selected $d_m$. Best
reported literature values are quoted for comparison, wherever available. For first two $d_m$, present energies with 
SCHO basis, practically coincide with ITP results \cite{roy15}, for all digits reported., while those from 
\cite{campoy02} show good agreement with ours. For last two $d_m$, no reference result could be found. Thus
fresh ITP calculations were performed for them and cited, which, as expected, shows excellent agreement with present energies 
in all states. It is hoped that this SCHO basis may be useful in future for other confined and/or free potentials in 1D. 

\begingroup      
\squeezetable
\begin{table}
\centering
\caption{$S_{x}, S_{p}, S$ for $n \! = \! 0-2$ states of ACHO at five specific $d_m$. See text for details.} 
\centering
\begin{ruledtabular}
\begin{tabular}{c|ccc|ccc|ccc}
$d_m$ & $S_{x}^{0}$ & $S_{p}^{0}$ & $S^{0}$ & $S_{x}^{1}$ & $S_{p}^{1}$ & $S^{1}$ & $S_{x}^{2}$ & $S_{p}^{2}$ & $S^{2}$ \\
\hline
0.12 & 0.3783 & 1.8296 & 2.2079 & 0.3857 & 2.2212 & 2.6069 & 0.3862 & 2.3692 & 2.7554 \\
2.04 & 0.3428 &	1.8779 & 2.2208 & 0.3807 & 2.2512 & 2.6319 & 0.3850 & 2.3852 & 2.7703 \\
5.0 & 0.2184 & 2.0238 & 2.2422 & 0.3636 & 2.3390	& 2.7027 & 0.3782 &	2.4512 & 2.8295  \\
8.0 & 0.093 & 2.1576 & 2.2506 & 0.3360 & 2.4313 & 2.7674 & 0.3695 & 2.5429 &	2.9124  \\
10.0 & 0.0233 & 2.2294 & 2.2527 & 0.3108 & 2.4900 & 2.8009 & 0.3611 & 2.6057 & 2.9668 \\
\end{tabular}
\end{ruledtabular}
\end{table} 
\endgroup  

Our calculated ACHO energies of Table~VII are portrayed in Fig.~(8), for lowest six states with respect to $d_m$, 
in panels (a) through (f). Ranges of $\epsilon_n$ and $d_m$ vary for each $n$. Changes in $\epsilon_0$ is 
rather unique amongst all $\epsilon_n$, where, from an initial positive value at small $d_m$, ground-state energy 
continuously falls. In all excited states, however, it passes through a maximum; with increase in $n$, which  
shifts to higher values of $d_m$. Next Fig.~(9) depicts our computed wave functions for $n \! = \! 0-5$ states of 
ACHO at same four $d_m$. Clearly, the maximum, minimum and nodal positions of all $\psi_n$'s shift towards right with 
increase in $d_m$. Clearly, these are characterized by requisite number of nodes for a given $n$. These plots 
indicate that particle gets localized in $x$ space as $d_m$ increases.  

\begin{figure}                         
\begin{minipage}[c]{0.25\textwidth}\centering
\includegraphics[scale=0.45]{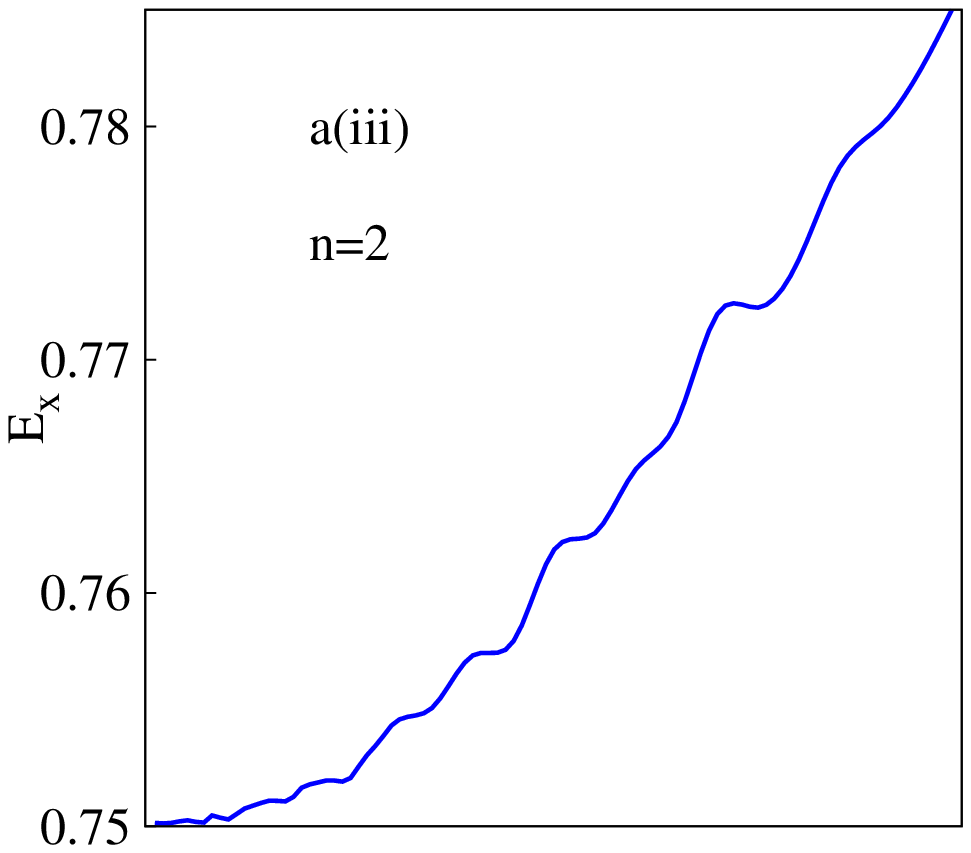}
\end{minipage}%
\hspace{0.45in}
\begin{minipage}[c]{0.25\textwidth}\centering
\includegraphics[scale=0.45]{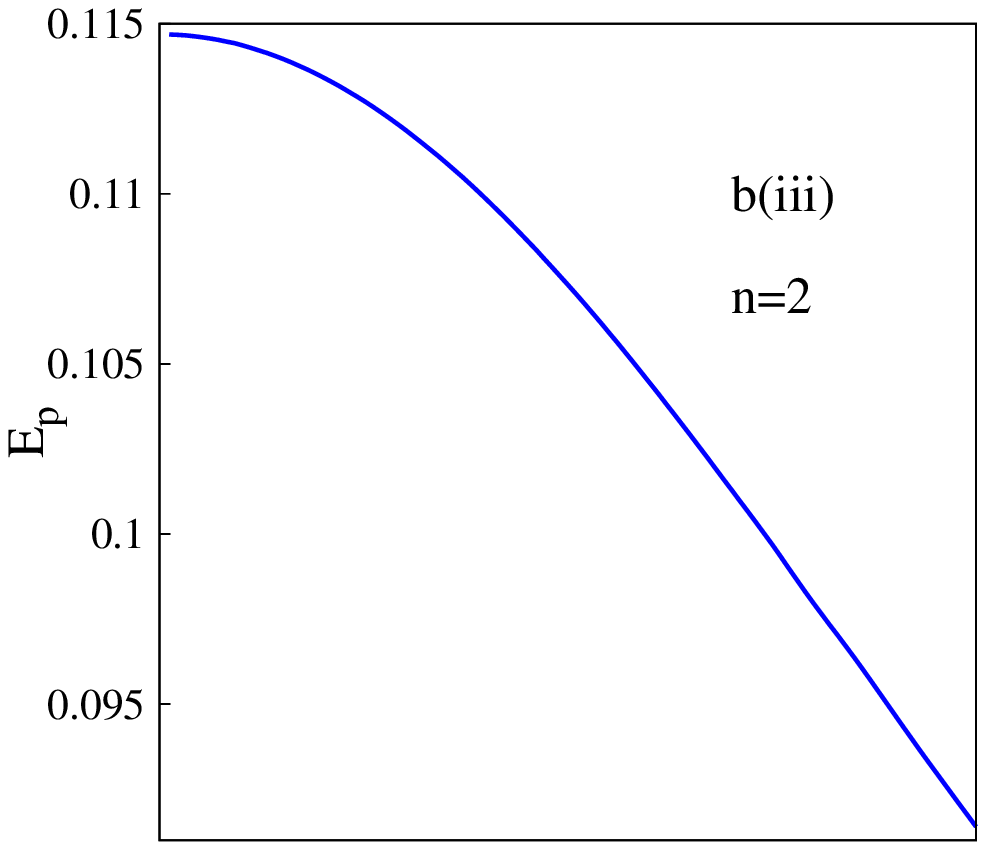}
\end{minipage}%
\hspace{0.45in}
\vspace{0.15in}
\begin{minipage}[c]{0.25\textwidth}\centering
\includegraphics[scale=0.45]{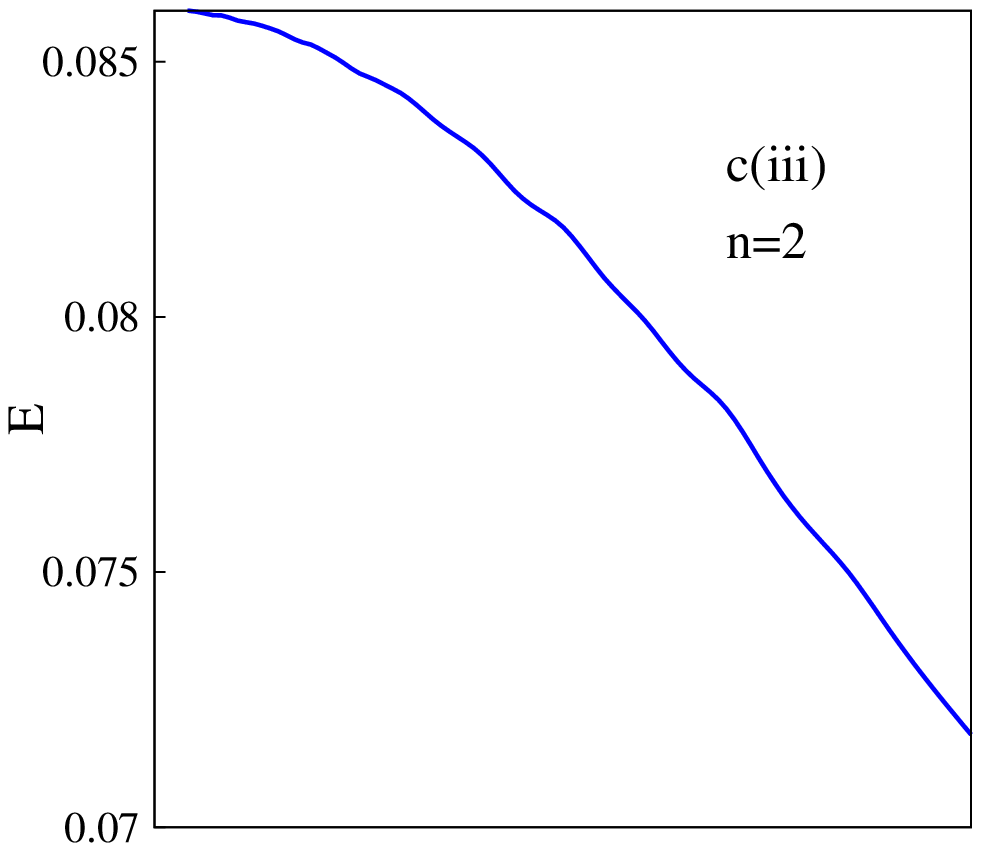}
\end{minipage}%
\hspace{0.45in}
\begin{minipage}[c]{0.25\textwidth}\centering
\includegraphics[scale=0.45]{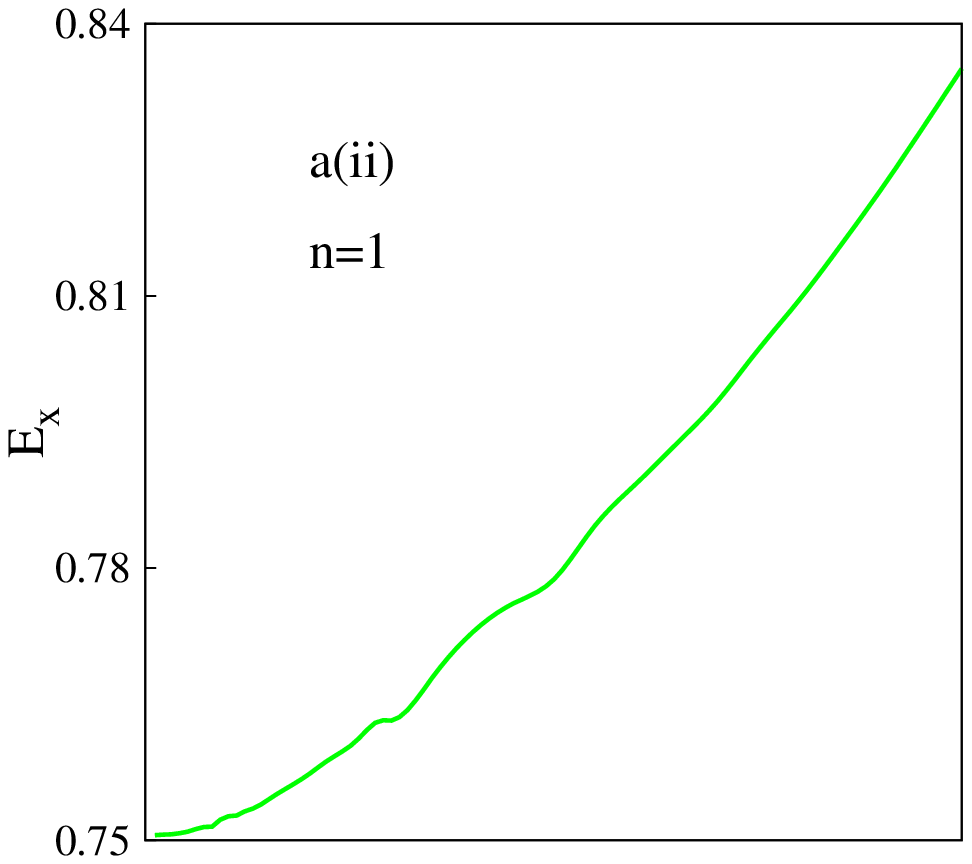}
\end{minipage}%
\hspace{0.45in}
\begin{minipage}[c]{0.25\textwidth}\centering
\includegraphics[scale=0.45]{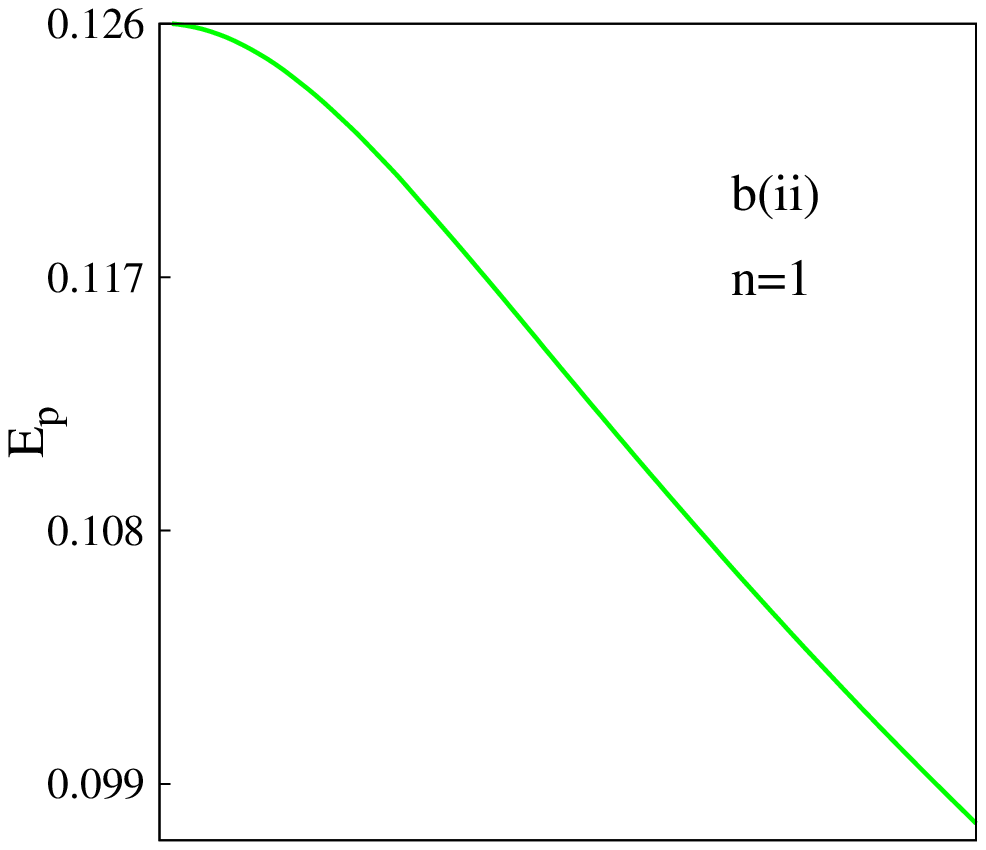}
\end{minipage}%
\hspace{0.45in}
\vspace{0.15in}
\begin{minipage}[c]{0.25\textwidth}\centering
\includegraphics[scale=0.45]{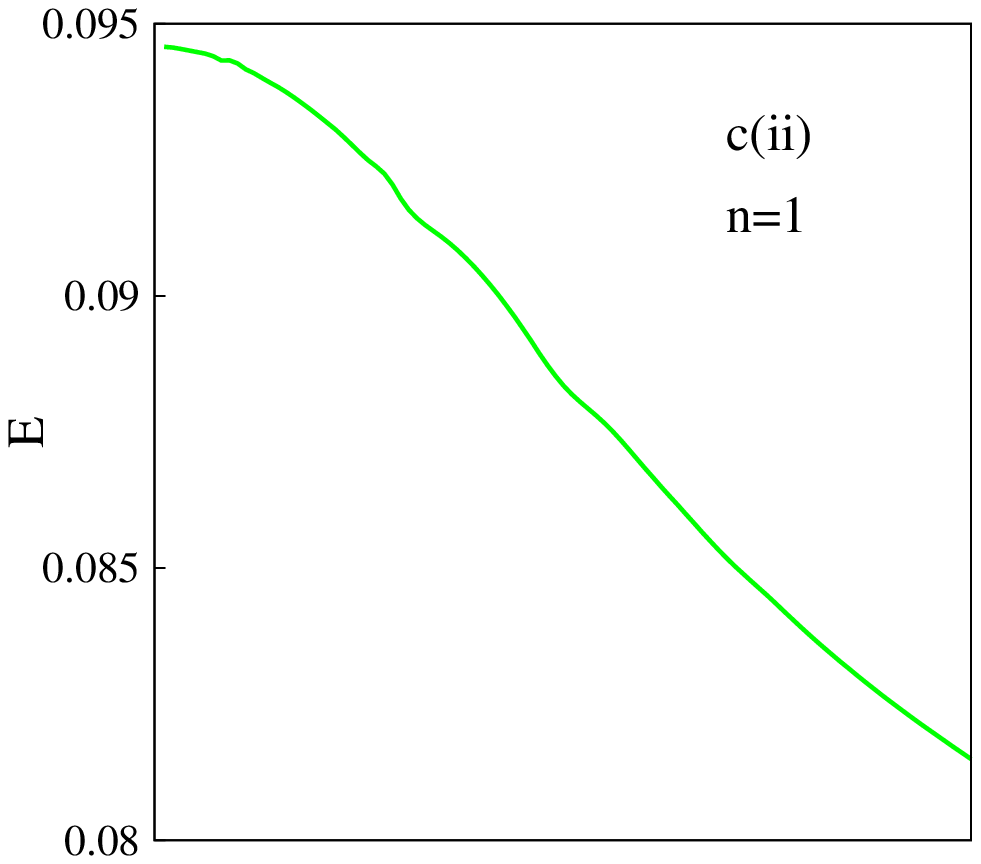}
\end{minipage}%
\hspace{0.45in}
\begin{minipage}[c]{0.25\textwidth}\centering
\includegraphics[scale=0.48]{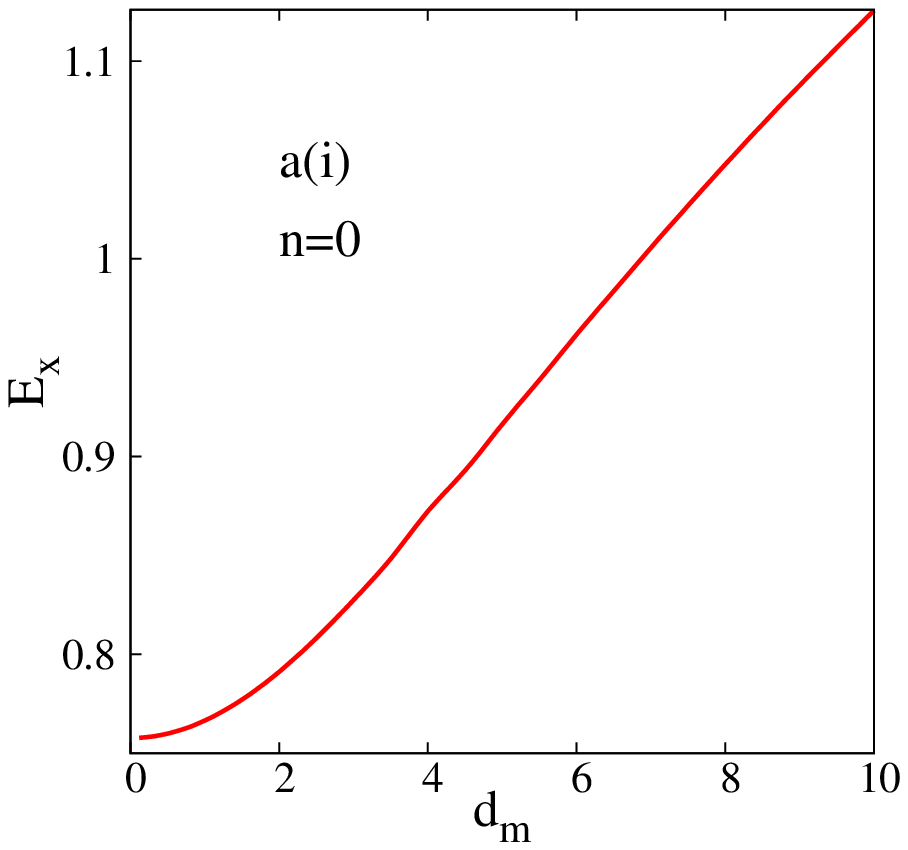}
\end{minipage}%
\hspace{0.45in}
\begin{minipage}[c]{0.25\textwidth}\centering
\includegraphics[scale=0.48]{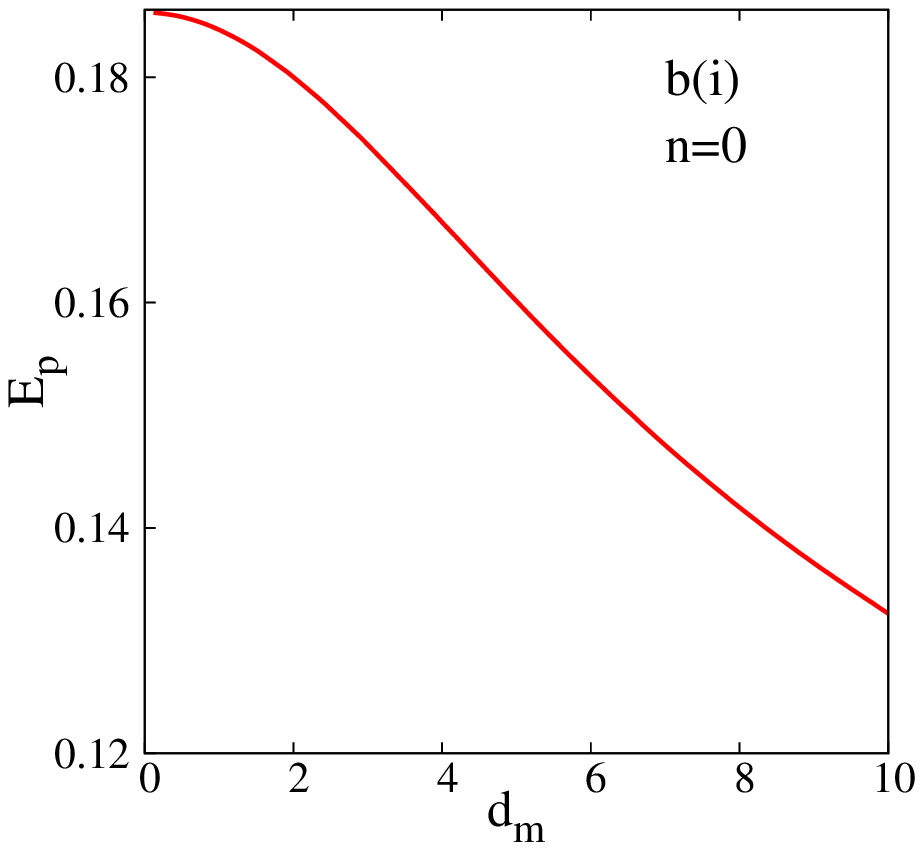}
\end{minipage}%
\hspace{0.45in}
\begin{minipage}[c]{0.25\textwidth}\centering
\includegraphics[scale=0.48]{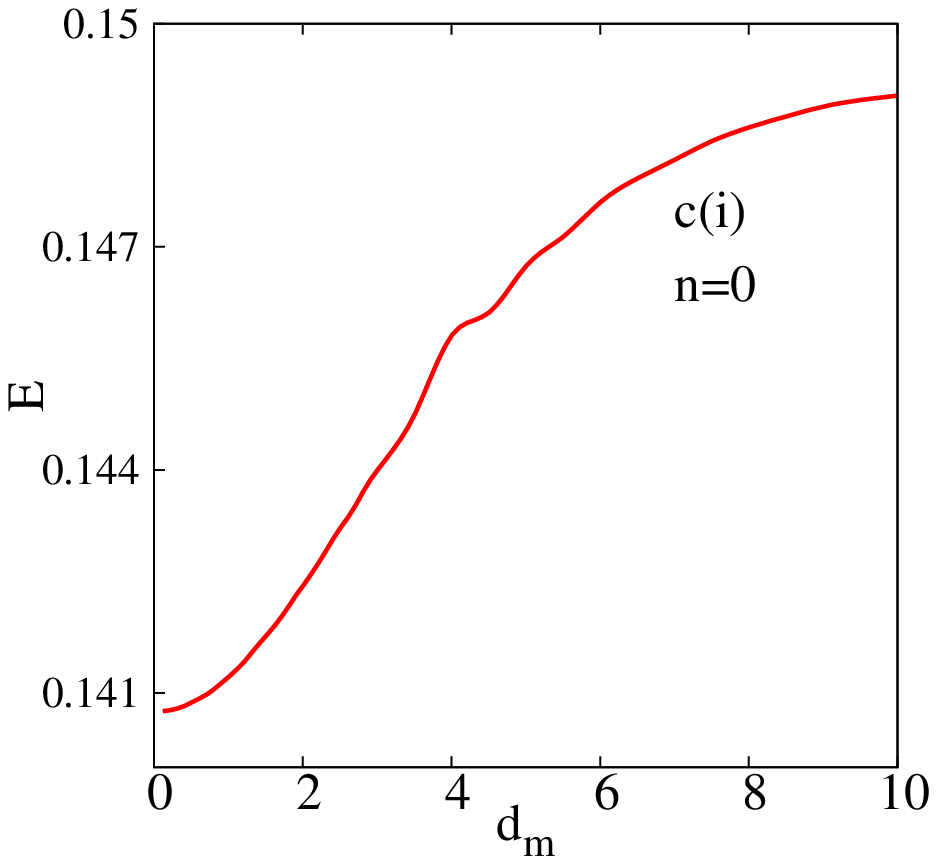}
\end{minipage}%
\caption{Plot of $E_x$, $E_p$, $E$ of first three states of ACHO potential, as function of $d_m$, in left (a), middle (b), 
right (c) columns; (i)--(iii) represent $n=0-2$ states. See text for details.}
\end{figure}

Now we move on to $I_x,I_p$ and $I$ variations with respect to $d_m$. These are shown graphically in Fig.~(10) ($n=0-2$) and 
Fig.~(S8) ($n=3,4$); left (a), middle (b) and right (c) panels display these in $x$, $p$ and composite space respectively.
For $n \! = \! 0-1$, $I_x$ tends to increase with $d_m$, on the whole, whereas for remaining states ($n \! \geq \! 2$), the  
same decreases. Thus, for $n=0,1$, it suggests localization in $x$ space and delocalization for higher states. A careful 
observation of panels b(i)--b(iii) of Fig.~(10) and b(i)--b(ii) of Fig.~(S8) reveals that, $I_{p}$, for all $n$ under 
investigation, consistently decreases with increase of $d_m$, signifying a delocalization of particle in $p$ space. Further 
exploration of net $I$ in segments c(i)--c(iii) of Fig.~(10) and c(i)--c(ii) of Fig.~(S8) 
shows that, except for $n \! = \! 0$, $I$ decreases as $d_m$ increases, indicating an overall delocalization in 
composite space. However, for $n \! = \! 0$, there appears a minimum in $I$, which could be possibly due to a
competition between localization-delocalization.     

\begingroup      
\squeezetable     
\begin{table}
\caption{$I_{x'},I_{p'}; S_{x'},S_{p'}; E_{x'},E_{p'}$ values for ground state of ACHO potential at three different $\eta$, 
namely 0.001,~0.01,~0.1, considering $d_m=1$.} 
\begin{ruledtabular}
\begin{tabular}{c|cc|cc|cc}
$\eta$  & $I_{x'}$ & $I_{p'}$ & $S_{x'}$ & $S_{p'}$ & $E_{x'}$ & $E_{p'}$   \\
   \hline
5 & 13.883826 & 0.32456 & 0.224258 & 1.997965 & 0.908473  & 0.182345 \\
10 &19.016519 & 0.224533 & 0.072132 & 2.162765 & 1.065115 & 0.179876 \\
20 &26.833658 & 0.172067 & -0.099461 & 2.346781 & 1.265513 & 0.168523 \\
\end{tabular}
\end{ruledtabular}
\end{table} 
\endgroup 

Next, it is necessary to consider changes in behavior of $S_x, S_p, S$ as function of $d_m$. These are offered in Fig.~(11) 
($n=0-2$) and Fig.~(S9) ($n=3,4$), adopting an identical presentation strategy as in previous figure. It is clear from left 
panels a(i)--a(iii) of Fig.~(11) and a(i)--a(ii) of Fig.~(S9) that, $S_x$ for all five $n$ generally decreases monotonically 
as $d_m$ grows, signifying localization of particle in $x$ space. Likewise, a scrutiny of mid-region plots b(i)--b(iii) of 
Fig.~(11) and b(i)--b(ii) of Fig.~(S9) leads to conclusion that particle is delocalized for all $n$ in $p$ space, for $S_p$ 
increases with $d_m$. Finally, five net $S$ plots, c(i)--c(iii) of Fig.~(11) and c(i)-c(ii) of Fig.~(S9), in right-side 
column maintains the same trend of $S_p$ with $d_m$. This could imply that, there is net delocalization in composite $x$, 
$p$ space (since extent of delocalization in $p$ space is more than extent of localization in $x$ space). A sample of 
$S_x, S_p$ and $S$ are reported in Table~VIII for $n \! = \! 0,1,2$ states at five particular $d_m$, i.e., 0.12, 2.04, 5, 
7 and 10. Evidently, these entries corroborate the findings of Fig.~(11).     
Now, we move on to an analysis of $E$, in Fig.~(12) ($n=0-2$) and Fig.~(S10) ($n=3,4$), adopting same presentation
format of $I$ and $S$. For all $n$ in a(i)--a(iii) of Fig.~(12) and a(i)--a(ii) of Fig.~(S10), there is an overall increase of 
$E_{x}$ with $d_m$ leading to localization in $x$ space. Panels b(i)--b(iii) of Fig.~(12) and b(i)--b(ii) of Fig.~(S10) in 
middle shows that, general trend of $E_{p}$ is a gradual decline with an increase in $d_m$; thus it is able to account for 
delocalization in $p$ space. At last, c(i)--c(iii) of Fig.~(12) and c(i)-c(ii) of Fig.~(S10) clearly show that, for all 
$n$ except zero, $E$ decreases with increasing $d_m$, interpreting a net delocalization in combined $x,p$ space. Interestingly, 
in the ground state, while $E_x, E_p$ maintain same pattern of other states, net measure $E$ for $n \! = \! 0$ deviates 
from that for excited states.  

On the basis of above discussion, it is clear that, only $S_{x}, S_{p}, S$ can satisfactorily explain 
localization-delocalization phenomena in an ACHO. Study of $E_{x}, E_{p}, E$ can also offer valuable 
insight in to the dual nature of $E$ in composite $x$, $p$ space, except for ground state. But, 
$I_{x}, I_{p}, I$ appear to be inadequate in explaining the contrasting phenomena in ACHO.

\begin{figure}                         
\begin{minipage}[c]{0.25\textwidth}\centering
\includegraphics[scale=0.45]{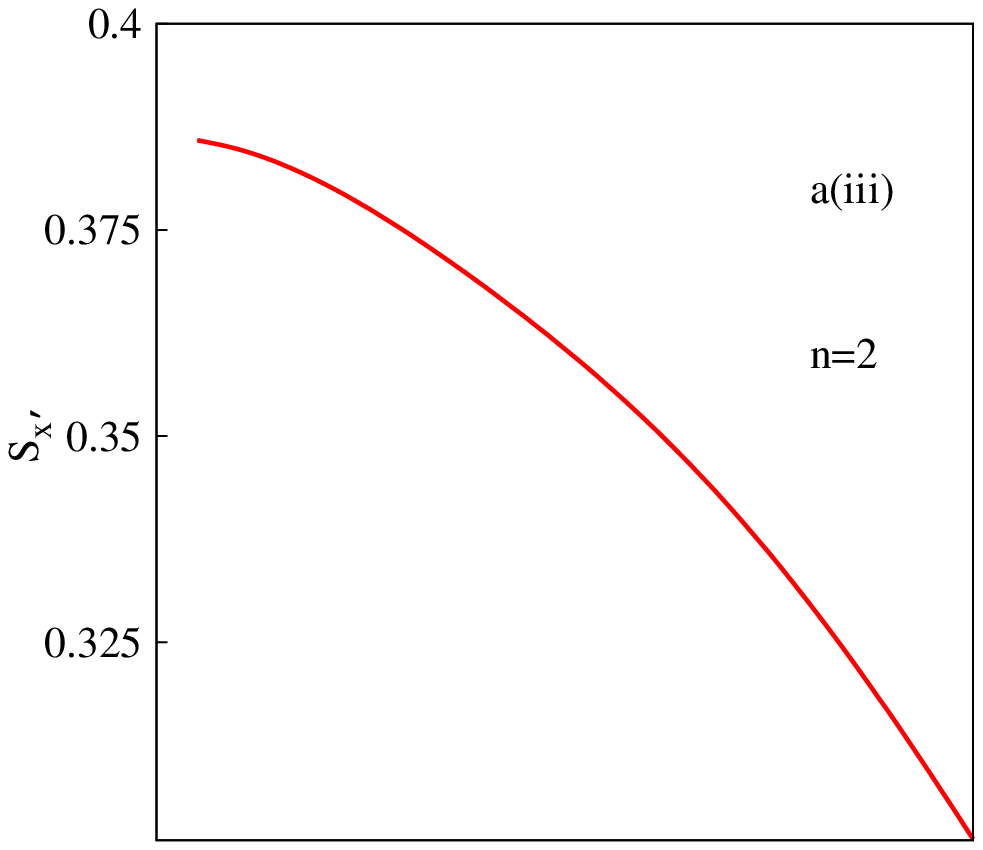}
\end{minipage}%
\hspace{0.45in}
\begin{minipage}[c]{0.25\textwidth}\centering
\includegraphics[scale=0.45]{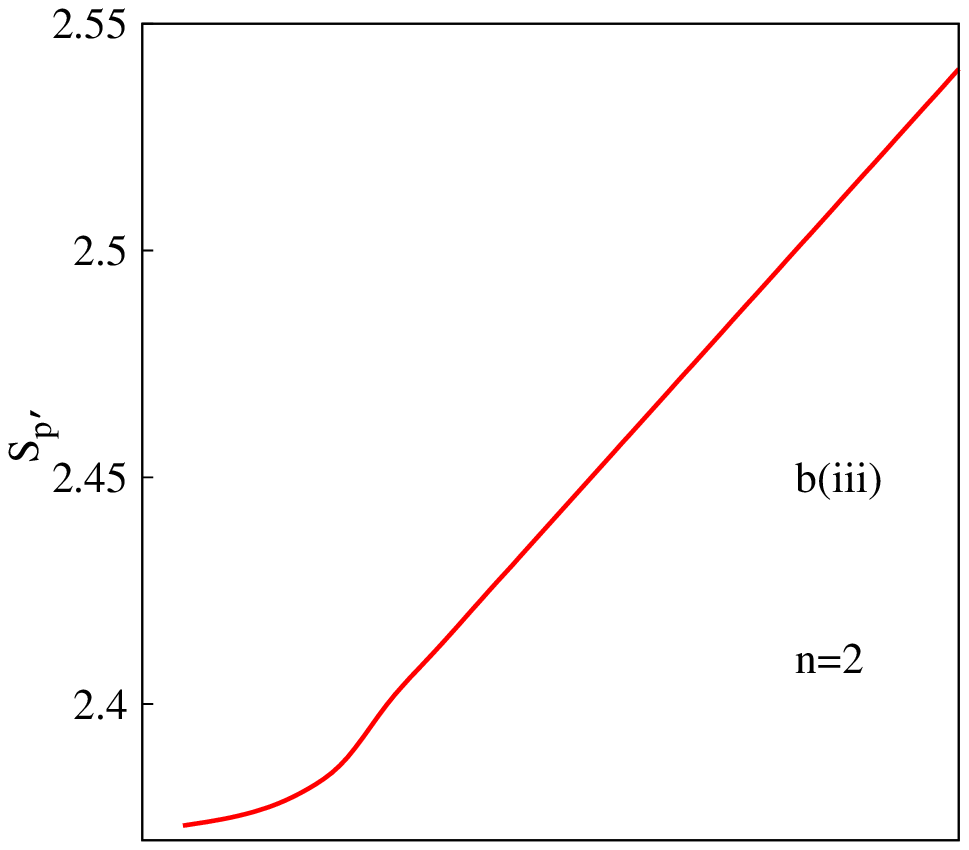}
\end{minipage}%
\hspace{0.45in}
\vspace{0.15in}
\begin{minipage}[c]{0.25\textwidth}\centering
\includegraphics[scale=0.45]{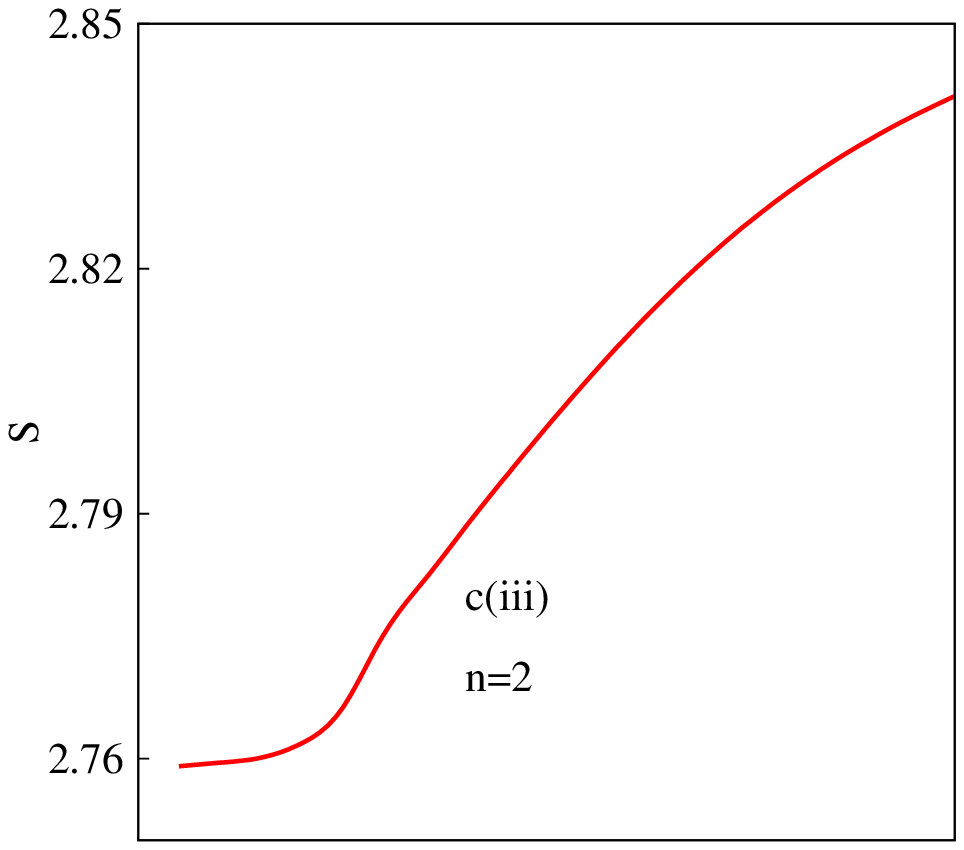}
\end{minipage}%
\hspace{0.45in}
\begin{minipage}[c]{0.25\textwidth}\centering
\includegraphics[scale=0.45]{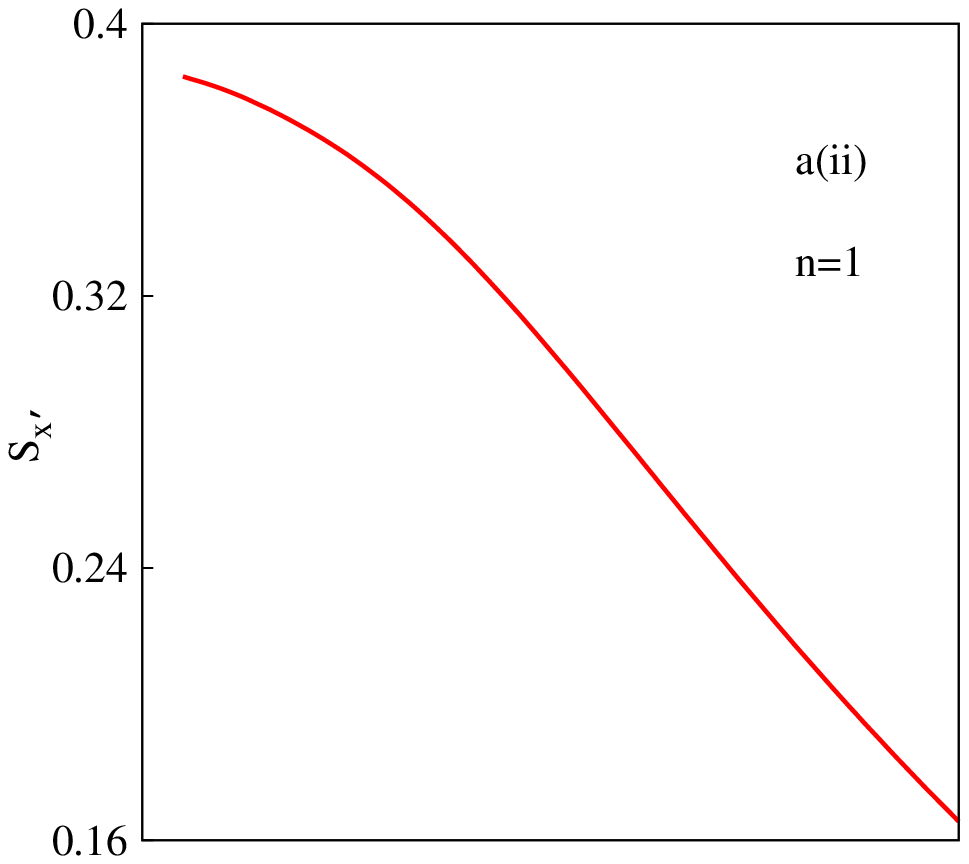}
\end{minipage}%
\hspace{0.45in}
\begin{minipage}[c]{0.25\textwidth}\centering
\includegraphics[scale=0.45]{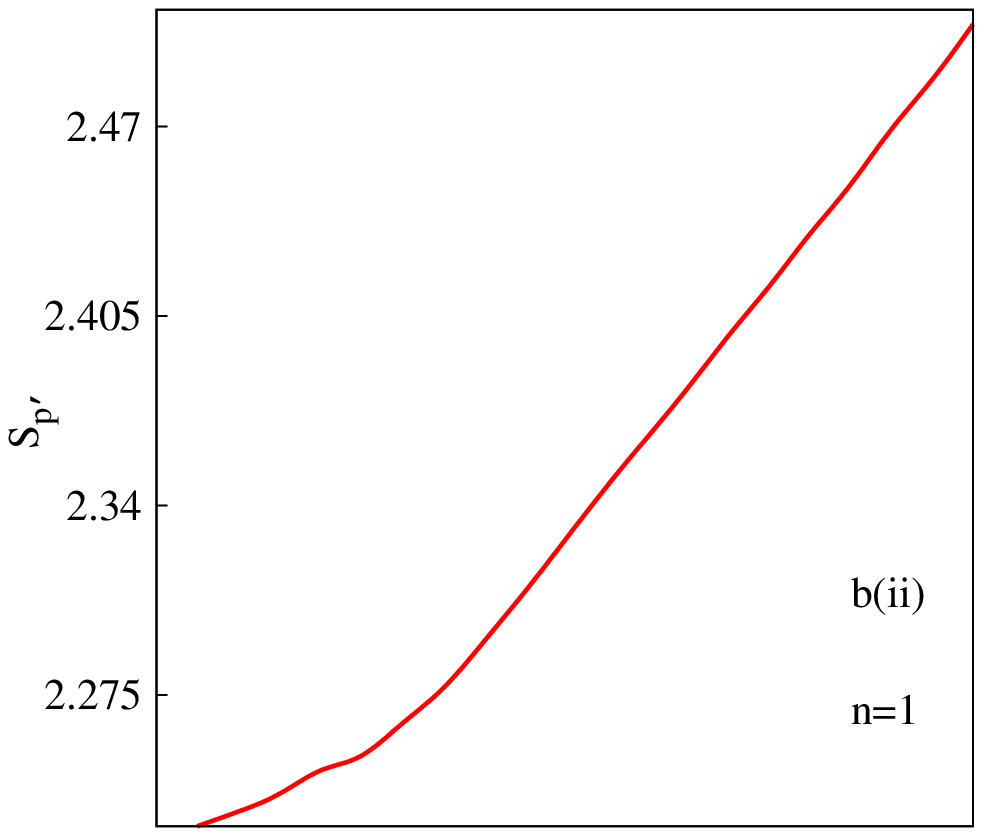}
\end{minipage}%
\hspace{0.45in}
\vspace{0.15in}
\begin{minipage}[c]{0.25\textwidth}\centering
\includegraphics[scale=0.45]{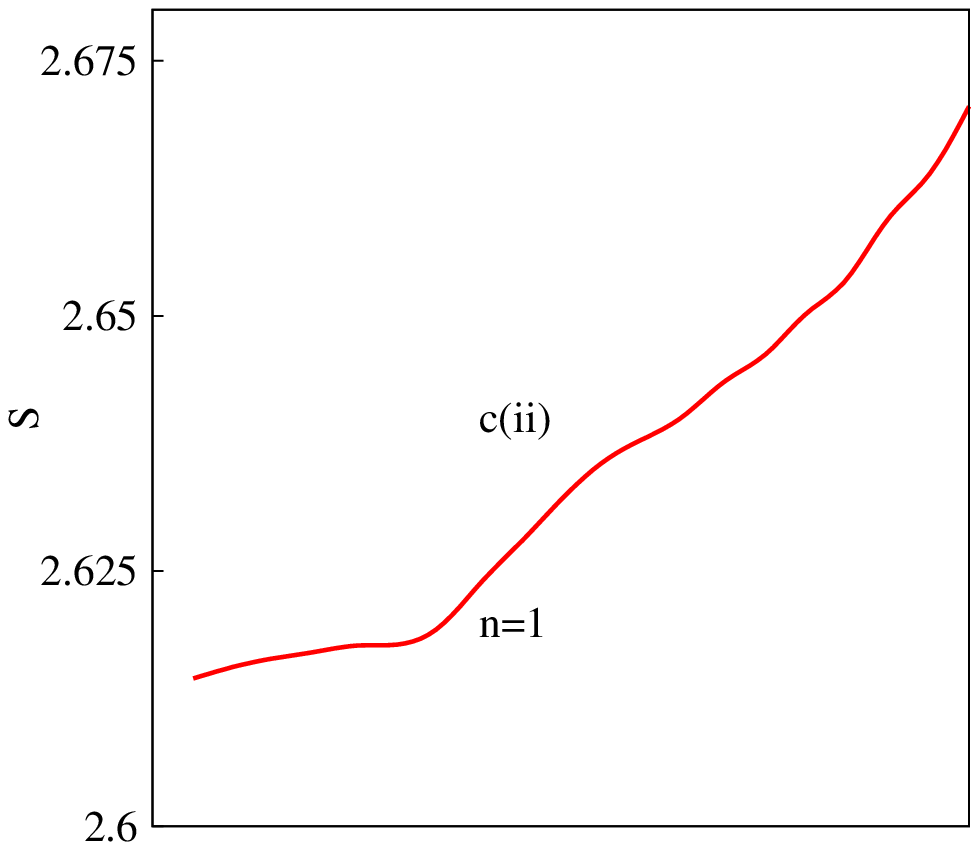}
\end{minipage}%
\hspace{0.45in}
\begin{minipage}[c]{0.25\textwidth}\centering
\includegraphics[scale=0.48]{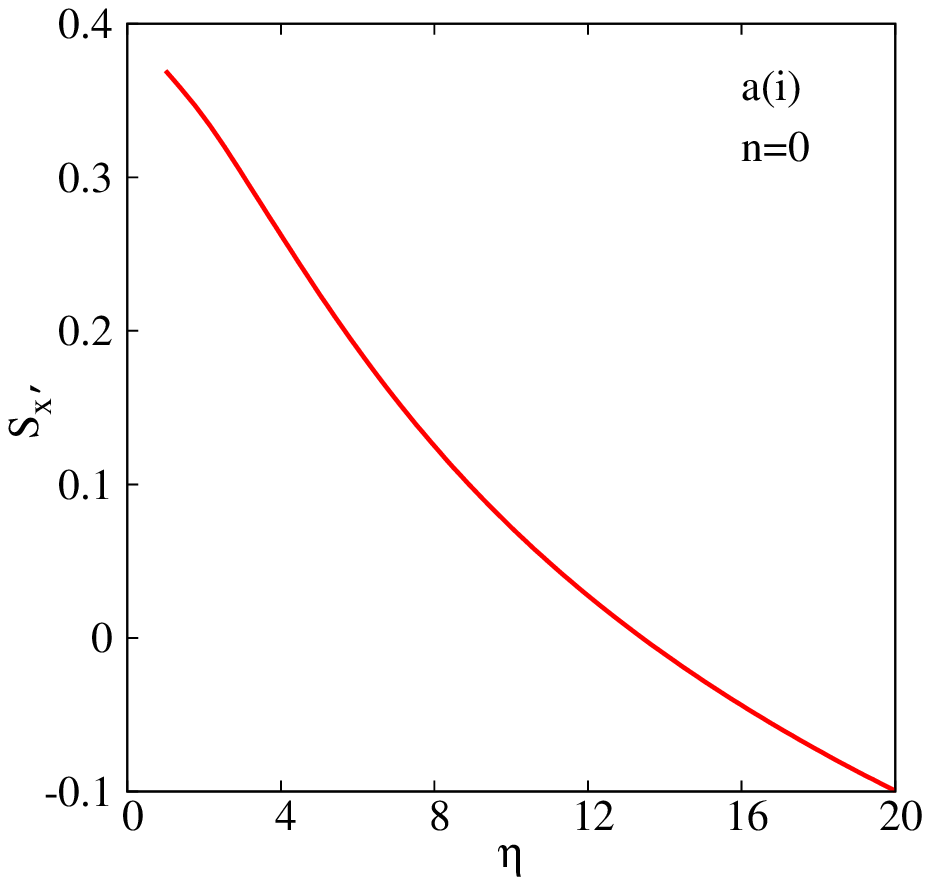}
\end{minipage}%
\hspace{0.45in}
\begin{minipage}[c]{0.25\textwidth}\centering
\includegraphics[scale=0.48]{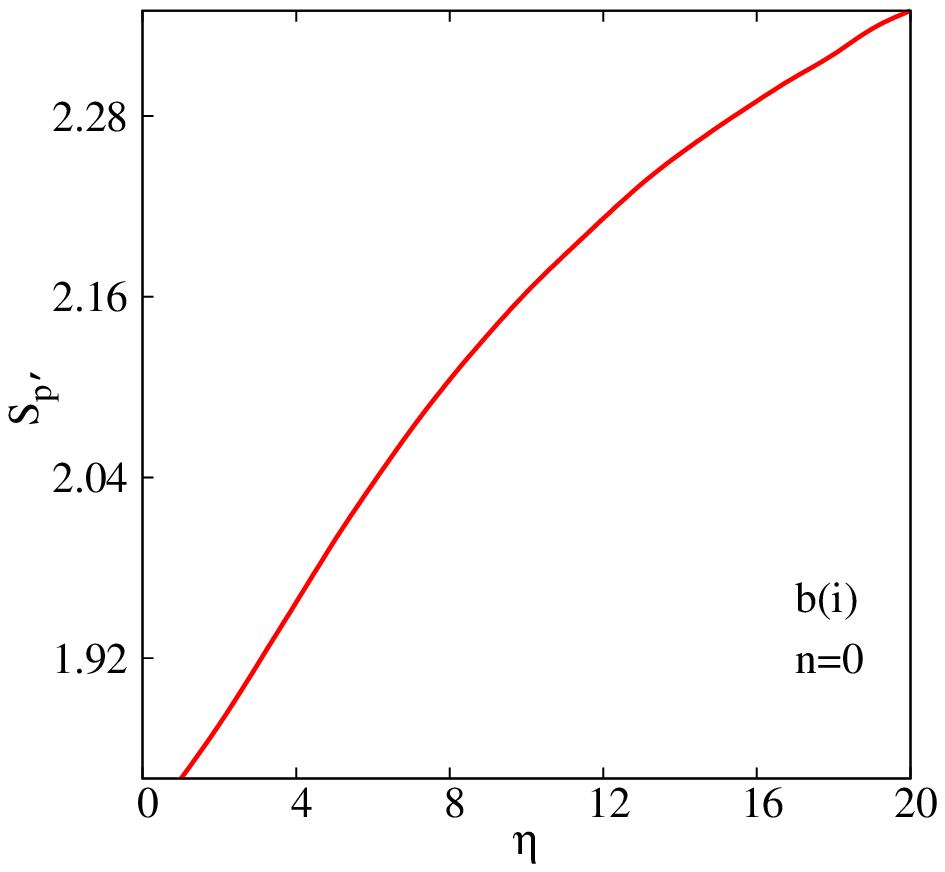}
\end{minipage}%
\hspace{0.45in}
\begin{minipage}[c]{0.25\textwidth}\centering
\includegraphics[scale=0.48]{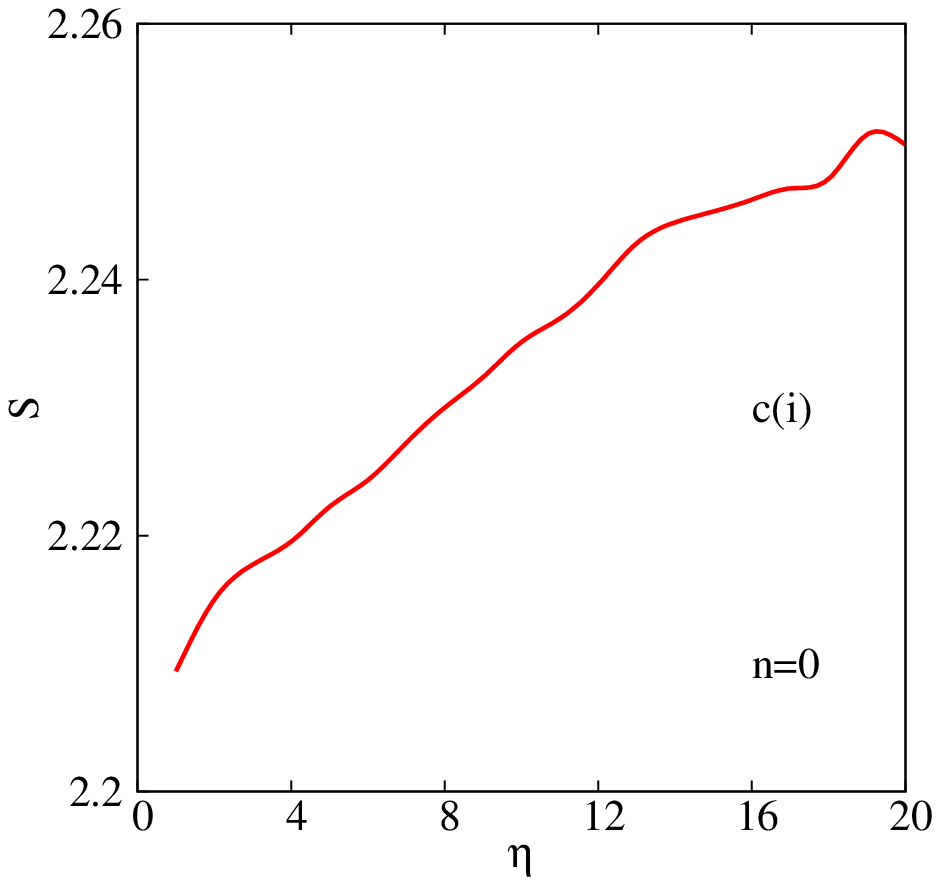}
\end{minipage}%
\caption{Plot of $S_{x'}$, $S_{p'}$, $S$ of first three states of ACHO potential, as function of $\eta$, in left (a), middle (b), 
right (c) columns; (i)--(iii) represent $n=0-2$ states. See text for details.}
\end{figure}

As done in SCHO case, before passing, we would like to mention a few words about the dependency of IE with dimensionless
variable $\eta$. Table~IX shows results for all three measures $I,S,E$, in both $x$ and $p$ space, in an 
ACHO potential, at three selected $\eta$, \emph{viz.,} 5, 10, 20. To get a more complete understanding, we consider 
lowest 5 states, varying $\eta$ from 1-20, which is sufficient for our purpose. Fig.~(13) gives Shannon entropy 
for $n=0-2$ states, including $S_{x'}, S_{p'}$ and $S$. The higher $n$ plots for $S$ are given in Fig.~(S11); analogous 
plots for $I$ and $E$ are offered in Figs.~(S12), (S13) respectively. 
In all these cases, $S_{x'}$ decreases with increase of $\eta$; which clearly indicates localization of particle
in $x$ space. It is interesting to note that at $\eta \rightarrow 0$ regime, $S_{x'}$ becomes independent of $n$, 
because of its PIB-like behavior. Similarly, it can be seen that, $S_{p'}$ increases with $\eta$ signifying de-localization
of particle in $p$ space. The net $S$ for all $n$ increases with $\eta$.    
As usual, $I_{x'}$ for $n=0-2$, increase with $\eta$, but for $n=3,4$ decrease. On the other side, $I_{p'}$ decreases for all $n$. 
There appears a minimum in $n=0$ for $I$, due to a tug-of-war between $x$ and $p$-space quantities. 
In case of $n=1-4$, an increase in $\eta$ causes lowering of $I$. In all cases, $E_{x'}$ increases with $\eta$, 
wheres $E_{p'}$ decreases. But net $E$ decreases for $n=0,1$, and increases for remaining $n$.     

\begin{figure}                         
\begin{minipage}[c]{0.3\textwidth}\centering
\includegraphics[scale=0.65]{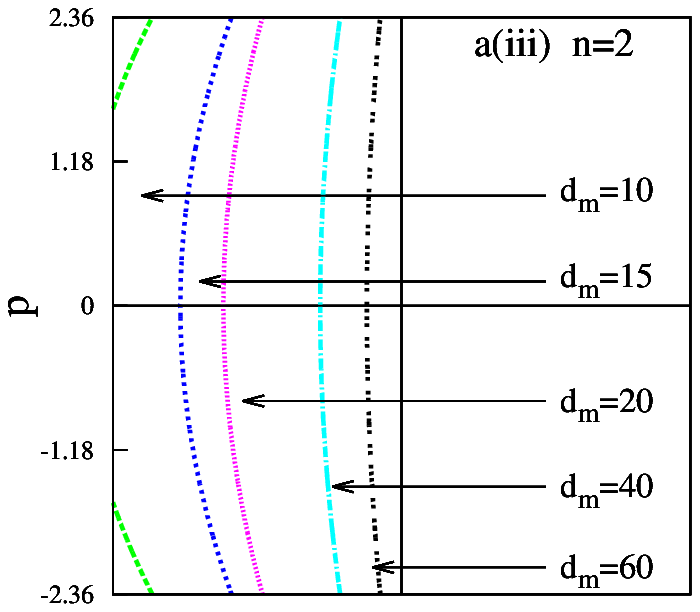}
\end{minipage}
\hspace{0.0in}
\begin{minipage}[c]{0.23\textwidth}\centering
\includegraphics[scale=0.45]{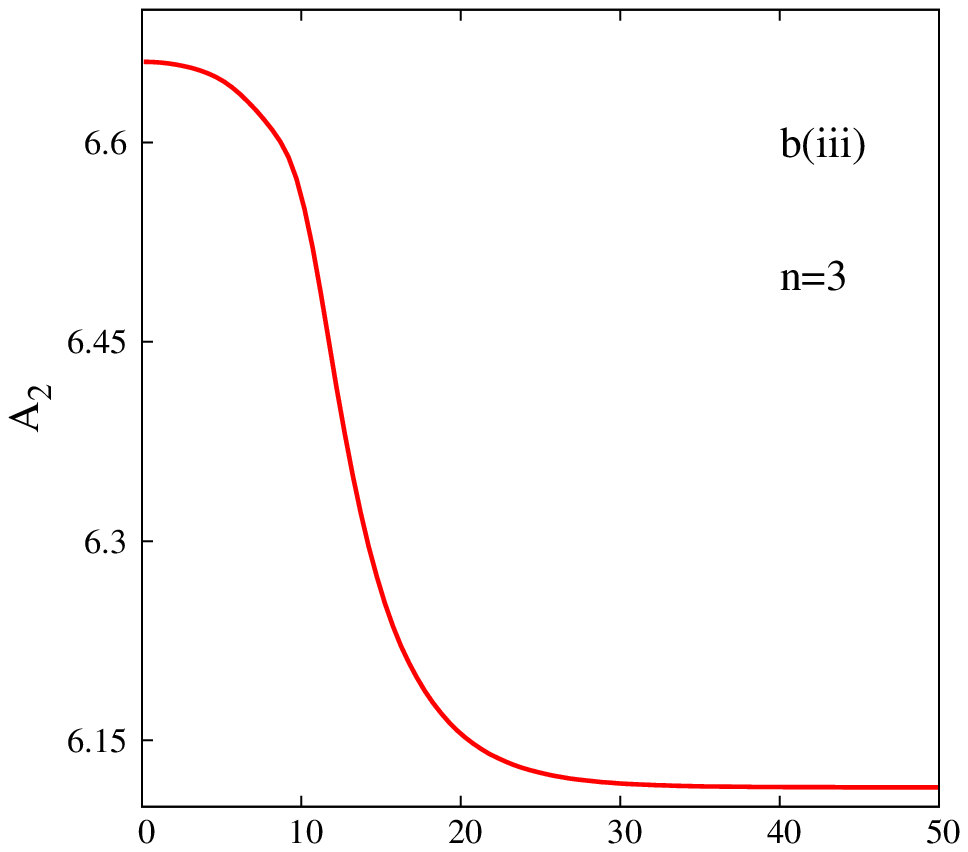}
\end{minipage}
\hspace{0.3in}
\vspace{0.15in}
\begin{minipage}[c]{0.23\textwidth}\centering
\includegraphics[scale=0.45]{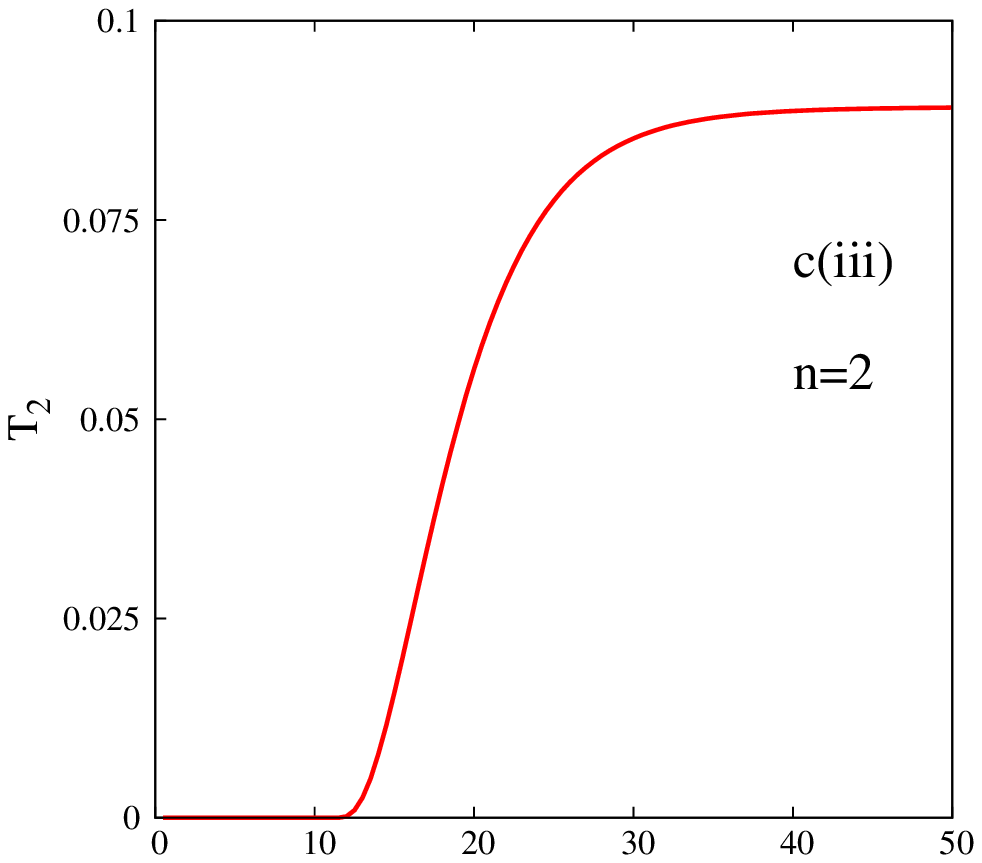}
\end{minipage}
\hspace{0.2in}                      
\begin{minipage}[c]{0.3\textwidth}\centering
\includegraphics[scale=0.65]{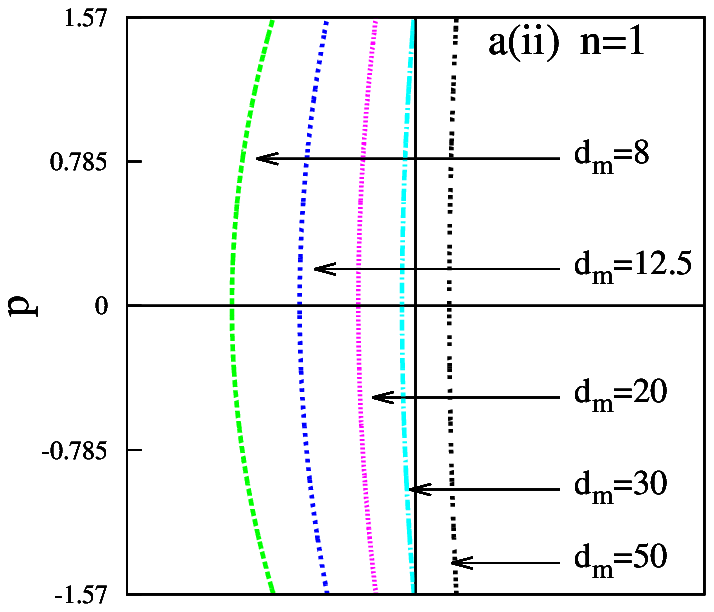}
\end{minipage}
\hspace{0.0in}
\begin{minipage}[c]{0.23\textwidth}\centering
\includegraphics[scale=0.45]{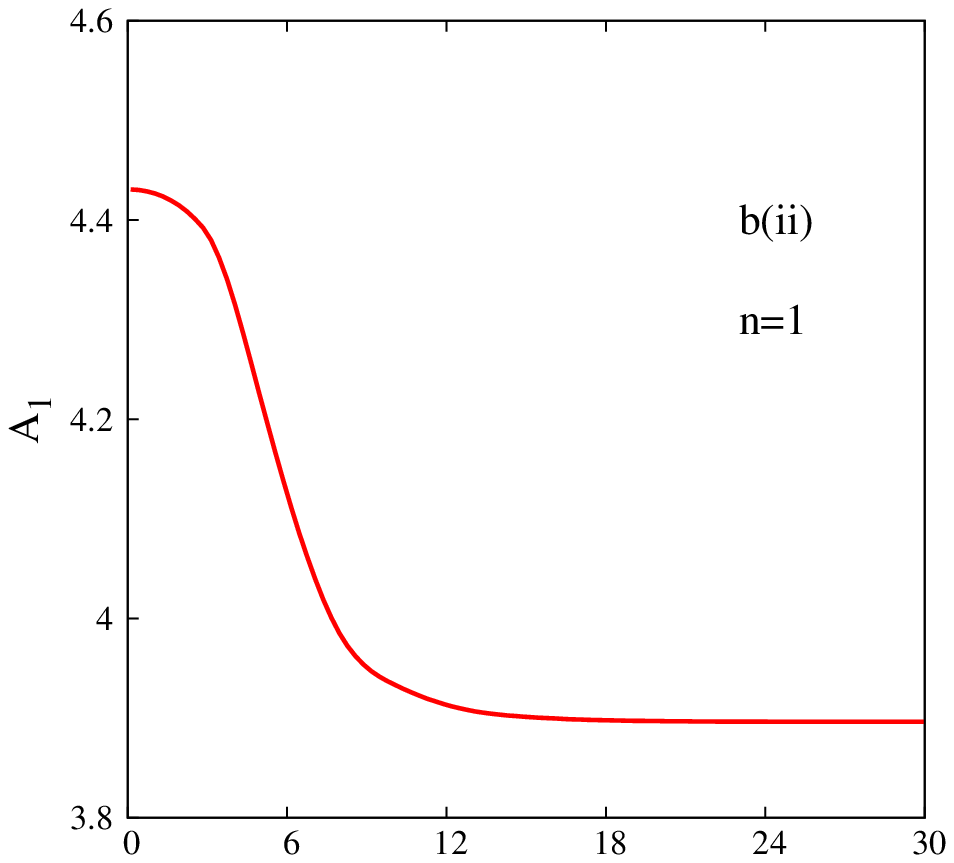}
\end{minipage}
\hspace{0.3in}
\vspace{0.15in}
\begin{minipage}[c]{0.23\textwidth}\centering
\includegraphics[scale=0.45]{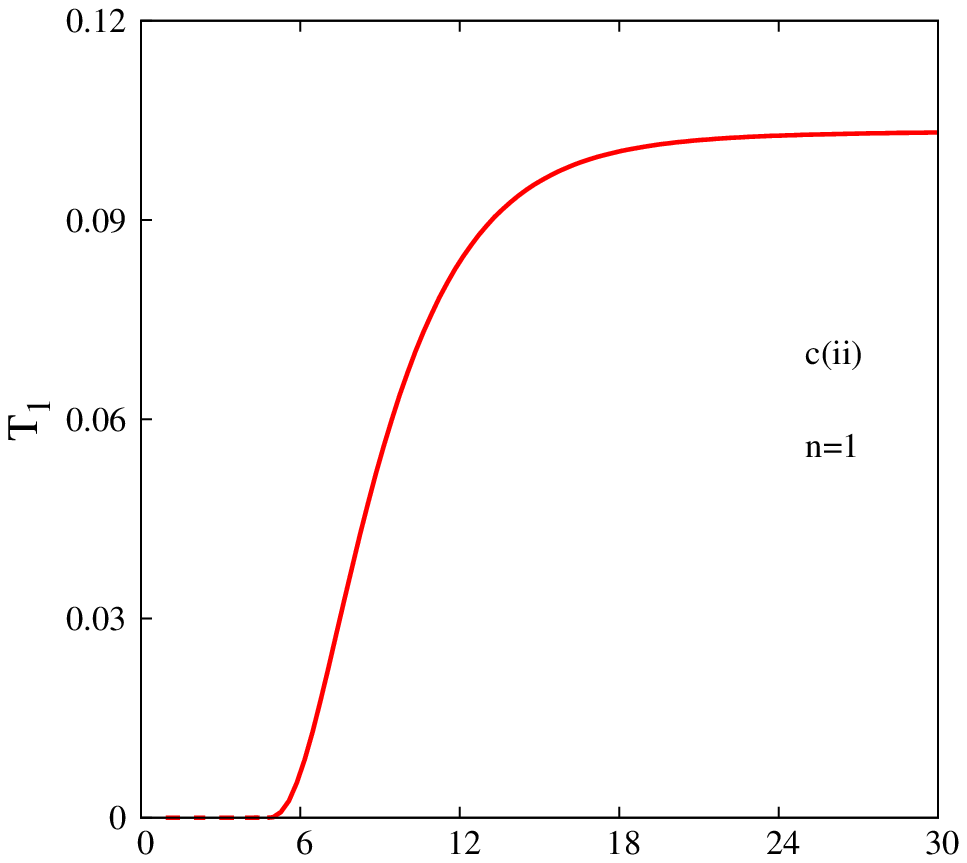}
\end{minipage}
\hspace{0.5in}                      
\begin{minipage}[c]{0.3\textwidth}\centering
\includegraphics[scale=0.65]{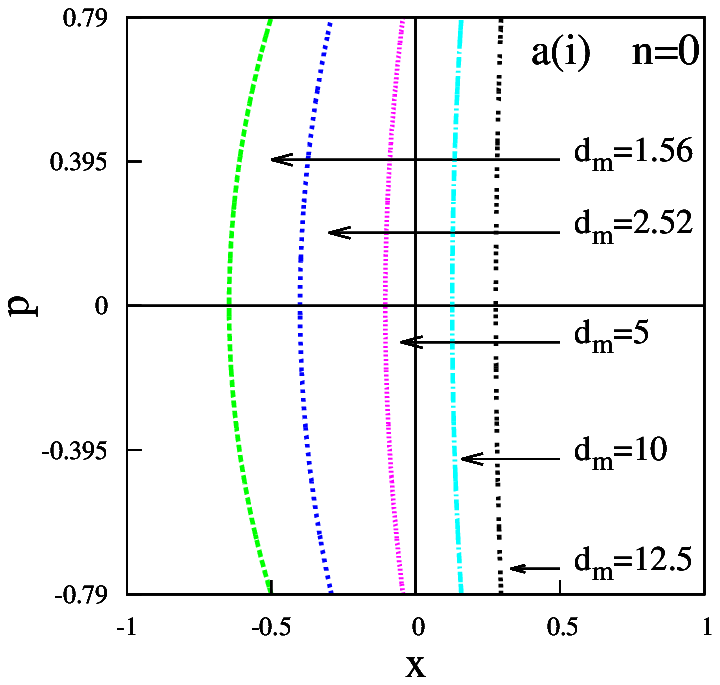}
\end{minipage}
\hspace{0.0in}
\begin{minipage}[c]{0.24\textwidth}\centering
\includegraphics[scale=0.48]{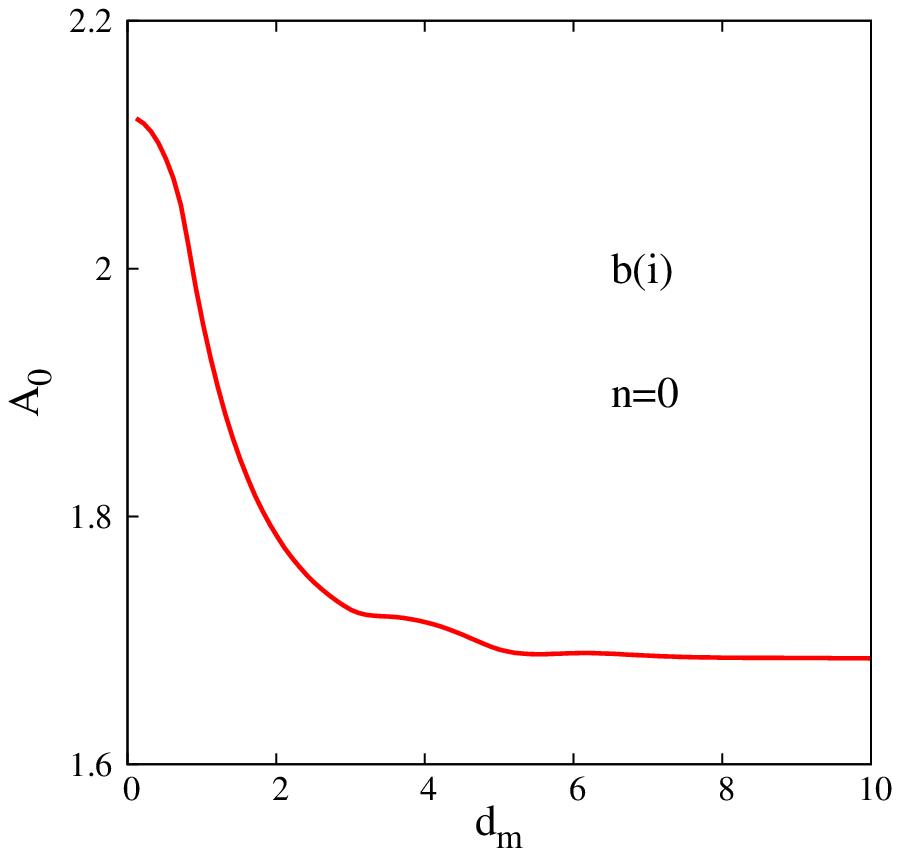}
\end{minipage}
\hspace{0.25in}
\begin{minipage}[c]{0.24\textwidth}\centering
\includegraphics[scale=0.48]{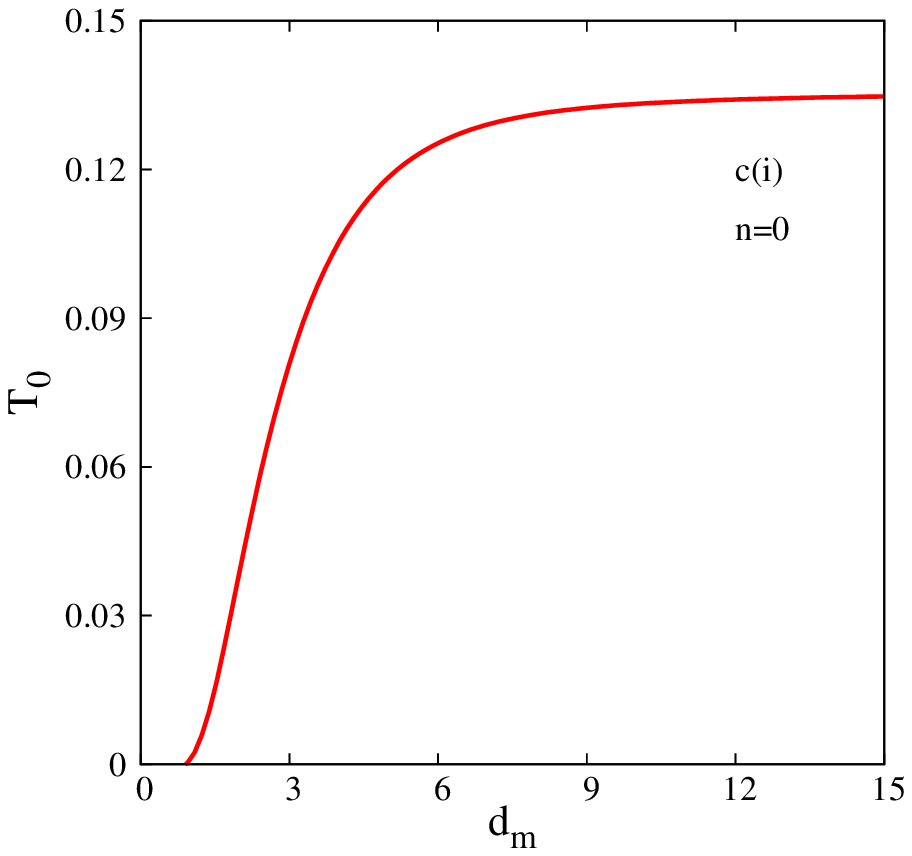}
\end{minipage}
\caption{Plots of phase space, $A_n$, $T_n$ of $n \! = \! 0-2$ states of ACHO potential as function of 
$d_m$, in left (a), middle (b) and right (c) columns; (i)--(iii) represent $n \! = \! 0-2~$ states. For details, see text.}
\end{figure}
  
Finally, Fig.~(14) and Fig.~(S14) offer a semi-classical viewpoint of ACHO potential from a phase-space analysis in terms of 
phase space characteristics, $A_n$ and $T_n$. Phase-space diagrams for $n \! = \! 0-2$ are displayed in panels a(i)--a(iii); 
in each case, five representative $d_m$ is chosen. Shapes of phase space changes with $d_m$. Third and fourth states are given in 
Fig.~(S14). Initially at small values of 
$d_m$, ACHO potential behaves as PIB. However, as $d_m$ increases, more localization tends to take place 
indicating deviation from a PIB situation. For $n \! = \! 0$, chosen values of $d_m$ are 1.56, 2.52, 5, 10 and 12.5;
in this case tunneling sets in at around $d_m \approx 0.9-1$. A similar analysis on other four states in 
a(ii)--a(iii) and Fig.~(S14) (a(i)--a(ii)) advocates that, bounded area decreases with $d_m$; moreover phase spaces 
gradually tend to assume rectangular shape. Tunneling for $n \! = \! 1-4$ occurs at around $d_m \approx 4.75, 11.5, 21$ and 33.5. 
From middle panels b(i)--b(iii) and Fig.~(S14) (b(i)--b(ii)), we observe that $A_n$ decreases as $d_m$ increases. $A_n$ in 
$n \! = \! 0$, jumps down from a higher to lower value when $d_m$ falls in the range 1.3--1.5. Equivalent jumps 
happen at approximate $d_m$ ranges of 3--5, 10--12.5, 20--22.5 and 32.5--35 respectively, for 
$n \! = \! 1-4$. These characteristic $d_m$ values mark the beginning of tunneling in these states. Lastly, from
$T_n$ plots in right-most panels, c(i)--c(iii) and Fig.~(S14) (c(i)-c(ii)), it is seen that, tunneling begins at a state-dependent 
threshold value and then increases as $d_m$ progresses. But at larger $d_m$ region, rate of tunneling falls off steadily showing 
very insignificant changes in $T_n$ plots.    

\section{Conclusion}
Information-based uncertainty measures like $I,S,E$ have been employed to understand the effect of confinement 
in SCHO and ACHO. SCHO may be treated as an interim model between PIB and a QHO, whereas, asymmetric confinement
in ACHO may be probed starting from a SCHO ($d_m \! = \! 0$). Accurate eigenvalues and eigenfunctions are obtained
from ITP method in both cases. For ACHO model, additionally, we also introduced a simple variation method utilizing the SCHO 
basis set, which offers excellent results. Nature of changes of $I_x, I_p, I$ complement the behavior of traditional
uncertainties, $\Delta x, \Delta p, \Delta x \Delta p$.  

In case of ACHO, an increase in $d_m$ value leads to localization in $x$ space and delocalization in $p$ space. 
While $S_{x}, S_{p}, S$ can completely explain the competing phenomena in ACHO, in contrast, $I_{x}, I_{p}, I$ 
proves insufficient for such situation. Further, $E_x, E_p, E$ offer important information which helps us 
for a better understanding of such systems. Besides, from a semi-classical perspective, it is found that, $A_n$ decreases 
with $x_c$ in SCHO and maintains the same trend with $d_m$ in ACHO. In the former, one observes delocalization, whereas 
localization occurs in the latter; in either situation, tunneling increases leading to a decrease in $A_n$. 
Further, in SCHO potential it is found that, $x_c$ and $\eta$ produce opposite effect on IE measures. However, in an ACHO,   
$\eta$ and $d_m$ cause similar changes in IE. 

\section{Acknowledgement} 
AG is grateful to UGC for a Junior Research Fellowship (JRF). NM acknowledges IISER Kolkata for a post-doctoral 
fellowship. The authors thank Dr.~A.~K.~Tiwari for useful discussion. Financial support from DST SERB 
(EMR/2014/000838) is gratefully acknowledged. Effort of Mr.~Suman K.~Chakraborty is appreciated, for his assistance 
in computers. Constructive comments from three anonymous referees have been very helpful.

\end{document}